\documentclass[aps, prx, reprint, superscriptaddress]{revtex4-2}
\pdfoutput=1
\usepackage{braket,dsfont,subfigure,amsfonts,amssymb,bm,graphicx,float,amsmath,amsthm,txfonts}
\usepackage[colorlinks]{hyperref}
\usepackage[dvipsnames]{xcolor}
\usepackage{enumitem}
\usepackage{multirow}
\usepackage{orcidlink}
\newcommand{\Tr}{{\rm Tr}}

\newcommand{\sslash}{\textbackslash \hspace{-1.3mm} \textbackslash}

\newcommand{\Z}{\mathbb{Z}}

\newcommand{\red}[1]{{\textcolor{red}{#1}}}
\newcommand{\blue}[1]{{\textcolor{blue}{#1}}}
\definecolor{pblue}{RGB}{1, 25, 147}
\definecolor{pgreen}{RGB}{0, 143, 0}
\definecolor{pred}{RGB}{148, 23, 81}
\newcommand{\pblue}[1]{{\textcolor{pblue}{#1}}}
\newcommand{\pgreen}[1]{{\textcolor{pgreen}{#1}}}
\newcommand{\pred}[1]{{\textcolor{pred}{#1}}}
\begin{document}

\title{Quantum criticality in open quantum systems from the purification perspective}
\author{Yuchen Guo~\orcidlink{0000-0002-4901-2737}}
\affiliation{State Key Laboratory of Low Dimensional Quantum Physics and Department of Physics, Tsinghua University, Beijing 100084, China}
\author{Shuo Yang~\orcidlink{0000-0001-9733-8566}}
\email{shuoyang@tsinghua.edu.cn}
\affiliation{State Key Laboratory of Low Dimensional Quantum Physics and Department of Physics, Tsinghua University, Beijing 100084, China}
\affiliation{Frontier Science Center for Quantum Information, Beijing 100084, China}
\affiliation{Hefei National Laboratory, Hefei 230088, China}

\begin{abstract}
Open quantum systems host mixed-state phases that go beyond the symmetry-protected topological and spontaneous symmetry-breaking paradigms established for closed, pure-state systems. 
Developing a unified and physically transparent classification of such phases remains a central challenge. 
In this work, we introduce a purification-based framework that systematically characterizes all mixed-state phases in one-dimensional systems with $\mathbb{Z}_2^{\sigma} \times \mathbb{Z}_2^{\tau}$ symmetry. 
By introducing an ancillary $\kappa$ chain and employing decorated domain-wall constructions, we derive eight purified fixed-point Hamiltonians labeled by topological indices $(\mu_{\sigma\tau},\mu_{\tau\kappa},\mu_{\kappa\sigma}) \in \{\pm1\}^3$.
Tracing out the ancilla recovers the full structure of mixed-state phases, including symmetric, strong-to-weak spontaneous symmetry breaking, average symmetry-protected topological phases, and their nontrivial combinations. 
Interpolations between the eight fixed points naturally define a three-dimensional phase diagram with a cube geometry. 
The edges correspond to elementary transitions associated with single topological indices, while the faces host intermediate phases arising from competing domain-wall decorations. 
Along the edges, we identify a class of critical behavior that connects distinct strong-to-weak symmetry-breaking patterns associated with distinct strong subgroups, highlighting a mechanism unique to mixed-state settings. 
Large-scale tensor-network simulations reveal a rich phase structure, including pyramid-shaped symmetry-breaking regions and a fully symmetry-broken phase at the cube center. 
Overall, our purification approach provides a geometrically transparent and physically complete classification of mixed-state phases, unified with a single $\mathbb{Z}_2^{\sigma} \times \mathbb{Z}_2^{\tau} \times \mathbb{Z}_2^{\kappa}$ model.
\end{abstract}

\maketitle

\tableofcontents

\section{Introduction}
In recent years, the study of quantum phase transitions~\cite{Sachdev1999, Vojta2003, Sachdev2011} has expanded to open quantum systems~\cite{Breuer2007, Weiss2012, Rivas2012}, where coupling to the environment induces decoherence and dissipation such that mixed states, rather than pure states, provide the natural description.  
This distinction has triggered a surge of activity in the classification of phases in open systems, including systematic studies of intrinsic topological order~\cite{Wang2024, Wang2025A, Sohal2025, Ellison2025, You2024, Luo2025}.
Moreover, in this broader setting, the notion of symmetry itself becomes enriched. 
In addition to conventional \emph{strong symmetry} $K\rho=e^{i\theta}\rho$, one may also encounter a \emph{weak symmetry}, defined by invariance of the density matrix under the simultaneous left and right actions of symmetry operators, $U\rho U^{\dagger} = \rho$~\cite{Buca2012, Ma2023, Huang2025A}, leading to the discovery of fundamentally new orders without pure-state counterparts.
Prominent examples are the so-called intrinsic average symmetry-protected topological (ASPT) order~\cite{Ma2023, Guo2025B, Xue2024, Ma2025A, Zhang2025}, which arises from nontrivial group extensions between strong and weak symmetry, and strong-to-weak spontaneous symmetry breaking (SWSSB)~\cite{Lee2023, Lessa2025, Sala2024, Gu2024, Huang2025, Orito2025, Zhang2025A, Sun2025A}, where a strong symmetry breaks into a weak one, leaving long-range order only in R\'enyi-$2$ or fidelity correlators~\cite{Liu2025B, Weinstein2025, Sun2025, Lee2025}.

A central question is how to characterize phase transitions between distinct phases of matter.
Several classes of phase transitions driven by dissipative dynamics have been systematically studied, including separability transitions of topologically ordered states~\cite{Lu2024, Fan2024, Chen2024, Sang2024, Sang2025A, Lee2025A} and decoherence-induced trivial-to-SWSSB transition~\cite{Lessa2025}.
In these cases, the Markov length diverges at the critical point, providing a powerful diagnostic of mixed-state phase transitions~\cite{Sang2025A, Sang2025B, Negari2025}.
Alternatively, one may adopt a \emph{bottom-up} approach, starting from a set of fixed-point density matrices representative of different phases and constructing imaginary-time Lindbladian evolutions that converge to these mixed states.
Subsequently, linear interpolations between such fixed points can be used to generate the full phase diagram and to reveal critical behavior between different phases through the gap closing of the imaginary-Liouville superoperator~\cite{Guo2025C, Guo2025A}.

In this work, we establish a unified purification-based framework for quantum phases in one-dimensional (1D) open systems.
Our main results can be summarized as follows:
\begin{itemize}
    \item \textbf{Purified fixed-point construction.}  
    We construct eight purified fixed-point Hamiltonians for 1D systems with $\Z_2^{\sigma}\times \Z_2^{\tau}\times \Z_2^{\kappa}$ symmetry, and show that tracing out the ancillary $\kappa$ degree of freedom generates all mixed-state phases of open systems with $\Z_2^{\sigma}\times \Z_2^{\tau}$ symmetry, including ASPT, SWSSB, and double ASPT phase (the coexistence of ASPT and SWSSB~\cite{Guo2025A, Kuno2025a}).
    A $\sigma\tau$-preserved phase deserves special attention, where both $\Z_2^{\sigma}$ and $\Z_2^{\tau}$ symmetries are broken, while the diagonal $\Z_2^{\sigma\tau}$ symmetry remains strong. 

    \item \textbf{Transitions between different SWSSB phases.}  
    We identify and characterize a class of transitions between distinct SWSSB phases, in which the identity of the remaining strong symmetry subgroup changes across the critical point.
    These transitions are detected by competing order parameters defined from R\'enyi-2 correlators, and are confirmed by both analytical arguments and numerical simulations.
    Enriched by topological decoration, we further reveal an intrinsic open-system phenomenon where the topological structure on one side is hidden, rather than eliminated, when crossing the critical point due to the transition of SWSSB.

    \item \textbf{Emergent symmetry breaking in higher-dimensional interpolations.}  
    Extending beyond edges, we map out the phase structure on the faces and in the bulk of the phase cube.
    We find pyramid-shaped SSB volumes emerging from six faces, as well as a fully symmetry-broken phase at the cube center, revealing a rich hierarchy of emergent symmetry breaking in open-system settings.
\end{itemize}

Together, these results demonstrate how purification provides a natural and systematic route to constructing, classifying, and diagnosing quantum phases and criticality in open quantum systems.

\section{Purification of mixed-state quantum phases}
In this section, we construct purified quantum states for the representative density matrices of several mixed-state quantum phases.
The physical degrees of freedom are represented by $\sigma$ and $\tau$ spins, while the ancillary $\kappa$ spins belong to an environment to be traced out
\begin{align}
    \rho = \Tr_{\kappa}\left[\ket{\psi}\bra{\psi}\right].
\end{align}

To begin with, we first discuss the realization of trivial weakly symmetric states and mixed states with SWSSB order, where the system only consists of a single $\Z_2$ degree of freedom.
The essential distinction between them lies in how the strong symmetry is broken, namely, explicitly or spontaneously.
This difference is revealed at the level of their purifications. 

\subsection{Trivial weakly symmetric states}
The maximally mixed state on $N$ sites satisfies the weak symmetry associated with any symmetry operator
\begin{align}
    &\rho^{\rm W}_{\rm \sigma} = \frac{1}{2^N}\prod_i \sigma_i^0,\label{Equ: W}\\
    &U\rho^{\rm W}_{\sigma} U^{\dagger} =\rho^{\rm W}_{\sigma},
\end{align}
where the density matrix no longer satisfies the strong symmetry, which is explicitly broken to a weak one.
A convenient purification of this state is given by a product of Bell pairs
\begin{align}
    \ket{\psi^{\rm Bell}_{\kappa\sigma}} = \frac{1}{2^{N/2}}\prod_i \left(\ket{\uparrow_{\sigma_i} \uparrow_{\kappa_i}}+\ket{\downarrow_{\sigma_i} \downarrow_{\kappa_i}}\right),
    \label{Equ: Bell}
\end{align}
where tracing out the environment ($\kappa$ spins) yields $\rho^{\rm W}$ that is weakly symmetric.
In this case, the invariance of the density operator is enforced by explicit local compensation, where the environment absorbs the system charge at each site, and no long-range order survives.
From this perspective, the environment functions as a local charge reservoir that keeps track of symmetry transformations.

\subsection{SWSSB}
\label{Sec: SWSSB}
The fixed-point density matrix characterizing the $\Z_2$ SWSSB phase is given by
\begin{align}
    \rho^{\rm SWSSB}_{\sigma} = \frac{1}{2^N}\left(\prod_i\sigma_i^0+\prod_i \sigma_i^x\right),
    \label{Equ: SWSSB}
\end{align}
where the strong symmetry is restored only after summing over different symmetry sectors, i.e., the spontaneous breaking of strong symmetry.
This mixed state can be purified to a 1D cluster state with nontrivial SPT order protected by $\Z_2^{\sigma} \times \Z_2^{\kappa}$ symmetry that follows the decorated domain-wall (DW) construction~\cite{Chen2014, Sala2024},
\begin{align}
\begin{aligned}
    \ket{\psi^{\rm SPT}_{\kappa\sigma}} &= \frac{1}{2^{N/2}}\sum_{\{\sigma_i\}}\ket{\cdots\uparrow_{\sigma}\rightarrow_{\kappa}\uparrow_{\sigma}\leftarrow_{\kappa}\downarrow_{\sigma}\leftarrow_{\kappa}\uparrow_{\sigma}\cdots}\\
    & = \frac{1}{2^{N/2}}\sum_{\{\sigma_i\}}\ket{\psi_{\kappa\sigma}^{\rm DW }}_{\{\sigma_i\}}
    \label{Equ: Cluster}
\end{aligned}
\end{align}
where for each configuration $\{\sigma_i\}$, excitations of $\kappa$ spins are placed on the domain wall of $\sigma$ spins, as denoted by $\ket{\psi_{\kappa\sigma}^{\rm DW }}_{\{\sigma_i\}}$.

\subsubsection{Physical picture}
In this cluster-state purification, the environment no longer compensates for symmetry charges locally.
Instead, it stores nonlocal domain-wall information that distinguishes different global symmetry sectors.
The reduced density matrix can then be viewed as a macroscopic mixture of these sectors, in which the strong symmetry is restored only after summation.
As a consequence, long-range order appears in fidelity or R\'enyi-$2$ correlators~\cite{Lessa2025, Guo2025A}, corresponding to the spontaneous breaking of a strong symmetry down to a weak one.

\subsubsection{Fidelity correlator}
The parent Hamiltonian of the cluster state in Eq.~\eqref{Equ: Cluster} is a stabilizer Hamiltonian
\begin{align}
    H^{\rm SPT}_{\kappa\sigma} = -\sum_i \left[\kappa_{i-1}^z\sigma_i^x\kappa_i^z+ \sigma_i^z\kappa_i^x\sigma_{i+1}^z\right],
    \label{Equ: SPT}
\end{align}
whose ground state is the simultaneous eigenstate of each local term with eigenvalue $1$.
By defining a string order parameter
\begin{align}
    S(O^1, O^2, i, j) \equiv O^1_i\!\left(\prod_{l=i}^{j-1}O^2_l\right)O^1_j,
\end{align}
we note that $\ket{\psi^{\rm SPT}_{\kappa\sigma}}$ exhibits a nonlocal string order~\cite{Pollmann2012}
\begin{align}
    \lim_{|i-j|\rightarrow \infty} |\braket{S(\sigma^z, \kappa^x, i, j)}| =1,
\end{align}
which serves as an indicator of the SPT order.
Regarding the reduced state $\rho^{\rm SWSSB}_{\sigma}$, we consider its fidelity correlator
\begin{align}
    \mathcal{C}^{\mathcal{F}}\left(O, i, j\right)\equiv \mathcal{F}\left(O_iO_j\rho O_iO_j, \rho\right),
\end{align}
where $\mathcal{F}$ denotes the fidelity between two mixed states
\begin{align}
    \mathcal{F}\left(\rho_1, \rho_2\right) = \left(\Tr\sqrt{\sqrt{\rho_1}\rho_2\sqrt{\rho_1}}\right)^2.
\end{align}

To establish the relationship between the string order of $\ket{\psi^{\rm SPT}_{\kappa\sigma}}$ and the fidelity correlator of $\rho^{\rm SWSSB}_{\sigma}$, we introduce two pure states
\begin{align}
    \ket{\psi_1} \equiv \sigma_i^z\sigma_j^z\ket{\psi^{\rm SPT}_{\kappa\sigma}}, \quad \ket{\psi_2} \equiv \prod_{l=i}^{j-1}\kappa_l^x\ket{\psi^{\rm SPT}_{\kappa\sigma}}.
\end{align}
In this way, the above string order parameter can be expressed using the fidelity between these two states, i.e.,
\begin{align}
    \begin{aligned}
        \left|\braket{S(\sigma^z, \kappa^x, i, j)}\right| &= \left|\braket{\psi_1|\psi_2}\right| = \sqrt{\mathcal{F}\left(\ket{\psi_1}\hspace{-0.5mm}\bra{\psi_1}, \ket{\psi_2}\hspace{-0.5mm}\bra{\psi_2}\right)}\\
         &\leq \sqrt{\mathcal{F}\left(\Tr_{\kappa}{[\ket{\psi_1}\hspace{-0.5mm}\bra{\psi_1}]}, \Tr_{\kappa}{[\ket{\psi_2}\hspace{-0.5mm}\bra{\psi_2}]}\right)}\\
         & = \sqrt{\mathcal{F}\left(\sigma_i^z\sigma_j^z \rho^{\rm SWSSB}_{\sigma}\sigma_i^z\sigma_j^z, \rho^{\rm SWSSB}_{\sigma}\right)}\\
         & = 
        \sqrt{\mathcal{C}^{\mathcal{F}}\left(\sigma^z, i, j\right)},
    \end{aligned}\label{Equ: String-Fidelity}
\end{align}
where the inequality follows from the monotonicity of fidelity under partial trace~\cite{Nielsen2009}.
Therefore, if the original pure state has a long-range string order
\begin{align}
    \lim _{|i-j|\rightarrow \infty}|\braket{S(\sigma^z, \kappa^x, i, j)}| \rightarrow O(1),
\end{align}
then the reduced state is long-range correlated in terms of the fidelity correlator
\begin{align}
    \lim _{|i-j|\rightarrow \infty}\mathcal{C}^{\mathcal{F}}\left(\sigma^z, i, j\right) \rightarrow O(1)
\end{align}
and thus exhibits SWSSB order~\cite{Lessa2025}.
It is worth noting that, due to the inequality relation in Eq.~\eqref{Equ: String-Fidelity}, the converse proposition cannot be proved.
This problem is solved with the following tensor-network analysis.

\subsubsection{R\'enyi-$2$ correlator}
The R\'enyi-$2$ correlator provides an alternative diagnostic of SWSSB order in mixed states that is tractable for large-scale numerical simulations and experimental detection~\cite{Lessa2025, Guo2025A}.
It is defined as 
\begin{align}
    \mathcal{C}^{(2)}(O, i, j)\equiv\frac{\Tr[\rho O_iO_j\rho O_iO_j]}{\Tr[\rho^2]} - \frac{\Tr[\rho O_i\rho O_i]}{\Tr[\rho^2]}\frac{\Tr[\rho O_j\rho O_j]}{\Tr[\rho^2]}.
\label{Equ: Renyi-2}
\end{align}

To analyze whether a mixed state exhibits long-range R\'enyi-$2$ correlations, we employ the tensor-network representation known as the locally purified density operator (LPDO)~\cite{Verstraete2004, Werner2016, Cheng2021, Guo2024A, Guo2024B}
\begin{align}
\scalebox{0.8}{
\begin{tikzpicture}[scale=0.75]
\tikzstyle{sergio}=[rectangle,draw=none]
\filldraw[fill=white, draw=black, rounded corners] (-0.25,-0.5)--(1.25,-0.5)--(1.25,0.5)--(-0.25,0.5)--cycle;
\filldraw[fill=white, draw=black, rounded corners] (2.25,-0.5)--(3.75,-0.5)--(3.75,0.5)--(2.25,0.5)--cycle;
\draw[line width=1pt] (1.25,0) -- (2.25,0);
\draw[line width=1pt] (-0.25,0) -- (-1.25,0);
\path (0.5,0) node [style=sergio]{\large $\mathsf{A}$};
\path (3,0) node [style=sergio]{\large $\mathsf{A}$};
\draw[line width=1pt] (3.75,0) -- (4.75,0);
\draw[line width=2pt, color=red] (0.5,0.5) -- (0.5,1);
\draw[line width=2pt, color=red] (0.5,-2.5) -- (0.5,-3);
\draw[line width=3pt, color=blue] (0.5,-1.5) -- (0.5,-0.5);
\path (0.2,1) node [style=sergio]{\large p};
\path (0.2,-1) node [style=sergio]{\large a};
\draw[line width=2pt, color=red] (3,-2.5) -- (3,-3);
\draw[line width=2pt, color=red] (3,0.5) -- (3,1);
\draw[line width=3pt, color=blue] (3,-1.5) -- (3,-0.5);
\path (2.7,1) node [style=sergio]{\large p};
\path (2.7,-1) node [style=sergio]{\large a};
\path (-1.75,-1) node [style=sergio]{$\cdots$};
\path (5.25,-1) node [style=sergio]{$\cdots$};
\path (-2.5,-1) node [style=sergio]{\large $\rho=$};
\filldraw[fill=white, draw=black, rounded corners] (2.25,-2.5)--(3.75,-2.5)--(3.75,-1.5)--(2.25,-1.5)--cycle;
\filldraw[fill=white, draw=black, rounded corners] (-0.25,-2.5)--(1.25,-2.5)--(1.25,-1.5)--(-0.25,-1.5)--cycle;
\path (0.2,-3) node [style=sergio]{\large p};
\path (2.7,-3) node [style=sergio]{\large p};
\path (3,-2) node [style=sergio]{\large $\mathsf{A}^*$};
\path (0.5,-2) node [style=sergio]{\large $\mathsf{A}^*$};
\draw[line width=1pt] (1.25,-2) -- (2.25,-2);
\draw[line width=1pt] (-0.25,-2) -- (-1.25,-2);
\draw[line width=1pt] (3.75,-2) -- (4.75,-2);
\end{tikzpicture}},
\label{Equ: LPDO}
\end{align}
which represents both the physical (p) and ancillary (a) degrees of freedom (Appendix~\ref{Sec: Appendix-A}).
An LPDO naturally encodes a purification as a matrix product state (MPS)~\cite{Perez2007, Orus2014, Cirac2021}
\begin{align}
\scalebox{0.8}{
\begin{tikzpicture}[scale=0.75]
\tikzstyle{sergio}=[rectangle,draw=none]
\filldraw[fill=white, draw=black, rounded corners] (-0.75,-0.5)--(0.75,-0.5)--(0.75,0.5)--(-0.75,0.5)--cycle;
\filldraw[fill=white, draw=black, rounded corners] (1.75,-0.5)--(3.25,-0.5)--(3.25,0.5)--(1.75,0.5)--cycle;
\draw[line width=1pt] (0.75,0) -- (1.75,0);
\draw[line width=1pt] (-0.75,0) -- (-1.75,0);
\draw[line width=1pt] (3.25,0) -- (4.25,0);
\draw[line width=2pt, color=red] (-0.25,0.5) -- (-0.25,1);
\draw[line width=3pt, color=blue] (0.25,0.5) -- (0.25,1);
\path (-0.5,1.25) node [style=sergio]{\large p};
\path (0.5,1.25) node [style=sergio]{\large a};
\draw[line width=2pt, color=red] (2.25,0.5) -- (2.25,1);
\draw[line width=3pt, color=blue] (2.75,0.5) -- (2.75,1);
\path (0,0) node [style=sergio]{\large $\mathsf{A}$};
\path (2.5,0) node [style=sergio]{\large $\mathsf{A}$};
\path (2,1.25) node [style=sergio]{\large p};
\path (3,1.25) node [style=sergio]{\large a};
\path (-2.25,0) node [style=sergio]{$\cdots$};
\path (4.75,0) node [style=sergio]{$\cdots$};
\path (-3.25,0) node [style=sergio]{\large $|\psi\rangle=$};
\end{tikzpicture}},
\label{Equ: Locally purified}
\end{align}
where each physical index is locally purified by an ancillary index, satisfying
\begin{align}
    \rho = \Tr_{\rm a}[\ket{\psi}\hspace{-0.5mm}\bra{\psi}].
\end{align}

Suppose that the symmetry of the purified state factorizes as a direct product of two finite Abelian groups $K\times G$, acting on different degrees of freedom ($\sigma$ and $\kappa$ spins in our setting), the symmetry transformation of the local tensor reads~\cite{Chen2011, Schuch2011}
\begin{align}
\begin{aligned}
&
\scalebox{0.8}{
\begin{tikzpicture}[scale=0.75]
\tikzstyle{sergio}=[rectangle,draw=none]
\filldraw[fill=white, draw=black, rounded corners] (-0.25,-0.5)--(1.25,-0.5)--(1.25,0.5)--(-0.25,0.5)--cycle;
\draw[line width=1pt] (1.25,0) -- (1.75,0);
\draw[line width=1pt] (-0.25,0) -- (-0.75,0);
\draw[line width=2pt, color=red] (0.25,0.5) -- (0.25,2);
\draw[line width=3pt, color=blue] (5,1) -- (5,0.5);
\path (0.5,0) node [style=sergio]{\large $\mathsf{A}$};
\filldraw[fill=white, draw=black] (0.25,1.25)circle (10pt);
\path (0.25,1.25) node [style=sergio]{$U_k$};
\path (2.125,0) node [style=sergio]{$=$};
\filldraw[fill=white, draw=black, rounded corners] (4,-0.5)--(5.5,-0.5)--(5.5,0.5)--(4,0.5)--cycle;
\draw[line width=1pt] (4,0) -- (2.5,0);
\filldraw[fill=white, draw=black] (3.25,0)circle (10pt);
\path (3.25,0) node [style=sergio]{$V^{-1}_k$};
\path (4.75,0) node [style=sergio]{\large $\mathsf{A}$};
\draw[line width=1pt] (5.5,0) -- (7,0);
\filldraw[fill=white, draw=black] (6.25,0)circle (10pt);
\path (6.25,0) node [style=sergio]{$V_k$};
\draw[line width=2pt, color=red] (4.5,0.5) -- (4.5,1);
\draw[line width=3pt, color=blue] (0.75,2) -- (0.75,0.5);
\path (5,1.25) node [style=sergio]{$\kappa$};
\path (4.5,1.25) node [style=sergio]{$\sigma$};
\end{tikzpicture}}\\
&
\scalebox{0.8}{
\begin{tikzpicture}[scale=0.75]
\tikzstyle{sergio}=[rectangle,draw=none]
\filldraw[fill=white, draw=black, rounded corners] (-0.25,-0.5)--(1.25,-0.5)--(1.25,0.5)--(-0.25,0.5)--cycle;
\draw[line width=1pt] (1.25,0) -- (1.75,0);
\draw[line width=1pt] (-0.25,0) -- (-0.75,0);
\draw[line width=2pt, color=red] (0.25,0.5) -- (0.25,2);
\draw[line width=3pt, color=blue] (0.75,2) -- (0.75,0.5);
\path (0.5,0) node [style=sergio]{\large $\mathsf{A}$};
\filldraw[fill=white, draw=black] (0.75,1.25)circle (10pt);
\path (0.75,1.25) node [style=sergio]{$U_g$};
\path (2.125,0) node [style=sergio]{$=$};
\filldraw[fill=white, draw=black, rounded corners] (4,-0.5)--(5.5,-0.5)--(5.5,0.5)--(4,0.5)--cycle;
\draw[line width=1pt] (4,0) -- (2.5,0);
\filldraw[fill=white, draw=black] (3.25,0)circle (10pt);
\path (3.25,0) node [style=sergio]{$V^{-1}_g$};
\path (4.75,0) node [style=sergio]{\large $\mathsf{A}$};
\draw[line width=1pt] (5.5,0) -- (7,0);
\filldraw[fill=white, draw=black] (6.25,0)circle (10pt);
\path (6.25,0) node [style=sergio]{$V_g$};
\draw[line width=2pt, color=red] (4.5,0.5) -- (4.5,1);
\draw[line width=3pt, color=blue] (5,1) -- (5,0.5);
\path (5,1.25) node [style=sergio]{$\kappa$};
\path (4.5,1.25) node [style=sergio]{$\sigma$};
\end{tikzpicture}},
\end{aligned}
\end{align}
where $(k, g)\in K\times G$.
For simplicity, we write $V_{(k,e)}\equiv V_k$ and $V_{(e,g)}\equiv V_g$, where $e$ denotes the identity element.
The symmetry actions $U_k$ and $U_g$ on the physical indices form linear representations of $K$ and $G$, respectively, whereas $V_{(k,g)}$ acting on the virtual indices generally forms a projective representation of $K\times G$~\cite{Chen2011, Schuch2011}.
Distinct phases correspond to inequivalent projective representations, which are classified by $\mathcal{H}^2[K\times G, U(1)]\cong \mathcal{H}^2[K, U(1)]\times \mathcal{H}^1\left[G,\mathcal{H}^1[K, U(1)]\right]\times \mathcal{H}^2[G, U(1)]$~\cite{Brown1982, Guo2025B}.
The first and third terms describe SPT phases solely protected by $K$ and $G$, respectively, while the second term captures phases induced by a mixed anomaly between $K$ and $G$ in the virtual space, i.e., exchanging $V_k$ and $V_g$ leads to an extra $U(1)$ phase~\cite{Chen2013, Witten2016} (Appendix~\ref{Sec: Appendix-B}).

Therefore, in the presence of such a nontrivial mixed anomaly, the virtual representation $V_g$ cannot be reduced to a one-dimensional form without destroying the projective structure between $V_k$ and $V_g$.
After tracing out the $\kappa$ spins, the local tensor therefore satisfies the following two relations
\begin{align}
\scalebox{0.8}{
\begin{tikzpicture}[scale=0.75]
\tikzstyle{sergio}=[rectangle,draw=none]
\filldraw[fill=white, draw=black, rounded corners] (-0.25,-1.5)--(1.25,-1.5)--(1.25,-0.5)--(-0.25,-0.5)--cycle;
\filldraw[fill=white, draw=black, rounded corners] (-0.25,-3)--(1.25,-3)--(1.25,-2)--(-0.25,-2)--cycle;
\draw[line width=1pt] (1.25,-1) -- (1.75,-1);
\draw[line width=1pt] (-0.25,-1) -- (-0.75,-1);
\draw[line width=1pt] (1.25,-2.5) -- (1.75,-2.5);
\draw[line width=1pt] (-0.25,-2.5) -- (-0.75,-2.5);
\draw[line width=2pt, color=red] (0.5,-0.5) -- (0.5,1);
\draw[line width=3pt, color=blue] (0.5,-1.5) -- (0.5,-2);
\draw[line width=2pt, color=red] (0.5,-3) -- (0.5,-4.5);
\path (0.5,-1) node [style=sergio]{\large $\mathsf{A}$};
\path (0.5,-2.5) node [style=sergio]{\large $\mathsf{A^*}$};
\path (2.25,-1.75) node [style=sergio]{$=$};
\filldraw[fill=white, draw=black, rounded corners] (4.25,-1.5)--(5.75,-1.5)--(5.75,-0.5)--(4.25,-0.5)--cycle;
\filldraw[fill=white, draw=black, rounded corners] (4.25,-3)--(5.75,-3)--(5.75,-2)--(4.25,-2)--cycle;
\draw[line width=1pt] (4.25,-1) -- (2.75,-1);
\filldraw[fill=white, draw=black] (3.5,-1)circle (10pt);
\path (3.5,-1) node [style=sergio]{$V_k^{-1}$};
\path (5,-1) node [style=sergio]{\large $\mathsf{A}$};
\draw[line width=1pt] (5.75,-1) -- (7.25,-1);
\filldraw[fill=white, draw=black] (6.5,-1)circle (10pt);
\path (6.5,-1) node [style=sergio]{$V_k$};
\draw[line width=1pt] (4.25,-2.5) -- (2.75,-2.5);
\filldraw[fill=white, draw=black] (3.5,-2.5)circle (10pt);
\path (3.5,-2.5) node [style=sergio]{$V_k^{*-1}$};
\path (5,-2.5) node [style=sergio]{\large $\mathsf{A^*}$};
\draw[line width=1pt] (5.75,-2.5) -- (7.25,-2.5);
\filldraw[fill=white, draw=black] (6.5,-2.5)circle (10pt);
\path (6.5,-2.5) node [style=sergio]{$V_k^*$};
\draw[line width=2pt, color=red] (5,-3) -- (5,-3.5);
\draw[line width=3pt, color=blue] (5,-1.5) -- (5,-2);
\draw[line width=2pt, color=red] (5,-0.5) -- (5,0);
\filldraw[fill=white, draw=black] (0.5,0.25)circle (10pt);
\path (0.5,0.25) node [style=sergio]{$U_k$};
\filldraw[fill=white, draw=black] (0.5,-3.75)circle (10pt);
\path (0.5,-3.75) node [style=sergio]{$U_k^{*}$};
\end{tikzpicture}},\\
\scalebox{0.8}{
\begin{tikzpicture}[scale=0.75]
\tikzstyle{sergio}=[rectangle,draw=none]
\filldraw[fill=white, draw=black, rounded corners] (-0.25,-1.5)--(1.25,-1.5)--(1.25,-0.5)--(-0.25,-0.5)--cycle;
\filldraw[fill=white, draw=black, rounded corners] (-0.25,-3)--(1.25,-3)--(1.25,-2)--(-0.25,-2)--cycle;
\draw[line width=1pt] (1.25,-1) -- (1.75,-1);
\draw[line width=1pt] (-0.25,-1) -- (-0.75,-1);
\draw[line width=1pt] (1.25,-2.5) -- (1.75,-2.5);
\draw[line width=1pt] (-0.25,-2.5) -- (-0.75,-2.5);
\draw[line width=2pt, color=red] (0.5,-0.5) -- (0.5,0);
\draw[line width=3pt, color=blue] (0.5,-1.5) -- (0.5,-2);
\draw[line width=2pt, color=red] (0.5,-3) -- (0.5,-3.5);
\path (0.5,-1) node [style=sergio]{\large $\mathsf{A}$};
\path (0.5,-2.5) node [style=sergio]{\large $\mathsf{A^*}$};
\path (2.25,-1.75) node [style=sergio]{$=$};
\filldraw[fill=white, draw=black, rounded corners] (4.25,-1.5)--(5.75,-1.5)--(5.75,-0.5)--(4.25,-0.5)--cycle;
\filldraw[fill=white, draw=black, rounded corners] (4.25,-3)--(5.75,-3)--(5.75,-2)--(4.25,-2)--cycle;
\draw[line width=1pt] (4.25,-1) -- (2.75,-1);
\filldraw[fill=white, draw=black] (3.5,-1)circle (10pt);
\path (3.5,-1) node [style=sergio]{$V_g^{-1}$};
\path (5,-1) node [style=sergio]{\large $\mathsf{A}$};
\draw[line width=1pt] (5.75,-1) -- (7.25,-1);
\filldraw[fill=white, draw=black] (6.5,-1)circle (10pt);
\path (6.5,-1) node [style=sergio]{$V_g$};
\draw[line width=1pt] (4.25,-2.5) -- (2.75,-2.5);
\filldraw[fill=white, draw=black] (3.5,-2.5)circle (10pt);
\path (3.5,-2.5) node [style=sergio]{$V_g^{*-1}$};
\path (5,-2.5) node [style=sergio]{\large $\mathsf{A^*}$};
\draw[line width=1pt] (5.75,-2.5) -- (7.25,-2.5);
\filldraw[fill=white, draw=black] (6.5,-2.5)circle (10pt);
\path (6.5,-2.5) node [style=sergio]{$V_g^*$};
\draw[line width=2pt, color=red] (5,-3) -- (5,-3.5);
\draw[line width=3pt, color=blue] (5,-1.5) -- (5,-2);
\draw[line width=2pt, color=red] (5,-0.5) -- (5,0);
\end{tikzpicture}}.
\label{Equ: Strong injectivity}
\end{align}
Specifically, Eq.~\eqref{Equ: Strong injectivity} means that the resulting LPDO violates the strong injectivity condition and leads to a long-range R\'enyi-$2$ correlator~\cite{Guo2025B} (Appendix~\ref{Sec: Appendix-A}).
On the other hand, for a purified state that is trivially symmetric, the virtual action only forms a linear representation of $K\times G$, which can be reduced to a one-dimensional representation as both $K$ and $G$ are finite Abelian groups.
Consequently, a purified SPT state with a mixed anomaly is not only sufficient but also necessary for the reduced mixed state to exhibit SWSSB order diagnosed by the R\'enyi-$2$ correlator, complementing the fidelity-based argument in the previous subsection.

\subsection{ASPT}
When entering the mixed-state regime, the classification of ASPT phases with $K \text{ strong (S)} \times G \text{ weak (W)}$ symmetry reduces to $\mathcal{H}^2[K, U(1)]\times \mathcal{H}^1\left[G,\mathcal{H}^1[K, U(1)]\right]$~\cite{Ma2023}.
The $\mathcal{H}^2[G, U(1)]$ sector is absent because a purely weak symmetry cannot support a nontrivial projective structure, where the virtual phases carried by the ket and the bra cancel in the density-operator representation~\cite{Guo2025B}.
Here, we focus on the construction of ASPT phases, where the system has two degrees of freedom carrying $K$ and $G$ symmetries, respectively.
There are two complementary routes to construct a mixed state with a nontrivial ASPT order protected by $K \text{(S)} \times G \text{(W)}$, characterized by a mixed anomaly between $K$ and $G$.

\subsubsection{Explicit weakening of $G$ from an SPT state}
One may start from a pure-state SPT protected by $K\times G$ with the desired topological structure and then explicitly reduce the strong $G$ symmetry to a weak one.
This can be realized by attaching a local charge reservoir to each physical degree of freedom that carries the $G$ symmetry.
On the other hand, if one spontaneously breaks $G$ into a weak one, the resulting state belongs to the double ASPT phase~\cite{Guo2025A, Kuno2025a}, which will be fully discussed later.

As an explicit example, consider an ASPT phase protected by $\Z_2^{\tau}\text{ (S)} \times \Z_2^{\sigma}\text{ (W)}$.
Starting from the standard cluster-state construction, we attach an ancillary $\kappa$ spin parallel to each $\sigma$ spin, namely
\begin{align}
\begin{aligned}
    &\ket{\psi^{\sigma-\tau\text{ SPT, }\sigma-\kappa\text{ Parallel}}_{\sigma \tau\kappa}}\\
    =&\, \frac{1}{2^{N/2}}\sum_{\{\sigma_i\}}\ket{\cdots\uparrow_{\kappa}\uparrow_{\sigma}\rightarrow_{\tau}\uparrow_{\kappa}\uparrow_{\sigma}\leftarrow_{\tau}\downarrow_{\kappa}\downarrow_{\sigma}\leftarrow_{\tau}\uparrow_{\kappa}\uparrow_{\sigma}\cdots}.
    \label{Equ: SPT-Parallel}
\end{aligned}
\end{align}
Tracing out the $\kappa$ spins yields the reduced mixed state
\begin{align}
    \rho^{\rm ASPT}_{\rm \sigma\tau} = \frac{1}{2^N}\sum_{\{\sigma_i\}}\ket{\psi_{\sigma\tau}^{\rm DW}}_{\{\sigma_i\}}\hspace{-0.5mm} \bra{\psi_{\sigma\tau}^{\rm DW}}_{\{\sigma_i\}},
\end{align}
which can equivalently be viewed as replacing the coherent superposition over $\{\sigma_i\}$ in the purified cluster construction in Eq.~\eqref{Equ: Cluster} (with $\sigma\tau$ spins) by a classical mixture~\cite{Ma2025A}.

\subsubsection{Decorated domain-wall construction from a trivial mixed state}
Alternatively, one can start from a trivial state with the same $K \text{(S)}\times G\text{(W)}$ symmetry, and then impose the mixed anomaly via the decorated domain-wall procedure.
For the $\Z_2^{\tau}\rm{(S)} \times \Z_2^{\sigma}\rm{(W)}$ example, we first introduce the DW duality map composed of controlled-$Z$ (CZ) gates~\cite{Li2022}
\begin{align}
    U_{\sigma\tau}^{\rm DW} \equiv \prod_i \text{CZ}_{\tau_{i-1}\sigma_i}\text{CZ}_{\sigma_i\tau_i},
\end{align}
which exchanges the trivial symmetric state and the cluster state,
\begin{align}
    U_{\sigma\tau}^{\rm DW}\ket{\psi_{\sigma\tau}^{\rm Trivial}} = \ket{\psi_{\sigma\tau}^{\rm SPT}}, \quad U_{\sigma\tau}^{\rm DW}\ket{\psi_{\sigma\tau}^{\rm SPT}} = \ket{\psi_{\sigma\tau}^{\rm Trivial}}.
    \label{Equ: Duality}
\end{align}
Here, the trivial symmetric state is defined as
\begin{align}
    \ket{\psi_{\sigma\tau}^{\rm Trivial}} &= \ket{\psi_{\sigma}^{\rm Trivial}}\otimes \ket{\psi_{\tau}^{\rm Trivial}}, \label{Equ: Multi-Trivial}\\
    \ket{\psi_{\sigma(\tau)}^{\rm Trivial}} &= \prod_i\ket{\rightarrow_{\sigma(\tau)_i}}.
    \label{Equ: Trivial}
\end{align}
Notably, this duality map is unitary under both open and periodic boundary conditions, i.e.,
\begin{align}
    U_{\sigma\tau}^{\rm DW\dagger} U_{\sigma\tau}^{\rm DW} = I.
\end{align}

In this way, we can construct the ASPT state protected by $\Z_2^{\tau}\text{(S)}\times\Z_2^{\sigma}\text{(W)}$ by applying the DW duality map to the product of Eq.~\eqref{Equ: W} and \eqref{Equ: Trivial}, i.e.,
\begin{align}
    \rho_{\sigma\tau}^{\rm ASPT} = U_{\sigma\tau}^{\rm DW}\left(\rho_{\sigma}^{\rm W}\otimes \ket{\psi_{\tau}^{\rm Trivial}}\hspace{-0.5mm}\bra{\psi_{\tau}^{\rm Trivial}}\right)U_{\sigma\tau}^{\rm DW\dagger}.
\end{align}
As the DW map acts only on $\sigma$ and $\tau$ spins, it commutes with the operation of tracing out $\kappa$ spins.
Therefore, the corresponding purification can be written as
\begin{align}
    \ket{\psi^{\sigma-\tau\text{ SPT, }\sigma-\kappa\text{ Parallel}}_{\sigma \tau\kappa}} = U_{\sigma\tau}^{\rm DW}\left(\ket{\psi_{\kappa\sigma}^{\rm Bell}}\otimes\ket{\psi_{\tau}^{\rm Trivial}}\right),
\end{align}
which is equivalent to Eq.~\eqref{Equ: SPT-Parallel}.

\subsection{Double ASPT}
\label{Sec: Double ASPT}
A recent study proposed an intrinsic open-system quantum phase referred to as the double ASPT phase, which is characterized by the coexistence of SWSSB and ASPT order~\cite{Guo2025A}.
In this phase, one symmetry subgroup undergoes SWSSB, but still protects a nontrivial ASPT order together with the other subgroup that remains strong.
Such a coexistence of SWSSB and ASPT is forbidden in pure-state systems, where protecting symmetries must remain unbroken.
While this phase has been previously studied using imaginary-time Lindbladian evolution~\cite{Guo2025C}, here we propose an explicit fixed-point construction from the purification perspective.

The main difference from the ASPT phase discussed above is that, in the double ASPT phase, the breaking of a strong symmetry to a weak one occurs spontaneously rather than explicitly.
Nevertheless, purification can still be constructed following a similar strategy, and we again outline two complementary routes.

\subsubsection{SWSSB from an SPT state}
Beginning with a pure-state SPT protected by $K\times G$ symmetry, we need to introduce an additional SPT structure between the system spins carrying $G$ symmetry and the environment spins.
For the $\Z_2^{\tau} \text{(S)} \times \Z_2^{\sigma}\text{(SWSSB)}$ example, the procedure can be interpreted as
\begin{align}
\begin{aligned}
    \ket{\psi_{\sigma\tau\kappa}^{\sigma-\tau\text{ SPT, }\sigma-\kappa\text{ SPT}}} &= U_{\kappa\sigma}^{\rm DW}\left(\ket{\psi_{\sigma\tau}^{\rm SPT}}\otimes\ket{\psi_{\kappa}^{\rm Trivial}}\right)\\
    &= U_{\kappa\sigma}^{\rm DW}U_{\sigma\tau}^{\rm DW}\ket{\psi_{\sigma\tau\kappa}^{\rm Trivial}},\label{Equ: 2-SPT}
\end{aligned}
\end{align}
where $\ket{\psi_{\kappa}^{\rm Trivial}}$ and $\ket{\psi_{\sigma\tau\kappa}^{\rm Trivial}}$ are defined similar to Eqs.~\eqref{Equ: Multi-Trivial} and \eqref{Equ: Trivial}, and we have adopted the duality relation in Eq.~\eqref{Equ: Duality}.

\subsubsection{Decorated domain-wall construction from an SWSSB state}
Alternatively, one may start from a trivial (i.e., without topological properties) state with the same symmetry properties $K\text{(S)}\times G\text{(SWSSB)}$ and introduce the topological structure directly at the mixed-state level.
This leads to the double ASPT density matrix
\begin{align}
    \rho_{\sigma\tau}^{\sigma\text{-double ASPT}} = U_{\sigma\tau}^{\rm DW}\left(\rho_{\sigma}^{\rm SWSSB}\otimes \ket{\psi_{\tau}^{\rm Trivial}}\hspace{-0.5mm}\bra{\psi_{\tau}^{\rm Trivial}}\right)U_{\sigma\tau}^{\rm DW\dagger}.
\end{align}
The corresponding purification can be constructed analogously
\begin{align}
\begin{aligned}
    \ket{\psi_{\sigma\tau\kappa}^{\sigma-\tau\text{ SPT, }\sigma-\kappa\text{ SPT}}} &= U_{\sigma\tau}^{\rm DW}\left(\ket{\psi_{\kappa\sigma}^{\rm SPT}}\otimes\ket{\psi_{\tau}^{\rm Trivial}}\right)\\
    &= U_{\sigma\tau}^{\rm DW}U_{\kappa\sigma}^{\rm DW}\ket{\psi_{\sigma\tau\kappa}^{\rm Trivial}},
\end{aligned}
\end{align}
which is equivalent to Eq.~\eqref{Equ: 2-SPT} since CZ gates commute with each other even when they overlap on the same lattice site.

\section{Eight mixed-state phases from a minimal $\Z_2^{\sigma}\times\Z_2^{\tau}\times\Z_2^{\kappa}$ purified model}
\label{Sec: Fixed-point}
In the previous section, we constructed purified models for several representative fixed-point density matrices, where the additional $\Z_2^{\kappa}$ symmetry enriches the phase structure of the original $\Z_2^{\sigma}\times \Z_2^{\tau}$ system.
Here, we extend our construction to a unified model that incorporates all different SPT phases protected by three $\Z_2$ symmetries and analyze the corresponding mixed-state phases obtained after tracing out $\kappa$ spins.

\subsection{Purified model}
\begin{figure}
    \centering
    \includegraphics[width=0.7\linewidth]{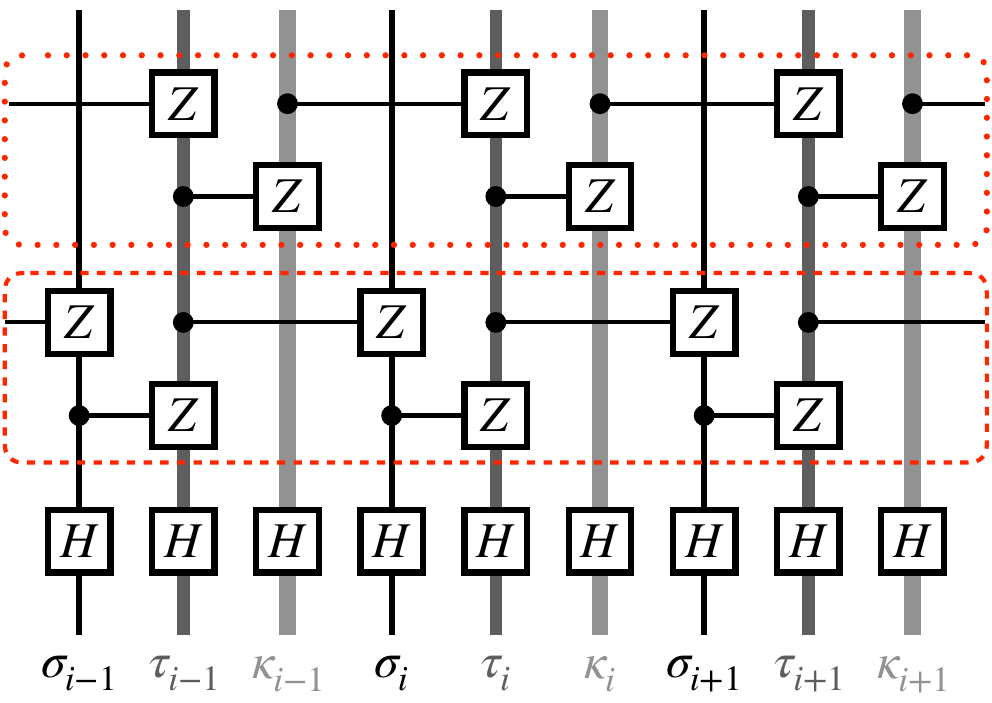}
    \caption{Quantum circuit to generate $\ket{\psi_{\sigma\tau\kappa}^{(-1, -1, +1)}}$.
    Starting from the product state $\prod_i \ket{\uparrow}_i$, one layer of Hadamard gates ($H$) generates the symmetric trivial state $\ket{\psi_{\sigma\tau\kappa}^{(+1, +1, +1)}}=\prod_i \ket{\rightarrow}_i$, then two layers of CZ gates $U_{\sigma\tau}^{\rm DW}$ (dashed line) and $U_{\tau\kappa}^{\rm DW}$ (dotted line) subsequently impose the mixed anomaly between corresponding pairs, leading to $\ket{\psi_{\sigma\tau\kappa}^{(-1, -1, +1)}}$.}
    \label{Fig: Circuit}
\end{figure}
We consider three $\Z_2$ degrees of freedom $\sigma_i$, $\tau_{i}$, and $\kappa_i$ in each unit cell, and assume that the total symmetry group is given by their direct product.
The pure-state SPT phases protected by $\Z_2^{\sigma}\times\Z_2^{\tau}\times\Z_2^{\kappa}$ symmetry are classified by $\mathcal{H}^2\left[\Z_2^3, \rm{U}(1)\right] = \Z_2^3$ and can be labeled by a triple of $\Z_2$ indices $\mu=(\mu_{\sigma\tau},\mu_{\tau\kappa},\mu_{\kappa\sigma})\in\{\pm1\}^3$.
A negative $\mu_{\alpha\beta}$ indicates the presence of a mixed anomaly between $\alpha$ and $\beta$ degrees of freedom, with $\alpha, \beta \in \{\sigma, \tau, \kappa\}$.

The fixed-point wavefunctions of these eight SPT phases can be constructed by decorating domain walls with controlled-$Z$ gates~\cite{Chen2014} in a unified manner
\begin{align}
    \ket{\psi^{\mu}_{\sigma\tau\kappa}} = \left(U_{\sigma\tau}^{\rm DW}\right)^{\frac{1-\mu_{\sigma\tau}}{2}}\left(U_{\tau\kappa}^{\rm DW}\right)^{\frac{1-\mu_{\tau\kappa}}{2}}\left(U_{\kappa\sigma}^{\rm DW}\right)^{\frac{1-\mu_{\kappa\sigma}}{2}}
    \ket{\psi_{\sigma\tau\kappa}^{\rm Trivial}},
    \label{Equ: psi_mu}
\end{align}
where $\ket{\psi_{\sigma\tau\kappa}^{\rm Trivial}}$ denotes the trivial symmetric product state.
As an illustrative example, the quantum circuit to generate $\ket{\psi^{(-1, -1, +1)}_{\sigma\tau\kappa}}$ is depicted in Fig.~\ref{Fig: Circuit}.

The stabilizer Hamiltonian for the trivial state reads as
\begin{align}
    H_{\sigma\tau\kappa}^{\rm Trivial} = -\sum_i(\sigma_i^x+\tau_i^x+\kappa_i^x),
\end{align}
while those for the other states can be generated by conjugating with the corresponding DW maps
\begin{align}
\begin{aligned}
    H_{\sigma\tau\kappa}^{\mu} = \left(U_{\sigma\tau}^{\rm DW}\right)^{\frac{1-\mu_{\sigma\tau}}{2}}\left(U_{\tau\kappa}^{\rm DW}\right)^{\frac{1-\mu_{\tau\kappa}}{2}}\left(U_{\kappa\sigma}^{\rm DW}\right)^{\frac{1-\mu_{\kappa\sigma}}{2}} H_{\sigma\tau\kappa}^{\rm Trivial}\\ \left[\left(U_{\sigma\tau}^{\rm DW}\right)^{\frac{1-\mu_{\sigma\tau}}{2}}\left(U_{\tau\kappa}^{\rm DW}\right)^{\frac{1-\mu_{\tau\kappa}}{2}}\left(U_{\kappa\sigma}^{\rm DW}\right)^{\frac{1-\mu_{\kappa\sigma}}{2}}\right]^{\dagger}.
\end{aligned}
\end{align}
The Hamiltonian then takes a compact form
\begin{align}
    H^{\mu}_{\sigma\tau\kappa}=-\sum_i\left(S_{\sigma_i}^{\mu}+S_{\tau_i}^{\mu}+S_{\kappa_i}^{\mu}\right),
    \label{Equ: Ham}
\end{align}
where the stabilizers read as
\begin{align}
    S_{\sigma_i}^{\mu}&=\sigma_i^x
        \left(\tau_{i-1}^z\tau_{i}^z\right)^{\frac{1-\mu_{\sigma\tau}}{2}}
        \left(\kappa_{i-1}^z\kappa_i^z\right)^{\frac{1-\mu_{\kappa\sigma}}{2}}\\
    S_{\tau_i}^{\mu}&=\tau_{i}^x
        \left(\sigma_i^z\sigma_{i+1}^z\right)^{\frac{1-\mu_{\sigma\tau}}{2}}
        \left(\kappa_{i-1}^z\kappa_{i}^z\right)^{\frac{1-\mu_{\tau\kappa}}{2}}\\
    S_{\kappa_i}^{\mu}&=\kappa_i^x
        \left(\tau_{i}^z\tau_{i+1}^z\right)^{\frac{1-\mu_{\tau\kappa}}{2}}
        \left(\sigma_{i}^z\sigma_{i+1}^z\right)^{\frac{1-\mu_{\kappa\sigma}}{2}}
\end{align}
using the relations
\begin{align}
\begin{aligned}
    U_{\sigma\tau}^{\rm DW}\sigma_i^xU_{\sigma\tau}^{{\rm DW}\dagger} &= \tau_{i-1}^z\sigma_i^x\tau_{i}^z\\
    U_{\sigma\tau}^{\rm DW}\tau_i^xU_{\sigma\tau}^{{\rm DW}\dagger} &= \sigma_{i}^z\tau_i^x\sigma_{i+1}^z
\end{aligned}
\end{align}
together with analogous relations for the other two sets of DW maps.

Depending on the values of $\mu$, the stabilizers involve single-site terms or three- /five-body interactions.
For instance, $S_{\tau_i}^{\mu}$ becomes a five-body operator when both $\mu_{\sigma\tau}=-1$ and $\mu_{\tau\kappa}=-1$.
Thus Eq.~\eqref{Equ: Ham} provides a single compact Hamiltonian formalism that generates all eight pure fixed-point states.

\subsection{Mixed-state phases}
Tracing out the $\kappa$ degrees of freedom yields mixed states
\begin{align}
    \rho_{\sigma\tau}^{\mu}=\Tr_{\kappa}\left[\ket{\psi^{\mu}_{\sigma\tau\kappa}}\bra{\psi^{\mu}_{\sigma\tau\kappa}}\right],
\end{align}
which realize the eight distinct mixed-state phases summarized in Table~\ref{Tab: Phases} (No.~1-8).

\begin{table*}
\caption{All phases in the phase cube.
The symmetry (two system $\Z_2^{\sigma}/\Z_2^{\tau}$ and one ancilla $\Z_2^{\kappa}$) and topological properties (three $\mu$ indices) of the purified states are listed, together with the resulting symmetry and topological characteristics of the mixed states after tracing out the $\kappa$ spins.
In the purified state, a minus sign occurs in $\mu_{\alpha\beta}$ if both $\Z_2$ symmetries remain unbroken and there is a mixed anomaly between them in the virtual space.
After tracing out $\kappa$ spins, the mixed anomaly between ancilla and system will result in SWSSB of the corresponding system spins. 
In particular, the simultaneous couplings $\mu_{\tau\kappa}=\mu_{\kappa\sigma}=-1$ will break both $\Z_2$ symmeties into weak ones, while keeping the diagonal $\Z_2^{\sigma\tau}$ strong.
Moreover, if one system $\Z_2$ symmetry is broken in the purified state, it will also undergo SNSSB in the reduced mixed state, leading to a simultaneous SNSSB of the diagonal $\Z_2^{\sigma\tau}$ symmetry.
}
\begin{tabular}{c|cccccccc|ccccc}
\hline\hline
\multirow{2}{*}{No.} & \multicolumn{8}{c}{Pure-state phase} & \multicolumn{5}{|c}{Mixed-state phase}\\\cline{2-14}
& $\Z_2^{\sigma}$ & $\Z_2^{\tau}$ & $\Z_2^{\kappa}$ & $\rm{Hatch}_{|\braket{K}|}$ & $\mu_{\sigma\tau}$ & $\mu_{\tau\kappa}$ & $\mu_{\kappa\sigma}$ & $\rm{Hatch}_{\mu}$ & $\Z_2^{\sigma}$ & $\Z_2^{\tau}$ &$\Z_2^{\sigma\tau}$ & $\mu_{\sigma\tau}$ & Phase name\\
\hline
1 & Symmetric & Symmetric & Symmetric & None & $+1$ & $+1$ & $+1$ & None & Symmetric & Symmetric & Symmetric & $+1$ & Trivial \\
2 & Symmetric & Symmetric & Symmetric & None & $-1$ & $+1$ & $+1$ & \pblue{/} & Symmetric & Symmetric & Symmetric & $-1$ & SPT \\
3 & Symmetric & Symmetric & Symmetric & None & $+1$ & $-1$ & $+1$ & \pgreen{\textbackslash} & Symmetric & SWSSB & SWSSB & $+1$ & $\tau$-SWSSB \\
4 & Symmetric & Symmetric & Symmetric & None & $+1$ & $+1$ & $-1$ & \pred{$|$} & SWSSB & Symmetric & SWSSB & $+1$ & $\sigma$-SWSSB \\
5 & Symmetric & Symmetric & Symmetric & None & $-1$ & $-1$ & $+1$ & \pblue{/} \pgreen{\textbackslash} & Symmetric & SWSSB & SWSSB & $-1$ & $\tau$-double ASPT \\
6 & Symmetric & Symmetric & Symmetric & None & $-1$ & $+1$ & $-1$ & \pblue{/} \pred{$|$} & SWSSB & Symmetric & SWSSB & $-1$ & $\sigma$-double ASPT \\
7 & Symmetric & Symmetric & Symmetric & None & $+1$ & $-1$ & $-1$ & \pgreen{\textbackslash} \pred{$|$} & SWSSB & SWSSB & Symmetric & $+1$ & $\sigma\tau$-preserved SWSSB \\
8 & Symmetric & Symmetric & Symmetric & None & $-1$ & $-1$ & $-1$ & \pblue{/} \pgreen{\textbackslash} \pred{$|$} & SWSSB & SWSSB & Symmetric & $-1$ & $\sigma\tau$-preserved double ASPT \\
9 & SSB & Symmetric & Symmetric & \red{//} & $+1$ & $+1$ & $+1$ & None & SNSSB & Symmetric & SNSSB & $+1$ & $\sigma$-SNSSB \\
10 & SSB & Symmetric & Symmetric & \red{//} & $+1$ & $-1$ & $+1$ & \pgreen{\textbackslash} & SNSSB & SWSSB & SNSSB & $+1$ & $\sigma$-SNSSB $\tau$-SWSSB\\
11 & Symmetric & SSB & Symmetric & \blue{\sslash} & $+1$ & $+1$ & $+1$ & None & Symmetric & SNSSB & SNSSB & $+1$ & $\tau$-SNSSB \\
12 & Symmetric & SSB & Symmetric & \blue{\sslash} & $+1$ & $+1$ & $-1$ & \pred{$|$} & SWSSB & SNSSB & SNSSB & $+1$ & $\sigma$-SWSSB $\tau$-SNSSB\\
13 & Symmetric & Symmetric & SSB & = & $+1$ & $+1$ & $+1$ & None & Symmetric & Symmetric & Symmetric & $+1$ & Trivial \\
14 & Symmetric & Symmetric & SSB & = & $-1$ & $+1$ & $+1$ & \pblue{/} & Symmetric & Symmetric & Symmetric & $-1$ & SPT \\
15 & SSB & SSB & Symmetric & \red{//} \blue{\sslash} & $+1$ & $+1$ & $+1$ & None & SNSSB & SNSSB & SNSSB & $+1$ & $\sigma$-SNSSB $\tau$-SNSSB \\
16 & Symmetric & SSB & SSB & \blue{\sslash} = & $+1$ & $+1$ & $+1$ & None & Symmetric & SNSSB & SNSSB & $+1$ & $\tau$-SNSSB \\
17 & SSB & Symmetric & SSB & \red{//} = & $+1$ & $+1$ & $+1$ & None & SNSSB & Symmetric & SNSSB & $+1$ & $\sigma$-SNSSB\\
18 & SSB & SSB & SSB & \red{//} \blue{\sslash} = & $+1$ & $+1$ & $+1$ & None & SNSSB & SNSSB & SNSSB & $+1$ & $\sigma$-SNSSB $\tau$-SNSSB\\
\hline\hline
\end{tabular}
\label{Tab: Phases}
\end{table*}

When no domain-wall decoration involves $\kappa$ spins, the reduced density matrix retains two strong $\Z_2$ symmetries and therefore remains a pure state (No.~1 and 2).
Introducing a $\sigma$–$\tau$ DW decoration yields a nontrivial SPT state protected by $\Z_2^\sigma\times\Z_2^\tau$, which is just the conventional cluster state (No.~2).

If the DW map involves either $\kappa$–$\sigma$ or $\tau$–$\kappa$ link, the corresponding reduced density matrix no longer supports two strong symmetries.
Instead, one of the two $\Z_2$ subgroups undergoes SWSSB (No.~3 and 4).
For example, $\mu_{\kappa\sigma}=-1$ produces a $\sigma$-SWSSB phase (here we explicitly identify which symmetry undergoes SWSSB in the phase name), where $\sigma$ is only weakly preserved while $\tau$ remains strong (No.~4).
Furthermore, when both $\kappa$–$\sigma$ and $\sigma$–$\tau$ links are present, the system exhibits nontrivial topological properties protected by $\Z_2^{\tau}\text{(S)}\times\Z_2^{\sigma}\text{(SWSSB)}$~\cite{Guo2025A}, giving rise to the $\sigma$-double ASPT phase (No.~6).
These phases have been discussed in detail in Sec.~\ref{Sec: SWSSB} and \ref{Sec: Double ASPT}.

A particularly intriguing set of phases arises when both $\kappa$–$\sigma$ and $\tau$–$\kappa$ links are present (No.~7 and 8).
In this case, neither $\Z_2^{\sigma}$ nor $\Z_2^{\tau}$  survives as an individual strong symmetry.
Instead, only the diagonal subgroup
\begin{align}
    \Z_2^{\sigma\tau}:\quad K_{\rm diag}=\prod_i \sigma_i^x\tau_i^x
\end{align} remains strong, while the individual $\sigma$ and $\tau$ symmetries undergo SWSSB.
For example, for $(\mu_{\sigma\tau},\mu_{\tau\kappa},\mu_{\kappa\sigma})=(+1,-1,-1)$ (No.~7), the reduced state takes the form
\begin{align}
\begin{aligned}
    \rho_{\sigma\tau}^{\sigma\tau\text{-P SWSSB}} & =\frac{1}{2^N}\prod_i\left(\sigma_i^0\tau_i^0+\sigma_i^x\tau_i^x\right) + \frac{1}{2^N}\prod_i\left(\sigma_i^0\tau_i^x+\sigma_i^x\tau_i^0\right)\\
    & \equiv O_1+O_2,
    \label{Equ: OD-SWSSB}
\end{aligned}
\end{align}
where the superscript $\sigma \tau \text{-P}$ denotes $\sigma \tau \text{-preserved}$ hereafter.
In analogy to Eq.~\eqref{Equ: SWSSB}, this density matrix preserves the diagonal $\Z_2^{\sigma\tau}$ symmetry strongly $K_{\rm diag}O_1=O_1$, $K_{\rm diag}O_2=O_2$, while $\Z_2^\sigma$ and $\Z_2^\tau$ survive merely in the SWSSB sense.

Although this phase may appear to involve the breaking of two symmetry generators, it still represents a simple SWSSB pattern.
Indeed, the symmetry group $\Z_2^{\sigma}\times \Z_2^{\tau}$ can be equivalently generated by $\Z_2^{\sigma}\times \Z_2^{\sigma\tau}$.
From this perspective, the phase consists of the diagonal symmetry that $\Z_2^{\sigma\tau}$ remains strong, and each component $\Z_2^{\sigma}/\Z_2^{\tau}$ undergoes SWSSB.
We therefore refer to this phase as the $\sigma\tau$-preserved SWSSB phase (No.~7).

As in Sec.~\ref{Sec: SWSSB}, the string order of the purified model guides the construction of a suitable mixed-state order parameter.
The parent Hamiltonian for $\mu=(+1,-1,-1)$ reads as
\begin{align}
    H_{\sigma\tau\kappa}^{(+1, -1, -1)}&=-\sum_i \left[\kappa_{i-1}^{z}\tau_i^{x}\kappa_{i}^{z}+\kappa_{i-1}^{z}\sigma_i^{x}\kappa_{i}^{z}+\sigma_i^{z}\tau_i^{z}\kappa_i^{x}\sigma_{i+1}^{z}\tau_{i+1}^{z}\right],
\end{align}
whose ground state features the dressed $\kappa$-string order
\begin{align}
    \lim_{|i-j|\rightarrow \infty} |\braket{S(\sigma^z\tau^z, \kappa^x, i, j)}| =1.
\end{align}
Therefore, tracing out $\kappa$ spins leads to a long-range order in the corresponding fidelity correlator
\begin{align}
\begin{aligned}
    &\mathcal{C}^{\mathcal{F}}\left(\sigma^z\tau^z, i, j\right)\\
    =&\,\mathcal{F}\left[(\sigma_i^{z}\tau_i^{z}\sigma_j^{z}\tau_j^{z})\rho_{\sigma\tau}^{\sigma\tau\text{-P SWSSB}}(\sigma_i^{z}\tau_i^{z}\sigma_j^{z}\tau_j^{z}),\rho_{\sigma\tau}^{\sigma\tau\text{-P SWSSB}}\right].
\end{aligned}
\end{align}
This construction further demonstrates that the purification perspective provides a natural route to defining mixed-state order parameters, inherited directly from the structure of the purified state.

If the $\sigma$–$\tau$ link is further introduced, the resulting mixed state becomes
\begin{align}
    \rho_{\sigma\tau}^{\sigma\tau\text{-P double ASPT}}=U_{\sigma\tau}^{\rm DW}\rho_{\sigma\tau}^{\sigma\tau\text{-P SWSSB}}U_{\sigma\tau}^{\rm DW\dagger},
    \label{Equ: OD-Double ASPT}
\end{align}
representing the last phase in our unified model (No.~8).
In this case, both $\Z_2^{\sigma}$ and $\Z_2^{\tau}$ are preserved only weakly, which cannot support any nontrivial ASPT phase~\cite{Ma2023}.
Nevertheless, a double ASPT phase jointly protected by $\Z_2^{\sigma\tau}\text{(S)}\times\Z_2^{\sigma}\text{(SWSSB)}$ remains possible.
Therefore, the central problem is whether $U_{\sigma\tau}^{\rm DW}$ has the capacity to generate such a $\sigma\tau$-preserved double ASPT phase.
In the following, we will show that this last piece of the puzzle can be fully resolved based on tensor-network analysis.

\subsection{Tensor network construction and mixed anomaly}
We first decompose the CZ gate into local tensors
\begin{align}
\scalebox{0.8}{
\begin{tikzpicture}[scale=0.8]
\tikzstyle{sergio}=[rectangle,draw=none]
\filldraw[fill=white, draw=black, rounded corners] (-7.5,-1.25)--(-5.5,-1.25)--(-5.5,-0.5)--(-7.5,-0.5)--cycle;
\draw[line width=2pt, color=red] (-7,-0.5) -- (-7,0);
\draw[line width=2pt, color=red] (-7,-1.75) -- (-7,-1.25);
\draw[line width=2pt, color=red] (-6,-1.25) -- (-6,-1.75);
\draw[line width=2pt, color=red] (-6,-0.5) -- (-6,0);
\path (-6.5,-0.875) node [style=sergio]{\large CZ};
\path (-5.125,-0.875) node [style=sergio]{$=$};
\draw[line width=1pt] (-3.5,-0.875) -- (-4,-0.875);
\filldraw[fill=white, draw=black, rounded corners] (-4.75,-1.25)--(-4,-1.25)--(-4,-0.5)--(-4.75,-0.5)--cycle;
\filldraw[fill=white, draw=black, rounded corners] (-3.5,-1.25)--(-2.75,-1.25)--(-2.75,-0.5)--(-3.5,-0.5)--cycle;
\draw[line width=2pt, color=red] (-4.375,-0.5) -- (-4.375,0);
\draw[line width=2pt, color=red] (-4.375,-1.75) -- (-4.375,-1.25);
\draw[line width=2pt, color=red] (-3.125,-1.25) -- (-3.125,-1.75);
\draw[line width=2pt, color=red] (-3.125,-0.5) -- (-3.125,0);
\path (-4.375,-0.875) node [style=sergio]{\large $\mathsf{A}$};
\path (-3.125,-0.875) node [style=sergio]{\large $\mathsf{B}$};
\end{tikzpicture}},
\end{align}
whose tensor elements are
\begin{align}
\scalebox{0.8}{
\begin{tikzpicture}[scale=0.8]
\tikzstyle{sergio}=[rectangle,draw=none]
\path (0.75,1.5) node [style=sergio]{$=$};
\draw[line width=1pt] (0.375,1.5) -- (-0.125,1.5);
\filldraw[fill=white, draw=black, rounded corners] (-0.875,1.125)--(-0.125,1.125)--(-0.125,1.875)--(-0.875,1.875)--cycle;
\draw[line width=2pt, color=red] (-0.5,1.875) -- (-0.5,2.375);
\draw[line width=2pt, color=red] (-0.5,0.625) -- (-0.5,1.125);
\path (-0.5,1.5) node [style=sergio]{\large $\mathsf{A}$};
\path (0.25,1.875) node [style=sergio]{1};
\path (1.75,1.5) node [style=sergio]{$\left[\begin{matrix} 0 & 0 \\ 0 & 1 \end{matrix}\right]$};
\tikzstyle{sergio}=[rectangle,draw=none]
\path (-3.25,1.5) node [style=sergio]{$=$};
\draw[line width=1pt] (-3.625,1.5) -- (-4.125,1.5);
\filldraw[fill=white, draw=black, rounded corners] (-4.875,1.125)--(-4.125,1.125)--(-4.125,1.875)--(-4.875,1.875)--cycle;
\draw[line width=2pt, color=red] (-4.5,1.875) -- (-4.5,2.375);
\draw[line width=2pt, color=red] (-4.5,0.625) -- (-4.5,1.125);
\path (-4.5,1.5) node [style=sergio]{\large $\mathsf{A}$};
\path (-3.75,1.875) node [style=sergio]{0};
\path (-2.25,1.5) node [style=sergio]{$\left[\begin{matrix} 1 & 0 \\ 0 & 0 \end{matrix}\right]$};
\path (-1.5,1) node [style=sergio]{,};
\path (2.5,1) node [style=sergio]{,};
\path (-1.5,-1.5) node [style=sergio]{,};
\path (2.5,-1.5) node [style=sergio]{.};
\path (0.75,-1) node [style=sergio]{$=$};
\draw[line width=1pt] (-0.375,-1) -- (-0.875,-1);
\filldraw[fill=white, draw=black, rounded corners] (-0.375,-1.375)--(0.375,-1.375)--(0.375,-0.625)--(-0.375,-0.625)--cycle;
\draw[line width=2pt, color=red] (0,-0.625) -- (0,-0.125);
\draw[line width=2pt, color=red] (0,-1.875) -- (0,-1.375);
\path (0,-1) node [style=sergio]{\large $\mathsf{B}$};
\path (-0.75,-0.625) node [style=sergio]{1};
\path (1.75,-1) node [style=sergio]{$\left[\begin{matrix} 1 & 0 \\ 0 & -1 \end{matrix}\right]$};
\path (-3.25,-1) node [style=sergio]{$=$};
\draw[line width=1pt] (-4.875,-1) -- (-4.375,-1);
\filldraw[fill=white, draw=black, rounded corners] (-4.375,-1.375)--(-3.625,-1.375)--(-3.625,-0.625)--(-4.375,-0.625)--cycle;
\draw[line width=2pt, color=red] (-4,-0.625) -- (-4,-0.125);
\draw[line width=2pt, color=red] (-4,-1.875) -- (-4,-1.375);
\path (-4,-1) node [style=sergio]{\large $\mathsf{B}$};
\path (-4.75,-0.625) node [style=sergio]{0};
\path (-2.25,-1) node [style=sergio]{$\left[\begin{matrix} 1 & 0 \\ 0 & 1 \end{matrix}\right]$};
\end{tikzpicture}}
\end{align}
To understand how this gate constitutes the DW duality map that generates nontrivial SPT order, we consider the following symmetry transformations
\begin{align}
\scalebox{0.8}{
\begin{tikzpicture}[scale=0.8]
\tikzstyle{sergio}=[rectangle,draw=none]
\draw[line width=1pt] (-12.625,1.5) -- (-13.125,1.5);
\filldraw[fill=white, draw=black, rounded corners] (-13.875,1.125)--(-13.125,1.125)--(-13.125,1.875)--(-13.875,1.875)--cycle;
\draw[line width=2pt, color=red] (-13.5,1.875) -- (-13.5,3.125);
\draw[line width=2pt, color=red] (-13.5,0.625) -- (-13.5,1.125);
\path (-13.5,1.5) node [style=sergio]{\large $\mathsf{A}$};
\filldraw[fill=white, draw=black] (-13.5,2.5)circle (10pt);
\path (-13.5,2.5) node [style=sergio]{\large $X$};
\draw[line width=1pt] (-9.875,1.5) -- (-11.125,1.5);
\filldraw[fill=white, draw=black, rounded corners] (-11.875,1.125)--(-11.125,1.125)--(-11.125,1.875)--(-11.875,1.875)--cycle;
\draw[line width=2pt, color=red] (-11.5,1.875) -- (-11.5,2.375);
\draw[line width=2pt, color=red] (-11.5,-0.125) -- (-11.5,1.125);
\path (-11.5,1.5) node [style=sergio]{\large $\mathsf{A}$};
\filldraw[fill=white, draw=black] (-10.5,1.5)circle (10pt);
\path (-10.5,1.5) node [style=sergio]{\large $X$};
\filldraw[fill=white, draw=black] (-11.5,0.5)circle (10pt);
\path (-11.5,0.5) node [style=sergio]{\large $X$};
\path (-12.25,1.5) node [style=sergio]{$=$};
\path (-9.75,0) node [style=sergio]{,};
\path (-2.75,0) node [style=sergio]{,};
\draw[line width=1pt] (-7.625,1.5) -- (-8.125,1.5);
\filldraw[fill=white, draw=black, rounded corners] (-8.875,1.125)--(-8.125,1.125)--(-8.125,1.875)--(-8.875,1.875)--cycle;
\draw[line width=2pt, color=red] (-8.5,1.875) -- (-8.5,3.125);
\draw[line width=2pt, color=red] (-8.5,0.625) -- (-8.5,1.125);
\path (-8.5,1.5) node [style=sergio]{\large $\mathsf{A}$};
\filldraw[fill=white, draw=black] (-8.5,2.5)circle (10pt);
\path (-8.5,2.5) node [style=sergio]{\large $Z$};
\draw[line width=1pt] (-2.875,1.5) -- (-3.375,1.5);
\filldraw[fill=white, draw=black, rounded corners] (-4.125,1.125)--(-3.375,1.125)--(-3.375,1.875)--(-4.125,1.875)--cycle;
\draw[line width=2pt, color=red] (-3.75,1.875) -- (-3.75,2.375);
\draw[line width=2pt, color=red] (-3.75,-0.125) -- (-3.75,1.125);
\path (-3.75,1.5) node [style=sergio]{\large $\mathsf{A}$};
\filldraw[fill=white, draw=black] (-3.75,0.5)circle (10pt);
\path (-3.75,0.5) node [style=sergio]{\large $Z$};
\path (-7.25,1.5) node [style=sergio]{$=$};
\draw[line width=1pt] (-4.875,1.5) -- (-6.125,1.5);
\filldraw[fill=white, draw=black, rounded corners] (-6.875,1.125)--(-6.125,1.125)--(-6.125,1.875)--(-6.875,1.875)--cycle;
\draw[line width=2pt, color=red] (-6.5,1.875) -- (-6.5,2.375);
\draw[line width=2pt, color=red] (-6.5,0.625) -- (-6.5,1.125);
\path (-6.5,1.5) node [style=sergio]{\large $\mathsf{A}$};
\filldraw[fill=white, draw=black] (-5.5,1.5)circle (10pt);
\path (-5.5,1.5) node [style=sergio]{\large $Z$};
\path (-4.5,1.5) node [style=sergio]{$=$};
\end{tikzpicture}}\label{Equ: CZ-A}
\end{align}
and
\begin{align}
\scalebox{0.8}{
\begin{tikzpicture}[scale=0.8]
\tikzstyle{sergio}=[rectangle,draw=none]
\draw[line width=1pt] (-12.375,1.5) -- (-12.875,1.5);
\filldraw[fill=white, draw=black, rounded corners] (-12.375,1.125)--(-11.625,1.125)--(-11.625,1.875)--(-12.375,1.875)--cycle;
\draw[line width=2pt, color=red] (-12,1.875) -- (-12,3.125);
\draw[line width=2pt, color=red] (-12,0.625) -- (-12,1.125);
\path (-12,1.5) node [style=sergio]{\large $\mathsf{B}$};
\filldraw[fill=white, draw=black] (-12,2.5)circle (10pt);
\path (-12,2.5) node [style=sergio]{\large $X$};
\path (-11.25,1.5) node [style=sergio]{$=$};
\draw[line width=1pt] (-9.625,1.5) -- (-10.875,1.5);
\filldraw[fill=white, draw=black, rounded corners] (-9.625,1.125)--(-8.875,1.125)--(-8.875,1.875)--(-9.625,1.875)--cycle;
\draw[line width=2pt, color=red] (-9.25,1.875) -- (-9.25,2.375);
\draw[line width=2pt, color=red] (-9.25,-0.125) -- (-9.25,1.125);
\path (-9.25,1.5) node [style=sergio]{\large $\mathsf{B}$};
\filldraw[fill=white, draw=black] (-10.25,1.5)circle (10pt);
\path (-10.25,1.5) node [style=sergio]{\large $Z$};
\filldraw[fill=white, draw=black] (-9.25,0.5)circle (10pt);
\path (-9.25,0.5) node [style=sergio]{\large $X$};
\draw[line width=1pt] (-7.375,1.5) -- (-7.875,1.5);
\filldraw[fill=white, draw=black, rounded corners] (-7.375,1.125)--(-6.625,1.125)--(-6.625,1.875)--(-7.375,1.875)--cycle;
\draw[line width=2pt, color=red] (-7,1.875) -- (-7,3.125);
\draw[line width=2pt, color=red] (-7,0.625) -- (-7,1.125);
\path (-7,1.5) node [style=sergio]{\large $\mathsf{B}$};
\filldraw[fill=white, draw=black] (-7,2.5)circle (10pt);
\path (-7,2.5) node [style=sergio]{\large $Z$};
\draw[line width=1pt] (-3.125,1.5) -- (-2.625,1.5);
\filldraw[fill=white, draw=black, rounded corners] (-2.625,1.125)--(-1.875,1.125)--(-1.875,1.875)--(-2.625,1.875)--cycle;
\draw[line width=2pt, color=red] (-2.25,1.875) -- (-2.25,2.375);
\draw[line width=2pt, color=red] (-2.25,-0.125) -- (-2.25,1.125);
\path (-2.25,1.5) node [style=sergio]{\large $\mathsf{B}$};
\filldraw[fill=white, draw=black] (-2.25,0.5)circle (10pt);
\path (-2.25,0.5) node [style=sergio]{\large $Z$};
\path (-6.25,1.5) node [style=sergio]{$=$};
\draw[line width=1pt] (-4.625,1.5) -- (-5.875,1.5);
\filldraw[fill=white, draw=black, rounded corners] (-4.625,1.125)--(-3.875,1.125)--(-3.875,1.875)--(-4.625,1.875)--cycle;
\draw[line width=2pt, color=red] (-4.25,1.875) -- (-4.25,2.375);
\draw[line width=2pt, color=red] (-4.25,0.625) -- (-4.25,1.125);
\path (-4.25,1.5) node [style=sergio]{\large $\mathsf{B}$};
\filldraw[fill=white, draw=black] (-5.25,1.5)circle (10pt);
\path (-5.25,1.5) node [style=sergio]{\large $X$};
\path (-3.5,1.5) node [style=sergio]{$=$};
\path (-8.75,0) node [style=sergio]{,};
\path (-1.75,0) node [style=sergio]{.};
\end{tikzpicture}}\label{Equ: CZ-B}
\end{align}

Next, we construct the matrix product operator (MPO) representation of the DW duality map $U^{\rm DW}$ by contracting the local tensors, i.e.,
\begin{align}
\scalebox{0.8}{
\begin{tikzpicture}[scale=0.8]
\tikzstyle{sergio}=[rectangle,draw=none]
\filldraw[fill=white, draw=black, rounded corners] (-9,-1.375)--(-7,-1.375)--(-7,-0.625)--(-9,-0.625)--cycle;
\draw[line width=2pt, color=red] (-8.5,-0.625) -- (-8.5,-0.125);
\draw[line width=2pt, color=red] (-8.5,-1.875) -- (-8.5,-1.375);
\draw[line width=2pt, color=red] (-7.5,-1.375) -- (-7.5,-1.875);
\draw[line width=2pt, color=red] (-7.5,-0.625) -- (-7.5,-0.125);
\path (-8,-1) node [style=sergio]{\large $\mathsf{U}$};
\draw[line width=1pt] (-5.25,-2.25) -- (-5.75,-2.25);
\filldraw[fill=white, draw=black, rounded corners] (-5.25,-1.375)--(-4.5,-1.375)--(-4.5,-0.625)--(-5.25,-0.625)--cycle;
\draw[line width=2pt, color=red] (-4.875,-0.625) -- (-4.875,-0.125);
\draw[line width=2pt, color=red] (-4.875,-1.875) -- (-4.875,-1.375);
\path (-4.875,-1) node [style=sergio]{\large $\mathsf{A}$};
\path (-6.125,-1) node [style=sergio]{$=$};
\draw[line width=1pt] (-2.75,0.25) -- (-3.25,0.25);
\filldraw[fill=white, draw=black, rounded corners] (-4,-0.125)--(-3.25,-0.125)--(-3.25,0.625)--(-4,0.625)--cycle;
\filldraw[fill=white, draw=black, rounded corners] (-5.25,-2.625)--(-4.5,-2.625)--(-4.5,-1.875)--(-5.25,-1.875)--cycle;
\draw[line width=2pt, color=red] (-3.625,0.625) -- (-3.625,1.125);
\draw[line width=2pt, color=red] (-4.875,-2.625) -- (-4.875,-3.125);
\draw[line width=2pt, color=red] (-3.625,-0.625) -- (-3.625,-0.125);
\path (-3.625,0.25) node [style=sergio]{\large $\mathsf{A}$};
\path (-4.875,-2.25) node [style=sergio]{\large $\mathsf{B}$};
\filldraw[fill=white, draw=black, rounded corners] (-4,-1.375)--(-3.25,-1.375)--(-3.25,-0.625)--(-4,-0.625)--cycle;
\draw[line width=2pt, color=red] (-3.625,-1.375) -- (-3.625,-1.875);
\path (-3.625,-1) node [style=sergio]{\large $\mathsf{B}$};
\draw[line width=1pt] (-4,-1) -- (-4.5,-1);
\draw[line width=1pt] (-9,-1) -- (-9.5,-1);
\draw[line width=1pt] (-6.5,-1) -- (-7,-1);
\end{tikzpicture}}.
\label{Equ: MPO for DW}
\end{align}
Based on Eq.~\eqref{Equ: CZ-A} and \eqref{Equ: CZ-B}, one obtains the corresponding virtual symmetry actions on the MPO tensor of the DW map
\begin{align}
\scalebox{0.8}{
\begin{tikzpicture}[scale=0.8]
\tikzstyle{sergio}=[rectangle,draw=none]
\filldraw[fill=white, draw=black, rounded corners] (-4.5,-1.375)--(-2.5,-1.375)--(-2.5,-0.625)--(-4.5,-0.625)--cycle;
\draw[line width=2pt, color=red] (-4,-0.625) -- (-4,-0.125);
\draw[line width=2pt, color=red] (-4,-2.625) -- (-4,-1.375);
\draw[line width=2pt, color=red] (-3,-1.375) -- (-3,-1.875);
\draw[line width=2pt, color=red] (-3,-0.625) -- (-3,-0.125);
\path (-3.5,-1) node [style=sergio]{\large $\mathsf{U}$};
\path (-6.125,-1) node [style=sergio]{$=$};
\draw[line width=1pt] (-4.5,-1) -- (-5.75,-1);
\draw[line width=1pt] (-1.25,-1) -- (-2.5,-1);
\filldraw[fill=white, draw=black] (-4,-2)circle (10pt);
\path (-4,-2) node [style=sergio]{\large $X$};
\filldraw[fill=white, draw=black, rounded corners] (-9,-1.375)--(-7,-1.375)--(-7,-0.625)--(-9,-0.625)--cycle;
\draw[line width=2pt, color=red] (-8.5,-0.625) -- (-8.5,0.625);
\draw[line width=2pt, color=red] (-8.5,-1.875) -- (-8.5,-1.375);
\draw[line width=2pt, color=red] (-7.5,-1.375) -- (-7.5,-1.875);
\draw[line width=2pt, color=red] (-7.5,-0.625) -- (-7.5,-0.125);
\path (-8,-1) node [style=sergio]{\large $\mathsf{U}$};
\path (-6.125,-1) node [style=sergio]{$=$};
\draw[line width=1pt] (-9,-1) -- (-9.5,-1);
\draw[line width=1pt] (-6.5,-1) -- (-7,-1);
\filldraw[fill=white, draw=black] (-8.5,0)circle (10pt);
\path (-8.5,0) node [style=sergio]{\large $X$};
\filldraw[fill=white, draw=black] (-1.875,-1)circle (10pt);
\path (-1.875,-1) node [style=sergio]{\large $Z$};
\filldraw[fill=white, draw=black] (-5.125,-1)circle (10pt);
\path (-5.125,-1) node [style=sergio]{\large $Z$};
\end{tikzpicture}},
\label{Equ: virtual_sigma}
\end{align}
and
\begin{align}
\scalebox{0.8}{
\begin{tikzpicture}[scale=0.8]
\tikzstyle{sergio}=[rectangle,draw=none]
\filldraw[fill=white, draw=black, rounded corners] (-4.5,-1.375)--(-2.5,-1.375)--(-2.5,-0.625)--(-4.5,-0.625)--cycle;
\draw[line width=2pt, color=red] (-4,-0.625) -- (-4,-0.125);
\draw[line width=2pt, color=red] (-4,-1.875) -- (-4,-1.375);
\draw[line width=2pt, color=red] (-3,-1.375) -- (-3,-2.625);
\draw[line width=2pt, color=red] (-3,-0.625) -- (-3,-0.125);
\path (-3.5,-1) node [style=sergio]{\large $\mathsf{U}$};
\path (-6.125,-1) node [style=sergio]{$=$};
\draw[line width=1pt] (-4.5,-1) -- (-5.75,-1);
\draw[line width=1pt] (-1.25,-1) -- (-2.5,-1);
\filldraw[fill=white, draw=black] (-3,-2)circle (10pt);
\path (-3,-2) node [style=sergio]{\large $X$};
\filldraw[fill=white, draw=black, rounded corners] (-9,-1.375)--(-7,-1.375)--(-7,-0.625)--(-9,-0.625)--cycle;
\draw[line width=2pt, color=red] (-8.5,-0.625) -- (-8.5,-0.125);
\draw[line width=2pt, color=red] (-8.5,-1.875) -- (-8.5,-1.375);
\draw[line width=2pt, color=red] (-7.5,-1.375) -- (-7.5,-1.875);
\draw[line width=2pt, color=red] (-7.5,-0.625) -- (-7.5,0.625);
\path (-8,-1) node [style=sergio]{\large $\mathsf{U}$};
\path (-6.125,-1) node [style=sergio]{$=$};
\draw[line width=1pt] (-9,-1) -- (-9.5,-1);
\draw[line width=1pt] (-6.5,-1) -- (-7,-1);
\filldraw[fill=white, draw=black] (-7.5,0)circle (10pt);
\path (-7.5,0) node [style=sergio]{\large $X$};
\filldraw[fill=white, draw=black] (-1.875,-1)circle (10pt);
\path (-1.875,-1) node [style=sergio]{\large $X$};
\filldraw[fill=white, draw=black] (-5.125,-1)circle (10pt);
\path (-5.125,-1) node [style=sergio]{\large $X$};
\end{tikzpicture}}.
\label{Equ: virtual_tau}
\end{align}
Therefore, this DW duality map will introduce a nontrivial projective representation in the virtual space associated with two $\Z_2$ symmetries (here taken to be $\Z_2^{\sigma}$ and $\Z_2^{\tau}$, acting on the two sublattices).
This noncommutativity is precisely the mixed anomaly responsible for the resulting SPT order, which therefore requires joint protection by both symmetries.

Moreover, we consider the transformation induced by the corresponding diagonal symmetry $\Z_2^{\sigma\tau}$
\begin{align}
\scalebox{0.8}{
\begin{tikzpicture}[scale=0.8]
\tikzstyle{sergio}=[rectangle,draw=none]
\filldraw[fill=white, draw=black, rounded corners] (-4.5,-1.375)--(-2.5,-1.375)--(-2.5,-0.625)--(-4.5,-0.625)--cycle;
\draw[line width=2pt, color=red] (-4,-0.625) -- (-4,-0.125);
\draw[line width=2pt, color=red] (-4,-2.625) -- (-4,-1.375);
\draw[line width=2pt, color=red] (-3,-1.375) -- (-3,-2.625);
\draw[line width=2pt, color=red] (-3,-0.625) -- (-3,-0.125);
\path (-3.5,-1) node [style=sergio]{\large $\mathsf{U}$};
\path (-6.125,-1) node [style=sergio]{$=$};
\draw[line width=1pt] (-4.5,-1) -- (-5.75,-1);
\draw[line width=1pt] (-1.25,-1) -- (-2.5,-1);
\filldraw[fill=white, draw=black] (-4,-2)circle (10pt);
\path (-4,-2) node [style=sergio]{\large $X$};
\filldraw[fill=white, draw=black] (-3,-2)circle (10pt);
\path (-3,-2) node [style=sergio]{\large $X$};
\filldraw[fill=white, draw=black, rounded corners] (-9,-1.375)--(-7,-1.375)--(-7,-0.625)--(-9,-0.625)--cycle;
\draw[line width=2pt, color=red] (-8.5,-0.625) -- (-8.5,0.625);
\draw[line width=2pt, color=red] (-8.5,-1.875) -- (-8.5,-1.375);
\draw[line width=2pt, color=red] (-7.5,-1.375) -- (-7.5,-1.875);
\draw[line width=2pt, color=red] (-7.5,-0.625) -- (-7.5,0.625);
\path (-8,-1) node [style=sergio]{\large $\mathsf{U}$};
\path (-6.125,-1) node [style=sergio]{$=$};
\draw[line width=1pt] (-9,-1) -- (-9.5,-1);
\draw[line width=1pt] (-6.5,-1) -- (-7,-1);
\filldraw[fill=white, draw=black] (-7.5,0)circle (10pt);
\path (-7.5,0) node [style=sergio]{\large $X$};
\filldraw[fill=white, draw=black] (-8.5,0)circle (10pt);
\path (-8.5,0) node [style=sergio]{\large $X$};
\filldraw[fill=white, draw=black] (-1.875,-1)circle (10pt);
\path (-1.875,-1) node [style=sergio]{\large $Y$};
\filldraw[fill=white, draw=black] (-5.125,-1)circle (10pt);
\path (-5.125,-1) node [style=sergio]{\large $Y$};
\end{tikzpicture}},
\label{Equ: virtual_sigmatau}
\end{align}
which can be viewed as the product of the $\sigma$ and $\tau$ generators.
In this case, the diagonal generator inherits the same nontrivial projective structure in the virtual space, which means that it also exhibits a mixed anomaly with a single generator, e.g., $\Z_2^{\sigma}$.
This provides a tensor-network criterion for identifying mixed anomalies even when the strong symmetry is diagonal: $\rho_{\sigma\tau}^{\sigma\tau\text{-P double ASPT}}$ defined in Eq.~\eqref{Equ: OD-Double ASPT} indeed realizes a double ASPT phase jointly protected by $\Z_2^{\sigma\tau}\text{(S)}\times\Z_2^{\sigma}\text{(SWSSB)}$.

Thus, the $\sigma\tau$-preserved variants most clearly demonstrate the novelty of open-system phases.
They show that the strong protecting symmetry can migrate from an elementary generator to a diagonal subgroup, while the remaining weak or SWSSB symmetry still participates in protecting nontrivial topology.
Even in the simplest $\Z_2$ setting, the interplay among strong, weak, and SWSSB already produces a remarkably rich phase structure, underlying the unconventional criticality discussed below.

\section{Phase transitions}
\label{Sec: Edge}
\begin{figure}
    \centering
    \includegraphics[width=0.85\linewidth]{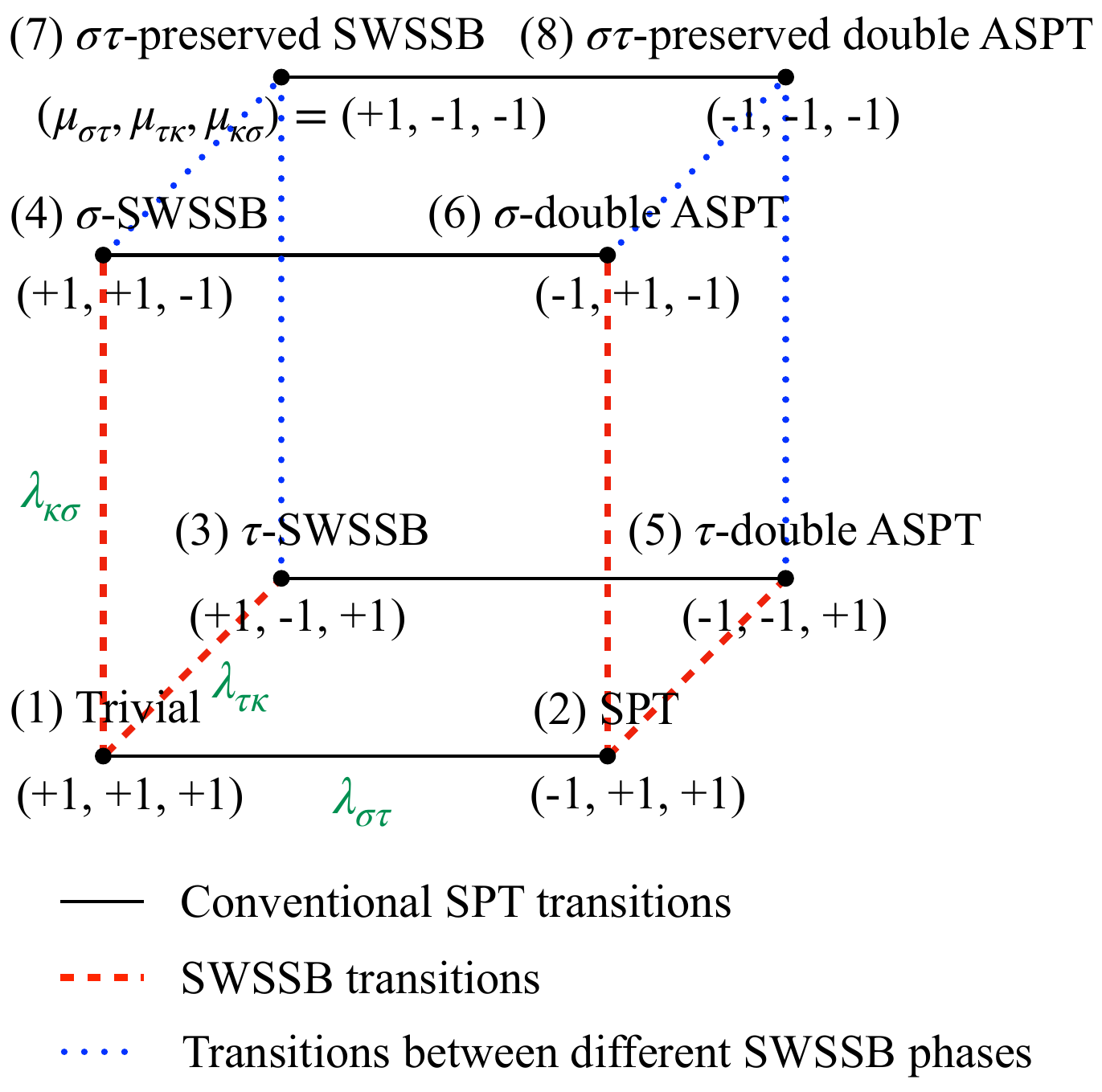}
    \caption{Phase cube of the three-$\Z_2$ construction.
    Each vertex corresponds to one of the eight mixed-state phases obtained by tracing out the $\kappa$ spins from the pure states specified by $(\mu_{\sigma\tau},\mu_{\tau\kappa},\mu_{\kappa\sigma})\in\{\pm1\}^3$.
    The vertices are labeled by their characteristic orders, including trivial, ASPT, SWSSB, double ASPT, and $\sigma\tau$-preserved variants.
    Edges of the cube represent single-flip transitions, i.e., the addition or removal of a single DW duality map in the pure-state picture.
    Conventional SPT-type transitions, such as trivial $\leftrightarrow$ SPT or SWSSB $\leftrightarrow$ ASPT, are shown in black.
    Red dashed edges denote SWSSB transitions, a phenomenon intrinsic to open systems.
    Blue dotted edges denote exchange between distinct SWSSB patterns with different strong subgroups.}
    \label{Fig: cube}
\end{figure}
In this section, we investigate the phase transitions between the above eight mixed-state phases from the purification perspective.
As illustrated schematically in Fig.~\ref{Fig: cube}, the eight phases are placed at the vertices of a cube according to three signs $(\mu_{\sigma\tau},\mu_{\tau\kappa},\mu_{\kappa\sigma})$.
Flipping a single sign corresponds to adding or removing one DW duality map in the underlying pure-state construction, realizing a conventional trivial $\leftrightarrow$ SPT transition in closed systems.
However, at the mixed-state level, this procedure gives rise to twelve inequivalent transitions, which we now classify in detail.

\subsection{Trivial $\leftrightarrow$ SPT / Trivial $\leftrightarrow$ SWSSB}
Flipping $\mu_{\sigma\tau}$ while keeping at least one strong $\Z_2$ symmetry drives a transition analogous to the conventional trivial $\leftrightarrow$ SPT transition of a closed system~\cite{Pollmann2010, Pollmann2012}, since $U_{\sigma\tau}^{\rm DW}$ commutes with the procedure of tracing out $\kappa$ spins.
As in the pure-state case, this transition can be further mapped to two decoupled critical Ising chains via a non-invertible Kennedy–Tasaki (KT) transformation~\cite{Li2022}, yielding a total central charge $c=1$.
The specific topological phase realized across the transition depends on which symmetry subgroup remains strong, namely $\Z_2^\sigma$ (No.~3$\leftrightarrow$5), $\Z_2^\tau$ (No.~4$\leftrightarrow$6), the diagonal $\Z_2^{\sigma\tau}$ (No. 7$\leftrightarrow$8), or both $\Z_2$ symmetries (No.~1$\leftrightarrow$2).

By contrast, flipping $\mu_{\tau\kappa}$ or $\mu_{\kappa\sigma}$ drives a qualitatively different transition in which one symmetry subgroup undergoes SWSSB.
At the level of correlators, this transition is marked by the emergence of long-range order in R\'enyi-$2$ or fidelity correlators.
The critical point belongs to the Ising universality class with central charge $c=1/2$ from the double space formalism~\cite{Ma2025B}, whose reduction compared to the purified model originates from the fact that the DW duality map generating this transition involves $\kappa$ spins to be traced out.
Concretely, the underlying unitary transformation effectively defines a quantum channel acting on the physical degrees of freedom,
\begin{align}
    \mathcal{E}(\rho_{\rm p})\equiv \Tr_{\rm a}\left[U\left(\rho_{\rm p}\otimes \ket{0}\hspace{-0.5mm}\bra{0}_{\rm a}\right)\right],
\end{align}
provided that the ancilla is refreshed at the beginning of the transformation.
Therefore, only the direction of trivial $\rightarrow$ SWSSB admits such a channel description.
In the following, we explicitly derive its tensor-network representation and illustrate how this channel generates a mixed state exhibiting SWSSB order through a local tensor equation.

Generalizing the LPDO structure to represent a quantum channel requires the introduction of an additional physical index at each site~\cite{Torlai2023}.
Consider the DW duality map $U_{\kappa\sigma}^{\rm DW}$ acting on the product state between $\ket{\psi_{\sigma\tau}}$ and $\ket{\psi_{\kappa}^{\rm Trivial}}$, the effective quantum channel after tracing out $\kappa$ spins can be represented as
\begin{align}
\scalebox{0.8}{
\begin{tikzpicture}[scale=0.8]
\tikzstyle{sergio}=[rectangle,draw=none]
\filldraw[fill=white, draw=black, rounded corners] (-12.75,-1.375)--(-12,-1.375)--(-12,-0.625)--(-12.75,-0.625)--cycle;
\draw[line width=2pt, color=red] (-12.375,-0.625) -- (-12.375,-0.125);
\draw[line width=2pt, color=red] (-12.375,-1.875) -- (-12.375,-1.375);
\draw[line width=3pt, color=blue] (-12,-1.1875) -- (-11.75,-1.1875) -- (-11.75,-1.875);
\path (-12.375,-1) node [style=sergio]{\large $\mathsf{E}$};
\path (-11,-1) node [style=sergio]{$=$};
\draw[line width=1pt] (-12.75,-1) -- (-13.25,-1);
\draw[line width=1pt] (-11.5,-1) -- (-12,-1);
\filldraw[fill=white, draw=black, rounded corners] (-10,-1.375)--(-8,-1.375)--(-8,-0.625)--(-10,-0.625)--cycle;
\draw[line width=2pt, color=red] (-9.5,-0.625) -- (-9.5,-0.125);
\draw[line width=2pt, color=red] (-9.5,-1.875) -- (-9.5,-1.375);
\draw[line width=2pt, color=red] (-8.5,-1.375) -- (-8.5,-1.875);
\draw[line width=3pt, color=blue] (-8.5,-0.625) -- (-8.5,-0.125) -- (-7.875,-0.125) -- (-7.875,-3);
\path (-9,-1) node [style=sergio]{\large $\mathsf{U}$};
\draw[line width=1pt] (-10,-1) -- (-10.5,-1);
\draw[line width=1pt] (-7.5,-1) -- (-8,-1);
\filldraw[fill=white, draw=black] (-8.875,-2.625)--(-8.125,-2.625)--(-8.125,-1.875)--(-8.875,-1.875)--cycle;
\path (-8.5,-2.25) node [style=sergio]{$\ket{\rightarrow}$};
\end{tikzpicture}},
\end{align}
where $\mathsf{U}$ denotes the MPO representation of the DW duality map, as defined in Eq.~\eqref{Equ: MPO for DW}.
The nonzero tensor elements of $\mathsf{E}$ are given by
\begin{align}
\scalebox{0.8}{
\begin{tikzpicture}[scale=0.8]
\tikzstyle{sergio}=[rectangle,draw=none]
\filldraw[fill=white, draw=black, rounded corners] (-12.875,-1.375)--(-12.125,-1.375)--(-12.125,-0.625)--(-12.875,-0.625)--cycle;
\draw[line width=2pt, color=red] (-12.5,-0.625) -- (-12.5,-0.125);
\draw[line width=2pt, color=red] (-12.5,-1.875) -- (-12.5,-1.375);
\draw[line width=3pt, color=blue] (-12.125,-1.1875) -- (-11.875,-1.1875) -- (-11.875,-1.875);
\path (-12.5,-1) node [style=sergio]{\large $\mathsf{E}$};
\path (-11.0625,-1) node [style=sergio]{$=I$};
\draw[line width=1pt] (-12.875,-1) -- (-13.375,-1);
\draw[line width=1pt] (-11.625,-1) -- (-12.125,-1);
\path (-13.375,-0.75) node [style=sergio]{$0$};
\path (-11.625,-0.75) node [style=sergio]{$0$};
\path (-11.625,-1.875) node [style=sergio]{$0$};
\filldraw[fill=white, draw=black, rounded corners] (-9.375,-1.375)--(-8.625,-1.375)--(-8.625,-0.625)--(-9.375,-0.625)--cycle;
\draw[line width=2pt, color=red] (-9,-0.625) -- (-9,-0.125);
\draw[line width=2pt, color=red] (-9,-1.875) -- (-9,-1.375);
\draw[line width=3pt, color=blue] (-8.625,-1.1875) -- (-8.375,-1.1875) -- (-8.375,-1.875);
\path (-9,-1) node [style=sergio]{\large $\mathsf{E}$};
\path (-7.5,-1) node [style=sergio]{$=Z$};
\draw[line width=1pt] (-9.375,-1) -- (-9.875,-1);
\draw[line width=1pt] (-8.125,-1) -- (-8.625,-1);
\path (-9.875,-0.75) node [style=sergio]{$0$};
\path (-8.125,-0.75) node [style=sergio]{$1$};
\path (-8.125,-1.875) node [style=sergio]{$1$};
\filldraw[fill=white, draw=black, rounded corners] (-12.875,-3.875)--(-12.125,-3.875)--(-12.125,-3.125)--(-12.875,-3.125)--cycle;
\draw[line width=2pt, color=red] (-12.5,-3.125) -- (-12.5,-2.625);
\draw[line width=2pt, color=red] (-12.5,-4.375) -- (-12.5,-3.875);
\draw[line width=3pt, color=blue] (-12.125,-3.6875) -- (-11.875,-3.6875) -- (-11.875,-4.375);
\path (-12.5,-3.5) node [style=sergio]{\large $\mathsf{E}$};
\path (-11,-3.5) node [style=sergio]{$=Z$};
\draw[line width=1pt] (-12.875,-3.5) -- (-13.375,-3.5);
\draw[line width=1pt] (-11.625,-3.5) -- (-12.125,-3.5);
\path (-13.375,-3.25) node [style=sergio]{$1$};
\path (-11.625,-3.25) node [style=sergio]{$0$};
\path (-11.625,-4.375) node [style=sergio]{$0$};
\filldraw[fill=white, draw=black, rounded corners] (-9.375,-3.875)--(-8.625,-3.875)--(-8.625,-3.125)--(-9.375,-3.125)--cycle;
\draw[line width=2pt, color=red] (-9,-3.125) -- (-9,-2.625);
\draw[line width=2pt, color=red] (-9,-4.375) -- (-9,-3.875);
\draw[line width=3pt, color=blue] (-8.625,-3.6875) -- (-8.375,-3.6875) -- (-8.375,-4.375);
\path (-9,-3.5) node [style=sergio]{\large $\mathsf{E}$};
\path (-7.5625,-3.5) node [style=sergio]{$=I$};
\draw[line width=1pt] (-9.375,-3.5) -- (-9.875,-3.5);
\draw[line width=1pt] (-8.125,-3.5) -- (-8.625,-3.5);
\path (-9.875,-3.25) node [style=sergio]{$1$};
\path (-8.125,-3.25) node [style=sergio]{$1$};
\path (-8.125,-4.375) node [style=sergio]{$1$};
\path (-10.625,-2) node [style=sergio]{,};
\path (-7.125,-2) node [style=sergio]{,};
\path (-10.625,-4.5) node [style=sergio]{,};
\path (-7.125,-4.5) node [style=sergio]{.};
\end{tikzpicture}}
\end{align}
Importantly, this channel tensor inherits the symmetry transformation from Eq.~\eqref{Equ: virtual_tau}, which means that
\begin{align}
\scalebox{0.8}{
\begin{tikzpicture}[scale=0.8]
\tikzstyle{sergio}=[rectangle,draw=none]
\filldraw[fill=white, draw=black, rounded corners] (-12.375,-1.375)--(-11.625,-1.375)--(-11.625,-0.625)--(-12.375,-0.625)--cycle;
\draw[line width=2pt, color=red] (-12,-0.625) -- (-12,-0.125);
\draw[line width=2pt, color=red] (-12,-1.875) -- (-12,-1.375);
\draw[line width=3pt, color=blue] (-11.625,-1.1875) -- (-11.375,-1.1875) -- (-11.375,-2.75);
\path (-12,-1) node [style=sergio]{\large $\mathsf{E}$};
\path (-10.75,-1) node [style=sergio]{$=$};
\draw[line width=1pt] (-12.375,-1) -- (-12.875,-1);
\draw[line width=1pt] (-11.125,-1) -- (-11.625,-1);
\filldraw[fill=white, draw=black] (-11.375,-2)circle (10pt);
\path (-11.375,-2) node [style=sergio]{\large $X$};
\filldraw[fill=white, draw=black, rounded corners] (-8.875,-1.375)--(-8.125,-1.375)--(-8.125,-0.625)--(-8.875,-0.625)--cycle;
\draw[line width=2pt, color=red] (-8.5,-0.625) -- (-8.5,-0.125);
\draw[line width=2pt, color=red] (-8.5,-1.875) -- (-8.5,-1.375);
\draw[line width=3pt, color=blue] (-8.125,-1.1875) -- (-7.875,-1.1875) -- (-7.875,-1.875);
\path (-8.5,-1) node [style=sergio]{\large $\mathsf{E}$};
\draw[line width=1pt] (-8.875,-1) -- (-10.375,-1);
\draw[line width=1pt] (-6.625,-1) -- (-8.125,-1);
\filldraw[fill=white, draw=black] (-7.375,-1)circle (10pt);
\path (-7.375,-1) node [style=sergio]{\large $X$};
\filldraw[fill=white, draw=black] (-9.625,-1)circle (10pt);
\path (-9.625,-1) node [style=sergio]{\large $X$};
\end{tikzpicture}}.
\end{align}
As a result, when this quantum channel acts on a short-range correlated state (i.e., a state that is both weakly and strongly injective), the output LPDO exhibits the symmetry transformation structure given in Eq.~\eqref{Equ: Strong injectivity}.
The breakdown of strong injectivity provides a direct tensor-network signature of the resulting SWSSB order.

By contrast, the reverse transition cannot be generated by a local quantum channel derived from the same unitary, since in the purified picture the $\kappa$ spins are entangled with the $\sigma$ or $\tau$ spins through a mixed anomaly.
More generally, a quantum channel $\mathcal{E}$ is a completely positive (CP) and trace-preserving map~\cite{Nielsen2009}, whereas its inverse map $\mathcal{E}^{-1}\circ\mathcal{E}=\mathcal{I}$ need not remain CP and therefore cannot always be physically implemented~
\footnote{$\mathcal{E}^{-1}$ cannot be coherently implemented and only admits a quasi-probability decomposition, which can be realized through classical postprocessing techniques developed in quantum error mitigation schemes~\cite{Endo2018, Guo2022, Cai2023, Guo2023}.}.
Therefore, this intrinsic asymmetry reflects the fundamentally non-invertible nature of the quantum channel driving the trivial $\rightarrow$ SWSSB transition, which replaces the Kramers–Wannier duality familiar from closed systems.

This asymmetry suggests that the phase cube does not merely organize symmetry-breaking patterns, but also encodes which topological structures remain locally recoverable under symmetric quantum channels~\cite{Sang2025A, Sang2025B}.
Although state-dependent recovery maps such as the Petz map can be constructed for specific input states~\cite{Petz2003}, there is no universal symmetric recovery channel that restores the entire mixed-state phase structure.
Therefore, the trivial and SWSSB phases differ not only in their symmetry properties but also in their channel-theoretic reversibility, echoing the notion of logical information stability in quantum error-correcting (QEC) codes~\cite{Knill1997, Dennis2002, Beny2010}.

When the initial state preserves both strong symmetries but is topologically trivial, this transition provides the simplest realizations of open-system criticality driven by SWSSB (No.~1$\leftrightarrow$3, No.~1$\leftrightarrow$4).
A closely related scenario arises when an SPT phase (i.e., with an additional decoration between $\sigma$ and $\tau$) is driven into a double ASPT (No.~2$\leftrightarrow$5, No.~2$\leftrightarrow$6).
In this case, the system evolves from a conventional SPT protected by two strong subgroups to a phase in which one subgroup undergoes SWSSB, while the remaining strong–weak pair continues protecting a nontrivial topology.
The resulting criticality is therefore topological in nature, yet is accompanied by the onset of SWSSB.
In analogy to the gapless SPT phases in closed systems~\cite{Scaffidi2017, Li2024, Yu2026}, we identify this transition as a gapless ASPT phase intrinsic to open systems~\cite{Guo2025A}.

\subsection{$\sigma$ ($\tau$) SWSSB $\leftrightarrow$ $\sigma\tau$-preserved SWSSB}
An interesting phenomenon occurs when the system interpolates between two different SWSSB patterns.
For instance, flipping $\mu_{\tau\kappa}$ in the presence of $\mu_{\kappa\sigma}=-1$ and $\mu_{\sigma\tau}=+1$ drives a transition between a $\sigma$-SWSSB phase and a $\sigma\tau$-preserved SWSSB phase (No.~4$\leftrightarrow$7).  
Although both phases exhibit SWSSB, they differ fundamentally in the identity of the symmetry subgroup that remains strong: $\Z_2^{\sigma}$ in the $\sigma$-SWSSB phase, and the diagonal subgroup $\Z_2^{\sigma\tau}$ in the $\sigma\tau$-preserved SWSSB phase.
At the level of density matrices, both phases can be written as macroscopic mixtures of projectors onto distinct global symmetry sectors.
However, the correspondence between these sectors involves a nonlocal redefinition of symmetry operators, reflecting the fact that the strong symmetry is realized on inequivalent subgroups.
At the critical point, the system unifies two inequivalent resolutions of the system-environment mixed anomaly, effectively exchanging which subgroup is restored globally.  

\subsubsection{Numerical simulation}
\begin{figure}
    \centering
    \includegraphics[width=\linewidth]{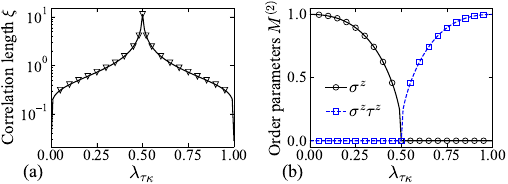}
    \caption{
    Numerical evidence for the $\sigma$-SWSSB $\leftrightarrow$ $\sigma\tau$-preserved SWSSB transition using the model in Eq.~\eqref{Equ: DQCP}.
    (a) Correlation length $\xi$.
    (b) Order parameters $M_{\sigma}^{(2)}$ and $M_{\sigma\tau}^{(2)}$ defined in Eq.~\eqref{Equ: Order Parameter}.}
    \label{Fig: Order Parameter}
\end{figure}

\begin{figure}
    \centering
    \includegraphics[width=\linewidth]{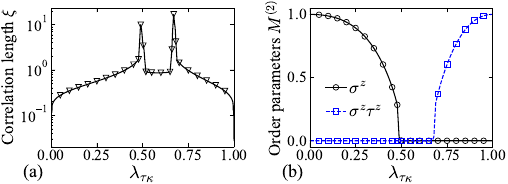}
    \caption{Phase diagram for the perturbed model in Eq.~\eqref{Equ: H_perturb}.
    (a) Correlation length $\xi$.
    (b) Two order parameters $M_{\sigma}^{(2)}$ and $M_{\sigma\tau}^{(2)}$.}
    \label{Fig: Perturb}
\end{figure}

To illustrate this transition, we numerically study the linear interpolation between two vertex Hamiltonians
\begin{align}
    H(\lambda_{\tau\kappa}) = (1-\lambda_{\tau\kappa})H_{\sigma\tau\kappa}^{(+1, +1, -1)}+\lambda_{\tau\kappa} H_{\sigma\tau\kappa}^{(+1, -1, -1)},\label{Equ: DQCP}
\end{align}
from which we obtain the mixed state by tracing out $\kappa$ spins.
This Hamiltonian satisfies a duality relation $\lambda_{\tau\kappa}\leftrightarrow 1-\lambda_{\tau\kappa}$ generated by the DW map $U_{\tau\kappa}^{\rm DW}$
\begin{align}
    U_{\tau\kappa}^{\rm DW}H(\lambda_{\tau\kappa})U_{\tau\kappa}^{\rm DW\dagger}=H(1-\lambda_{\tau\kappa}),
    \label{Equ: Duality of H}
\end{align}
which enforces a self-dual critical point at $\lambda_c=0.5$.
A central feature of this transition is the competition between two order parameters, each characterizing one of the SWSSB patterns
\begin{align}
\begin{aligned}
    M_{\sigma}^{(2)} &= \lim_{|i-j|\to\infty} \mathcal{C}^{(2)}(\sigma^z, i, j),\\
    M_{\sigma\tau}^{(2)} &= \lim_{|i-j|\to\infty} \mathcal{C}^{(2)}(\sigma^z\tau^z, i, j),
\label{Equ: Order Parameter}
\end{aligned}
\end{align}
which are formulated in terms of R\'enyi-$2$ correlators, more amenable to large-scale numerical simulations than fidelity correlators.

We compute the ground state using the time-dependent variational principle (TDVP) method~\cite{Haegeman2011}, represented as a uniform MPS~\cite{ZaunerStauber2018, Vanderstraeten2019} with physical dimension $d_p=8$ and bond dimension $D=24$.
As shown in Fig.~\ref{Fig: Order Parameter}(a), the correlation length $\xi$ diverges at $\lambda_c=0.5$, signaling a continuous phase transition.
Fig.~\ref{Fig: Order Parameter}(b) shows the behavior of the two order parameters after tracing out the $\kappa$ spins across the transition.
On the $\sigma$-SWSSB side ($\lambda_{\tau\kappa}<\lambda_c$), $M_{\sigma}^{(2)}$ is finite while $M_{\sigma\tau}^{(2)}$ vanishes, while on the $\sigma\tau$-preserved SWSSB side ($\lambda_{\tau\kappa}>\lambda_c$), the situation reverses.
Both order parameters vanish continuously at the critical point $\lambda_c$, directly demonstrating the interchange of distinct SWSSB patterns.

To further elucidate the nature of this transition, we introduce a small perturbation
\begin{align}
    H_{\sigma\tau\kappa\text{, Perturb}}^{(+1, +1, -1)}&=-\sum_i \left[2\tau_i^{x}+\kappa_{i-1}^{z}\sigma_i^{x}\kappa_{i}^{z}+\sigma_i^{z}\kappa_i^{x}\sigma_{i+1}^{z}\right]
\end{align}
and redefine the model as
\begin{align}
    H_{\rm Perturb}(\lambda_{\tau\kappa}) = (1-\lambda_{\tau\kappa})H_{\sigma\tau\kappa\text{, Perturb}}^{(+1, +1, -1)}+\lambda_{\tau\kappa} H_{\sigma\tau\kappa}^{(+1, -1, -1)}.
    \label{Equ: H_perturb}
\end{align}
The perturbed model exhibits two distinct critical points, with an intermediate trivial symmetric phase in which both order parameters vanish, as shown in Fig.~\ref{Fig: Perturb}.
As the perturbation does not affect the domain wall decoration formed between $\sigma$ and $\kappa$ spins, it leaves $M_{\sigma}^{(2)}$ unchanged, but suppresses $M_{\sigma\tau}^{(2)}$ order by enlarging the transverse field of $\tau$ spins.
This shows that the direct exchange of SWSSB patterns in the unperturbed model can be divided into two successive SWSSB transitions, yielding a total central charge $c=1$.
The self-duality in Eq.~\eqref{Equ: Duality of H} further constrains the locations of the critical points in the unperturbed case.
Both of the above two purified models can be analytically solved by mapping to two decoupled Ising chains using the KT transformation, which is illustrated in Appendix~\ref{Sec: Appendix-D}.

\subsubsection{Topological decoration}
The most direct manifestation of intrinsically open-system physics emerges when a double ASPT phase, e.g., protected by $\Z_2^\tau\text{(S)}\times \Z_2^\sigma\text{(SWSSB)}$, is transitioned to its $\sigma\tau$-preserved counterpart protected by $\Z_2^{\sigma\tau}\text{(S)}\times \Z_2^\sigma\text{(SWSSB)}$ (No.~6$\leftrightarrow$8).
Both phases host nontrivial topology, while crossing the critical line exchanges which subgroup is stronger.
Therefore, the central problem is the fate of different topological structures around the transitions.
From the tensor-network perspective, the virtual representations of $\Z_2^{\sigma}$, $\Z_2^{\tau}$, and $\Z_2^{\sigma\tau}$ are inherently noncommuting.
Although the anomaly between two weak symmetries is hidden due to the cancellation between ket and bra~\cite{Guo2025B}, the anomaly involving the strong subgroup remains visible and dominates the topological structure.
As the system is tuned across the transition, the dominant anomaly is reshuffled rather than destroyed, allowing the topological character to persist while its protecting symmetry changes.

Such behavior has no pure-state counterpart, as once a protecting subgroup undergoes SSB in closed systems, the associated projective representation is irreversibly lost.
Only in open systems, where the topological structure can be hidden under SWSSB rather than eliminated, can two distinct topological phases be continuously connected in this manner.
This exchange of topology between different protection subgroups constitutes one of the central results of our work.

Fig.~\ref{Fig: cube} summarizes all twelve transitions on the edges of the phase cube.
Conventional SPT-type transitions are shown in black solid lines, the trivial-SWSSB transitions are shown in red dashed lines, and transitions of different SWSSB patterns are shown in blue dotted lines.
This unified picture demonstrates that even the minimal three-$\Z_2$ construction suffices to realize both conventional and intrinsic quantum criticality in open systems.

\section{Phase cube}
The eight fixed-point Hamiltonians $H^{(\mu_{\sigma\tau},\mu_{\tau\kappa},\mu_{\kappa\sigma})}$ introduced in Sec.~\ref{Sec: Fixed-point} exhaust all possible decorated domain-wall configurations compatible with the symmetry group $\Z_2^\sigma\times\Z_2^\tau\times\Z_2^\kappa$.  
They therefore constitute the elementary building blocks of the phase structure of one-dimensional open systems.
In this section, we implement a trilinear interpolation between eight fixed-point Hamiltonians to construct a parameterized model
\begin{align}
\begin{aligned}
    H(\lambda_{\sigma\tau}, \lambda_{\tau\kappa}, \lambda_{\kappa\sigma}) &= (1-\lambda_{\sigma\tau})(1-\lambda_{\tau\kappa})(1-\lambda_{\kappa\sigma})H^{(+1, +1, +1)}_{\sigma\tau\kappa}\\
    & +\lambda_{\sigma\tau}(1-\lambda_{\tau\kappa})(1-\lambda_{\kappa\sigma})H^{(-1, +1 ,+1)}_{\sigma\tau\kappa}\\
    & +(1-\lambda_{\sigma\tau})\lambda_{\tau\kappa}(1-\lambda_{\kappa\sigma})H^{(+1, -1, +1)}_{\sigma\tau\kappa}\\
    & +(1-\lambda_{\sigma\tau})(1-\lambda_{\tau\kappa})\lambda_{\kappa\sigma} H^{(+1, +1, -1)}_{\sigma\tau\kappa}\\
    & +\lambda_{\sigma\tau}\lambda_{\tau\kappa}(1-\lambda_{\kappa\sigma}) H^{(-1, -1, +1)}_{\sigma\tau\kappa}\\
    & +\lambda_{\sigma\tau}(1-\lambda_{\tau\kappa})\lambda_{\kappa\sigma} H^{(-1, +1, -1)}_{\sigma\tau\kappa}\\
    & +( 1-\lambda_{\sigma\tau})\lambda_{\tau\kappa}\lambda_{\kappa\sigma} H^{(+1, -1, -1)}_{\sigma\tau\kappa}\\
    & +\lambda_{\sigma\tau}\lambda_{\tau\kappa}\lambda_{\kappa\sigma} H^{(-1,-1, -1)}_{\sigma\tau\kappa},
\end{aligned}
\end{align}
with $(\lambda_{\sigma\tau},\lambda_{\tau\kappa},\lambda_{\kappa\sigma})\in[0,1]^3$.
This construction generates the most general local Hamiltonian that is adiabatically connected to these eight fixed-point limits, while preserving the symmetry and the short-range nature of interactions.  
Geometrically, the parameter space defines a \emph{phase cube}, whose vertices correspond to the eight fixed-point phases, whose edges realize the twelve single-parameter transitions discussed in Sec.~\ref{Sec: Edge}, and whose faces encode the two-parameter interpolation between competing domain-wall decorations.  
In this sense, the phase cube provides a minimal complete embedding of all mixed-state phases arising from decorated domain-wall structures.

An important structural feature of the cube is the action of the three DW duality transformations,
\begin{align}
    U_{\sigma\tau}^{\rm DW}H(\lambda_{\sigma\tau}, \lambda_{\tau\kappa}, \lambda_{\kappa\sigma})U_{\sigma\tau}^{\rm DW\dagger} &= H(1-\lambda_{\sigma\tau}, \lambda_{\tau\kappa}, \lambda_{\kappa\sigma}),\\
    U_{\tau\kappa}^{\rm DW}H(\lambda_{\sigma\tau}, \lambda_{\tau\kappa}, \lambda_{\kappa\sigma})U_{\tau\kappa}^{\rm DW\dagger} &= H(\lambda_{\sigma\tau}, 1-\lambda_{\tau\kappa}, \lambda_{\kappa\sigma}),\\
    U_{\kappa\sigma}^{\rm DW}H(\lambda_{\sigma\tau}, \lambda_{\tau\kappa}, \lambda_{\kappa\sigma})U_{\kappa\sigma}^{\rm DW\dagger} &= H(\lambda_{\sigma\tau}, \lambda_{\tau\kappa}, 1-\lambda_{\kappa\sigma}).\label{Equ: DW3}
\end{align}
These dualities pair opposite faces and map fixed-point states into one another.
These dualities strongly constrain the internal structure of the phase cube and organize the classification of face phase diagrams and bulk regions.
In the following, we apply large-scale numerical simulations to explore the complete phase cube and to
identify the phase boundaries and emergent phases in its interior.

\subsection{Face phase diagrams}
We first focus on the phase diagrams defined on the six faces of the cube, i.e., $\lambda_{\sigma\tau}=0$, $\lambda_{\sigma\tau}=1$, $\lambda_{\tau\kappa}=0$, $\lambda_{\tau\kappa}=1$, $\lambda_{\kappa\sigma}=0$, and $\lambda_{\kappa\sigma}=1$.
We use several key physical diagnostics, including the correlation length, symmetry indicators, and topological invariants, to discuss the phase diagram.

\subsubsection{Correlation length}
\begin{figure}
    \centering
    \includegraphics[width=\linewidth]{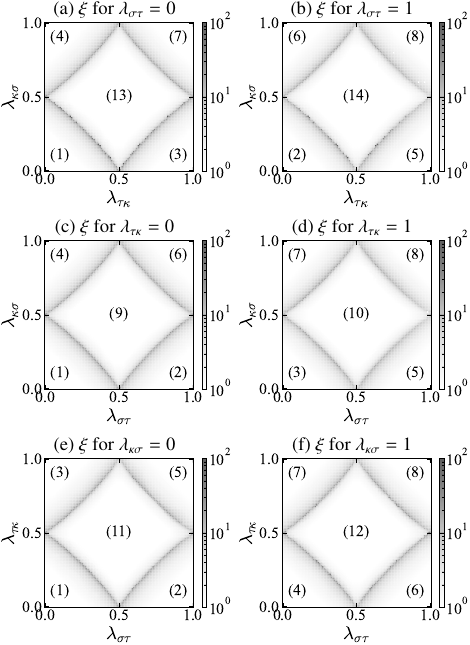}
    \caption{Correlation lengths $\xi$ for six faces of the cube.
    (a) $\lambda_{\sigma\tau}=0$.
    (b) $\lambda_{\sigma\tau}=1$.
    (c) $\lambda_{\tau\kappa}=0$.
    (d) $\lambda_{\tau\kappa}=1$.
    (e) $\lambda_{\kappa\sigma}=0$.
    (f) $\lambda_{\kappa\sigma}=1$.}
    \label{Fig: Xi}
\end{figure}
To obtain a first glance at the phase structure on each face, we plot the correlation lengths $\xi$ of the purified states in Fig.~\ref{Fig: Xi}. 
Divergences of correlation lengths mark continuous phase transitions and delineate the phase boundaries.
For all six faces, the phase diagrams are divided by critical lines into five regions.
In addition to the four corner phases inherited from the corner fixed-point states discussed before, an additional phase emerges at the center of the phase diagram on each face (No.~9-14).

\subsubsection{Symmetry indicators}
\begin{figure}
    \centering
    \includegraphics[width=0.874\linewidth]{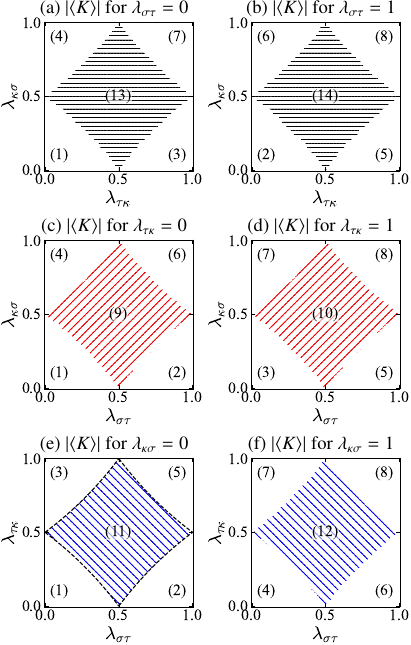}
    \caption{Symmetry indicators $|\braket{K}|$ for six faces of the cube.
    SSB of $\Z_2^{\sigma}$, $\Z_2^{\tau}$, and $\Z_2^{\kappa}$ are marked by \red{//}, \blue{\sslash}, =, respectively.
    (a) $\lambda_{\sigma\tau}=0$.
    (b) $\lambda_{\sigma\tau}=1$.
    (c) $\lambda_{\tau\kappa}=0$.
    (d) $\lambda_{\tau\kappa}=1$.
    (e) $\lambda_{\kappa\sigma}=0$.
    (f) $\lambda_{\kappa\sigma}=1$, where phase boundaries determined by Eq.~\eqref{Equ: analytic_boundary} are marked by dashed lines.}
    \label{Fig: Symmetry}
\end{figure}
To identify the nature of the intermediate phases, we next examine the symmetry properties by evaluating three symmetry indicators, $\left|\braket{\sigma_i^x}\right|$, $\left|\braket{\tau_i^x}\right|$, and $\left|\braket{\kappa_i^x}\right|$, which diagnose the breaking of the corresponding $\Z_2$ symmetries.
For clarity, we choose a threshold of $\left|\braket{K}\right|<0.99$ for the occurrence of SSB, and use the hatches of \red{//}, \blue{\sslash}, and = to mark the SSB of $\Z_2^\sigma$, $\Z_2^{\tau}$, and $\Z_2^{\kappa}$, respectively.
The results in Fig.~\ref{Fig: Symmetry} demonstrate that each intermediate phase breaks a single $\Z_2$ symmetry, as summarized in Table~\ref{Tab: Phases} (No.~9-14).
Notably, this depiction also allows for the coexistence of different SSB patterns (to be shown later), which can be simply illustrated by the overlap of different hatches, providing a compact and intuitive visualization.

\subsubsection{Intermediate SSB phases: physical picture and phase boundaries}
We illustrate the physical origin of the intermediate SSB phases by
focusing on the face $\lambda_{\kappa\sigma}=0$ (No.~11) as a representative example
[Fig.~\ref{Fig: Symmetry}(e)].
Roughly speaking, the $\tau$-SSB observed in the central region of the $(\lambda_{\sigma\tau},\lambda_{\tau\kappa})$ plane originates from a frustration mechanism between two incompatible decorated domain-wall couplings acting on the $\tau$ chain.
Each interpolation parameter controls a distinct decoration structure, including the $\sigma$–$\tau$ decoration ($\sigma_i^z\tau_{i}^x\sigma_{i+1}^z$) and the $\tau$–$\kappa$ decoration ($\kappa_{i-1}^z\tau_i^x\kappa_i^z$).
Therefore, when both decorations are simultaneously activated with comparable strengths ($\lambda_{\sigma\tau}\sim\lambda_{\tau\kappa}$), each $\tau$ spin is subjected to two competing dressing environments.
These two mechanisms attempt to align the $\tau$ spins along different correlated directions defined by $\sigma$ and $\kappa$, respectively.

Energetically, the system reduces its energy by choosing a definite configuration of $\tau^z$ (either all $+1$ or all $-1$), thereby removing the destructive interference between the two decoration mechanisms.
This spontaneously selects one of the two sectors related by the global $\Z_2^{\tau}$ symmetry and yields a two-fold degenerate ground state.
The central phase can thus be interpreted as a frustration-induced SSB phase surrounded by distinct SPT regimes.

The phase boundaries surrounding the $\tau$-SSB region can be understood analytically from the structure of the bilinear interpolation, where the Hamiltonian reads
\begin{align}
\begin{aligned}
    &H(\lambda_{\sigma\tau}, \lambda_{\tau\kappa}, 0) 
    = -\sum_i\left\{\sigma_i^x\left[(1-\lambda_{\sigma\tau})+\lambda_{\sigma\tau}\tau_{i-1}^z\tau_{i}^z\right]\right.\\
    +&\,\tau_i^x\left[(1-\lambda_{\sigma\tau})+\lambda_{\sigma\tau}\sigma_i^z\sigma_{i+1}^z\right]\left[(1-\lambda_{\tau\kappa})+\lambda_{\tau\kappa}\kappa_{i-1}^z\kappa_i^z\right]\\
    +&\,\left.\kappa_i^x\left[(1-\lambda_{\tau\kappa})+\lambda_{\tau\kappa}\tau_{i}^z\tau_{i+1}^z\right]\right\}.
\end{aligned}
\label{Equ: Bilinear}
\end{align}
In the regime where $\lambda_{\sigma\tau}$ and $\lambda_{\tau\kappa}$ are small (lower-left corner of the phase diagram), the $\sigma$ and $\kappa$ spins remain almost fully polarized along the $x$ direction, and the dominant low-energy degree of freedom is the $\tau$ chain.
Treating $\sigma$ and $\kappa$ as static backgrounds ($\sigma_i^x=\kappa_i^x=1$) yields 
\begin{align}    
\begin{aligned}
    H_{\text{eff}}^{(\tau)}
    & = -\sum_i\left[(2-\lambda_{\sigma\tau}-\lambda_{\tau\kappa})+(\lambda_{\sigma\tau}+\lambda_{\tau\kappa})\tau_i^z\tau_{i+1}^z\right.\\
    & + \left.(1-\lambda_{\sigma\tau})(1-\lambda_{\tau\kappa})\tau_i^x\right],
\end{aligned}
\end{align}
which is nothing but an effective 1D transverse-field Ising model (after ignoring the constant term) for $\tau$,
\begin{align}
    H_{\text{eff}}^{(\tau)} = -\sum_i \left[ h_x(\lambda_{\sigma\tau},\lambda_{\tau\kappa})\,\tau_i^x + J(\lambda_{\sigma\tau}, \lambda_{\tau\kappa})\tau_i^z\tau_{i+1}^z \right],
\label{Equ: Eff}
\end{align}
with the effective coefficients
\begin{align}
    h_x = (1-\lambda_{\sigma\tau})(1-\lambda_{\tau\kappa}),\qquad J = \lambda_{\sigma\tau} + \lambda_{\tau\kappa}.
    \label{Equ: Coeff}
\end{align}
The transition into the $\tau$-SSB phase is therefore expected when the transverse field and Ising coupling equal, i.e.,
\begin{align}
    h_x = J,
\end{align}
which leads to the analytic condition
\begin{align}
    (1-\lambda_{\sigma\tau})(1-\lambda_{\tau\kappa}) = \lambda_{\sigma\tau} + \lambda_{\tau\kappa},
    \label{Equ: analytic_boundary}
\end{align}
By duality, the same reasoning applies to the other three sides of the diamond-shaped intermediate SSB region.
The resulting nonlinear phase boundaries correspond to the slightly curved edges indicated by the dashed lines in Fig.~\ref{Fig: Symmetry}(e), in good agreement with numerical results.
All these transitions belong to the conventional Ising universality class with central charge $c=1/2$.

\subsubsection{topological invariants}
\begin{figure}
    \centering
    \includegraphics[width=0.874\linewidth]{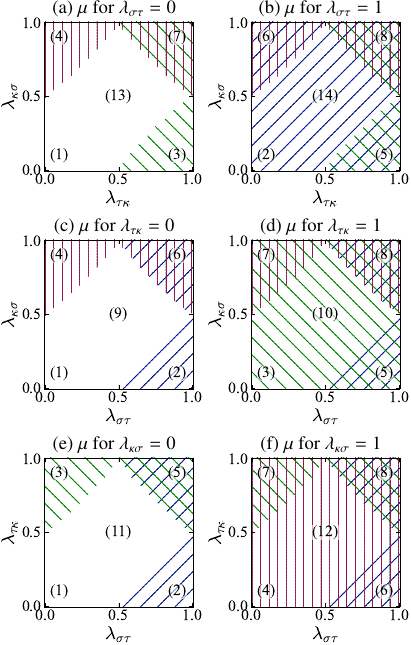}
    \caption{Topological invariants $\mu_{\sigma\tau}$, $\mu_{\tau\kappa}$, and $\mu_{\kappa\sigma}$ for six faces of the cube.
    We use \pblue{/}, \pgreen{\textbackslash}, and \pred{$|$} to represent the minus signs of $\mu_{\sigma\tau}$, $\mu_{\tau\kappa}$, and $\mu_{\kappa\sigma}$, respectively.
    (a) $\lambda_{\sigma\tau}=0$.
    (b) $\lambda_{\sigma\tau}=1$.
    (c) $\lambda_{\tau\kappa}=0$.
    (d) $\lambda_{\tau\kappa}=1$.
    (e) $\lambda_{\kappa\sigma}=0$.
    (f) $\lambda_{\kappa\sigma}=1$.}
    \label{Fig: TI}
\end{figure}
Finally, we evaluate the topological invariants characterizing the SPT structure of the purified states.
These invariants are defined through the relative phases arising from the commutation relations of the virtual symmetry representations, provided that the corresponding symmetries are not spontaneously broken
\begin{align}
    V_\sigma V_\tau &= \mu_{\sigma\tau}V_\tau V_\sigma \label{Equ: TI-1} \\
    V_\tau V_\kappa &= \mu_{\tau\kappa}V_\kappa V_\tau \\
    V_\kappa V_\sigma &= \mu_{\kappa\sigma}V_\sigma V_\kappa
    \label{Equ: TI-3}
\end{align}
This definition is consistent with the triple of indices that label eight fixed-point topological states.
For SSB phases, e.g., the intermediate $\tau$-SSB phase on the $\lambda_{\kappa\sigma}=0$ face, we directly set the corresponding $\mu_{\sigma\tau}$ and $\mu_{\tau\kappa}$ as $1$, while the remaining $\mu_{\kappa\sigma}$ is still well defined and diagnoses a possible mixed anomaly between the unbroken $\Z_2^{\kappa}$ and $\Z_2^{\sigma}$ symmetries.
The numerical approach to calculating these indices is illustrated in Appendix~\ref{Sec: Appendix-E}.

To visualize the results, we use three hatches \pblue{/}, \pgreen{\textbackslash}, and \pred{$|$} to represent the minus signs of $\mu_{\sigma\tau}$, $\mu_{\tau\kappa}$, and $\mu_{\kappa\sigma}$, respectively, as shown in Fig.~\ref{Fig: TI}.
In particular, the eight corner phases agree with the analytical classification, as summarized in Table~\ref{Tab: Phases}.
Moreover, comparing Fig.~\ref{Fig: TI}(e) and (f), we find that although both faces host intermediate phases with $\tau$-SSB order, they differ in their topological properties, where the entire $\lambda_{\kappa\sigma}=1$ face possesses an additional mixed anomaly $\mu_{\kappa\sigma}=-1$ (\pred{$|$}).
This difference originates from the additional DW duality transformation relating the two faces in Eq.~\eqref{Equ: DW3}.
The remaining four faces can be analyzed similarly.

\subsubsection{Mixed-state phases and phase transitions}
\begin{figure}
    \centering
    \includegraphics[width=0.87\linewidth]{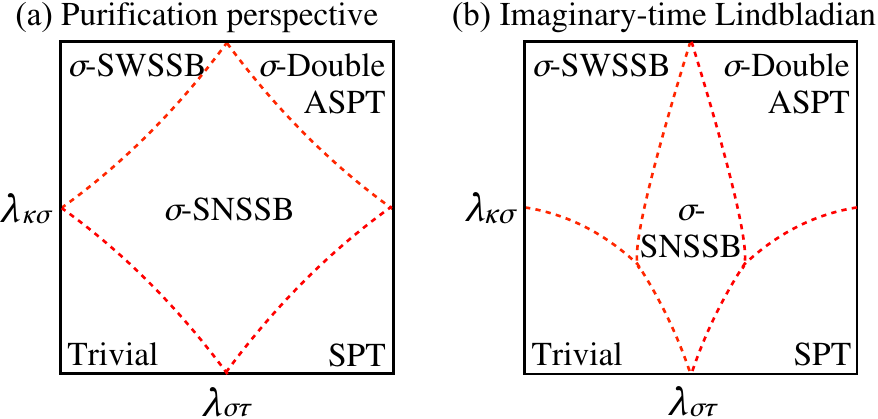}
    \caption{Comparison of the phase diagrams obtained from (a) the purification perspective in our study and (b) the imaginary-time Lindbladian evolution in Ref.~\cite{Guo2025A}.}
    \label{Fig: Compare}
\end{figure}
Here, we analyze the mixed-state phases after tracing out $\kappa$ spins for the intermediate SSB phases in six faces and summarize them in Table~\ref{Tab: Phases} (No.~9-14).
Firstly, for $\lambda_{\sigma\tau}=0$ and $\lambda_{\sigma\tau}=1$, the symmetry that undergoes SSB is $\Z_2^{\kappa}$, having no effect in the corresponding mixed states.
Moreover, the breaking of $\Z_2^{\kappa}$ excludes any mixed anomaly between physical and ancillary spins, preventing the occurrence of SWSSB.
Consequently, the mixed-state phases on these faces directly inherit the symmetry and topological properties of the $\sigma$ and $\tau$ sectors of the corresponding pure states.
Specifically, the intermediate phase on the $\lambda_{\sigma\tau}=0$ face is trivial (No.~13), while that on the $\lambda_{\sigma\tau}=1$ face belongs to an SPT phase (No.~14).

For the other four faces, the situation is different, where the SSB pattern persists after tracing out $\kappa$ spins.
For a mixed state, the conventional SSB is termed as strong-to-none spontaneous symmetry breaking (SNSSB) to avoid ambiguity~\cite{Guo2025A}.
Moreover, if a mixed anomaly exists between $\Z_2^{\kappa}$ and another unbroken symmetry (e.g., $\Z_2^{\sigma}$), tracing out $\kappa$ induces SWSSB in the latter, as discussed in Sec.~\ref{Sec: SWSSB}.
As a result, the intermediate phase on the $\lambda_{\tau\kappa}=0$ face belongs to the $\sigma$-SNSSB phase (No.~9), while that on the $\lambda_{\tau\kappa}=1$ face exhibits additional $\tau$-SWSSB order (No.~10).
The same reasoning applies to the $\lambda_{\kappa\sigma}=0$ and $\lambda_{\kappa\sigma}=1$ faces (No.~11 and 12).

Of particular interest is the appearance of multi-criticality points along the edges of the cube.
For example, the point $(\lambda_{\sigma\tau}, \lambda_{\kappa\sigma})=(0, 0.5)$ on the $\lambda_{\tau\kappa}=0$ face corresponds to a triple point between trivial, $\sigma$-SNSSB, and $\sigma$-SWSSB phases.
These three phases represent distinct symmetry-breaking patterns of a single $\Z_2^{\sigma}$ symmetry in mixed states, as illustrated in Fig.~\ref{Fig: Compare}(a).
At this triple point, the critical line between trivial and $\sigma$-SNSSB (full breaking of $\Z_2^{\sigma}$) intersects the line between $\sigma$-SWSSB and $\sigma$-SNSSB (partial breaking of $\Z_2^{\sigma}$), allowing a direct transition from the trivial phase to the $\sigma$-SWSSB phase.

This structure is qualitatively consistent with the previous results obtained using imaginary-time Lindbladian evolution~\cite{Guo2025A}, which are compared in Fig.~\ref{Fig: Compare}(b) (in the original reference, the SPT phase is denoted as an ASPT phase with $\Z_2^{\sigma}\text{(S)}\times\Z_2^{\tau}\text{(S)}$, which are equivalent).
In that framework, the steady state can be interpreted as the reduced state of a ground state defined on an enlarged purified system~\cite{Guo2025C}.
However, the formal derivation relies on repeated reinitialization of ancillae at each time step, i.e., an infinite-size reservoir with a fast relaxation process.
In contrast, our purified model only involves a single ancilla on each site, so the correspondence between the two formalisms is not exact.
As discussed above, the duality between trivial and SWSSB phases is lost at the mixed-state level when the physical and ancillary degrees of freedom are already entangled before the unitary transformation.
In other words, the left-right flipping symmetry of the phase diagram in Fig.~\ref{Fig: Compare}(b) is not preserved if we directly target the mixed-state phases using the imaginary-time Lindbladian framework.
Consequently, while there exists a one-to-one correspondence between the phases in the two approaches, the precise locations of the phase boundaries do not coincide.

\subsection{Inside the phase cube}
We now move beyond the six faces and investigate the internal structure of the phase cube.
Our next analysis addresses two closely related questions.
The first problem we answer is, how does each phase identified on the faces evolve as one moves into the bulk of the cube?
Second, do genuinely new phases emerge in the interior that are absent on the faces?

\subsubsection{Geometry of the SSB regions}
\begin{figure}
    \centering
    \includegraphics[width=0.881\linewidth]{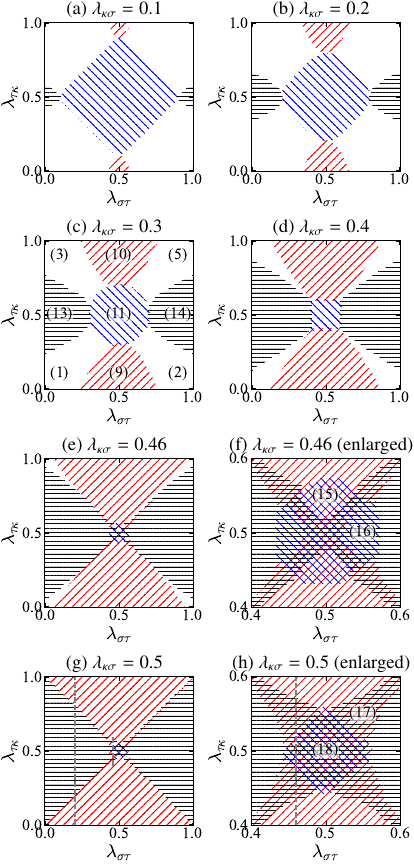}
    \caption{SSB patterns for (a) $\lambda_{\kappa\sigma}=0.1$, (b) $\lambda_{\kappa\sigma}=0.2$, (c) $\lambda_{\kappa\sigma}=0.3$, (d) $\lambda_{\kappa\sigma}=0.4$, (e-f) $\lambda_{\kappa\sigma}=0.46$, and (g-h) $\lambda_{\kappa\sigma}=0.5$.
    The hatches are used in the same way as Fig.~\ref{Fig: Symmetry}.}
    \label{Fig: SSB}
\end{figure}

We now analyze how the $\tau$-SSB region (\blue{\sslash}) on the $\lambda_{\kappa\sigma}=0$ surface evolves as the system enters the bulk ($\lambda_{\kappa\sigma}>0$).  
The key point is that the model contains three $\Z_2$ degrees of freedom $(\sigma,\tau,\kappa)$, each subject to two decoration-induced longitudinal couplings and one transverse field.  
In the parameter regime where the other two chains remain approximately $x$-polarized (i.e., in the symmetric phase), we also yield an effective transverse-field Ising model for the $\tau$ chain.
To reach this, we note that the full Hamiltonian reads as
\begin{align}
    H(\lambda_{\sigma\tau}, \lambda_{\tau\kappa}, \lambda_{\kappa\sigma}) = (1-\lambda_{\kappa\sigma})H(\lambda_{\sigma\tau}, \lambda_{\tau\kappa}, 0) + \lambda_{\kappa\sigma}H(\lambda_{\sigma\tau}, \lambda_{\tau\kappa}, 1).
\end{align}
In this case, both Hamiltonians can be expanded and approximated similarly to Eq.~\eqref{Equ: Bilinear}-\eqref{Equ: Eff}, leading to a reduced Ising coupling coefficient
\begin{align}
    J_\tau = (1-\lambda_{\kappa\sigma})(\lambda_{\sigma\tau}+\lambda_{\tau\kappa}),
\end{align}
while the strength of the transverse field remains unchanged, $h_x=(1-\lambda_{\sigma\tau})(1-\lambda_{\tau\kappa})$.
Increasing $\lambda_{\kappa\sigma}$ therefore suppresses the tendency of the $\tau$ chain to develop ferromagnetic order.
As a consequence, the $\tau$-SSB region gradually shrinks as the system moves into the bulk, in full agreement with the numerical results shown in Fig.~\ref{Fig: SSB}(a-d).
This analysis reveals a simple geometric picture, where each SSB region is rooted on one face of the cube and extends into the bulk along a single direction, occupying a pyramid-like volume.

\subsubsection{Complete symmetry breaking at the cube center}
The above effective model relies on the assumption that the spectator chains $\sigma$ and $\kappa$ remain unbroken, which applies well to small $\lambda_{\kappa\sigma}$, e.g., $\lambda_{\kappa\sigma}=0.1$ in Fig.~\ref{Fig: SSB}(a). 
For larger values of $\lambda_{\kappa\sigma}$ (Fig.~\ref{Fig: SSB}(b-d)), strong competition between $\tau$-SSB and $\sigma$-SSB (\red{//}) or $\kappa$-SSB (=) cuts away the four corners of the $\tau$-SSB diamond.
When further approaching the cube center, the competition between different SSB patterns may lead to phases with SSB of two or even all three $\Z_2$ symmetries.
For instance, on the $\lambda_{\kappa\sigma}=0.46$ plane in Fig.~\ref{Fig: SSB}(e, f), $\tau$-SSB and $\sigma$-SSB regions overlap near the center of the phase diagram, leading to the simultaneous SSB of both symmetries (\red{//} \blue{\sslash}, No.~15).
A similar argument applies to the coexistence of $\tau$- and $\kappa$-SSB marked by \blue{\sslash} = (No.~16).

The geometric center of the cube, $(\lambda_{\sigma\tau},\lambda_{\tau\kappa},\lambda_{\kappa\sigma})=(0.5, 0.5, 0.5)$, plays a special role due to the exact permutation symmetry among the three spin species $(\sigma,\tau,\kappa)$.
At this point, the Hamiltonian takes the form
\begin{align}
\begin{aligned}
    &H(0.5, 0.5, 0.5) =-\frac14\sum_i\left[\sigma_i^x(1+\tau_{i-1}^z\tau_i^z)(1+\kappa_{i-1}^z\kappa_i^z)\right.\\
    +&\left.\tau_i^x(1+\sigma_i^z\sigma_{i+1}^z)(1+\kappa_{i-1}^z\kappa_i^z)+\kappa_i^x(1+\sigma_i^z\sigma_{i+1}^z)(1+\tau_i^z\tau_{i+1}^z)\right],
\end{aligned}
\label{Equ: H_center}
\end{align}
where the transverse-field term for each chain is multiplicatively modulated by two Ising-string decorations from the other two chains.
As a result, ferromagnetic order in any chain enhances the effective transverse field experienced by the remaining two chains, and vice versa.
This mutual reinforcement leads to frustration between the three chains (similar to anti-ferromagnetic interaction on a triangular lattice), suggesting that the system may lower its energy by developing simultaneous SSB patterns in all three subsystems.

To confirm this intuition, we perform a variational mean-field analysis.
Assuming translation-invariant product states on each chain, we parameterize the longitudinal magnetizations as $m_\sigma=\braket{\sigma_i^z}$, $m_\tau=\braket{\tau_i^z}$, and $m_\kappa=\braket{\kappa_i^z}$, respectively.
Within this approximation, one obtains the mean-field energy per site
\begin{align}
\begin{aligned}
    &E_{\mathrm{MF}}=-\frac14\left[
    \sqrt{1-m_\sigma^2}(1+m_\tau^2)(1+m_\kappa^2)\right.\\
    +&\left.\sqrt{1-m_\tau^2}(1+m_\sigma^2)(1+m_\kappa^2)
    +\sqrt{1-m_\kappa^2}(1+m_\sigma^2)(1+m_\tau^2)\right],
\end{aligned}
\end{align}
which is minimized at $m_\sigma^2=m_\tau^2=m_\kappa^2=m^2$ by symmetry.
This reduces the energy to
\begin{align}
    E_{\mathrm{MF}}(m)=-\frac{3}{4}\sqrt{1-m^2}\left(1+m^2\right)^2.
\label{Equ: MF_energy}
\end{align}
Extremizing Eq.~(\ref{Equ: MF_energy}) yields two types of solutions: the paramagnetic state $m=0$ and a pair of symmetry-breaking states $m=\pm\sqrt{3/5}$.
Evaluating their energies shows that
\begin{align}
    E_{\mathrm{MF}}\left(\pm \sqrt{ \frac{3}{5}}\right) \approx -1.214 \quad<\quad E_{\mathrm{MF}}(0)=-\frac{3}{4},
\end{align}
demonstrating that the full symmetry-breaking solution is strongly favored.
We conclude that the cube center lies deep inside a phase with complete SSB of the full symmetry $\Z_2^{\sigma}\times\Z_2^{\tau}\times\Z_2^{\kappa}$, consistent with the tiny region (\red{//} \blue{\sslash} =) at the center of Fig.~\ref{Fig: SSB}(g, h).
The corresponding mixed-state phase is presented in Table~\ref{Tab: Phases} (No.~18).

\begin{figure}
    \centering
    \includegraphics[width=\linewidth]{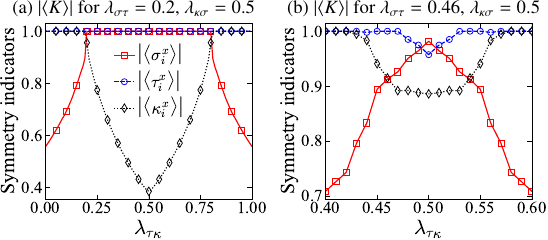}
    \caption{Symmetry indicators $\left|\braket{\sigma_i^x}\right|$, $\left|\braket{\tau_i^x}\right|$, and $\left|\braket{\kappa_i^x}\right|$ for (a) $(\lambda_{\sigma\tau}=0.2$, $\lambda_{\kappa\sigma}=0.5)$, and (b) $(\lambda_{\sigma\tau}=0.46$, $\lambda_{\kappa\sigma}=0.5)$.}
    \label{Fig: SSB_slice}
\end{figure}

To better visualize the SSB patterns inside the phase cube, especially around the cube center, we examine symmetry indicators along two representative cuts in the $\lambda_{\kappa\sigma}=0.5$ plane, namely the lines $(\lambda_{\sigma\tau}=0.2, \lambda_{\kappa\sigma}=0.5)$ and $(\lambda_{\sigma\tau}=0.46, \lambda_{\kappa\sigma}=0.5)$.
The locations of these two lines are marked in Fig.~\ref{Fig: SSB}(g, h).
The results for $(\lambda_{\sigma\tau}=0.2, \lambda_{\kappa\sigma}=0.5)$ are shown in Fig.~\ref{Fig: SSB_slice}(a).
Along this path, $\Z_2^{\tau}$ symmetry is preserved throughout, while the system undergoes a sequence of SSB transitions, where a $\Z_2^{\kappa}$-SSB phase (= in Fig.~\ref{Fig: SSB}(g, h)) is sandwiched between two $\Z_2^{\sigma}$-SSB phases (\red{//}).
In contrast, the cut at $(\lambda_{\sigma\tau}=0.46,\lambda_{\kappa\sigma}=0.5)$, shown in Fig.~\ref{Fig: SSB_slice}(b), reveals a qualitatively different evolution.
Starting from a $\Z_2^{\sigma}$-SSB phase (\red{//}) at small $\lambda_{\tau\kappa}$, the system first enters an intermediate region in which both $\Z_2^{\sigma}$ and $\Z_2^{\kappa}$ are broken (\red{//} =), and eventually reaches a fully SSB phase where all three $\Z_2$ symmetries are simultaneously broken (\red{//} \blue{\sslash} =).
These observations are fully consistent with the global phase structure shown in Fig.~\ref{Fig: SSB}(g, h), as well as with the theoretical analysis presented above.

\subsection{Summary of the phase cube}
The complete three-dimensional phase cube can now be understood within a unified framework.
The eight fixed-point Hamiltonians located at the cube corners generate, through a trilinear interpolation, six two-dimensional phase diagrams on the faces and a rich internal structure in the bulk.
Each face hosts an intermediate SSB phase arising from frustrated competition between two incompatible decorated domain-wall couplings, and the corresponding ordered region extends into the bulk along the remaining parameter direction, producing a pyramid-shaped SSB volume.

As these pyramidal SSB regions grow inward, they begin to intersect and compete.
For moderate values of the third parameter, the corners of each $\tau$-, $\sigma$-, or $\kappa$-SSB diamond are carved away by the competing instabilities of the other two chains, resulting in the characteristic deformation observed in Fig.~\ref{Fig: SSB}(a-d). 
At the geometric center $(\lambda_{\sigma\tau},\lambda_{\tau\kappa},\lambda_{\kappa\sigma})=(0.5, 0.5, 0.5)$, the permutation symmetry among $(\sigma,\tau,\kappa)$ implies that all three decorated domain-wall channels act with equal strength.  
A mean-field analysis confirms that this point lies deep inside a phase in which the entire symmetry group $\Z_2^\sigma\times\Z_2^\tau\times \Z_2^\kappa$ is spontaneously broken.  
This geometric picture provides a compact and intuitive global description of how symmetry breaking, topology, and open-system effects intertwine in the minimal three-$\Z_2$ construction.

\section{Conclusions and Discussions}
In this work, we develop a purification-based framework that embeds mixed-state phases of 1D open quantum systems with $\Z_2^\sigma\times\Z_2^\tau$ symmetry into an enlarged Hilbert space with extended symmetry $\Z_2^\sigma\times\Z_2^\tau\times\Z_2^\kappa$.
The resulting eight purified fixed points, labeled by the triple of topological invariants $(\mu_{\sigma\tau},\mu_{\tau\kappa},\mu_{\kappa\sigma})$, organize into a phase cube that geometrically encodes all mixed-state phases after tracing out the ancillary $\kappa$ spins.
This construction provides a unified picture of how SPT, SNSSB, SWSSB, and double-ASPT phases emerge from mixed anomalies and domain-wall decorations.

The cube structure further reveals how the competition between different domain-wall decorations organizes the global phase structure.
Edges correspond to single-index transitions, faces host intermediate phases arising from competing domain-wall decorations, and the cube interior exhibits pyramid-shaped symmetry-breaking regions whose mutual intersection leads to a fully symmetry-broken phase.
A particularly distinctive feature is the transition between SWSSB phases, where partial symmetry breaking is transferred between different symmetry subgroups while mixed anomalies persist, a phenomenon absent in closed systems.

Beyond providing a complete classification of mixed-state phases, our phase cube offers a geometric and physically intuitive picture of how SPT, various SSB patterns, and their combinations emerge, compete, and coexist.  
It also clarifies the relation to imaginary-time Lindbladian evolution, whose reinitialization of ancillae during the evolution leads to quantitative differences in phase boundaries.  
Nevertheless, both approaches share the same fixed-point phases and mixed anomaly structure.

From an information-theoretic perspective, the three $\mathbb{Z}_2$ invariants form a $\mathbb{Z}_2^3$ structure that is topologically isomorphic to three independent logical sectors.
Each vertex specifies which symmetry-protected structures remain stable, and each edge corresponds to the collapse of one such sector.
In this sense, the phase cube resembles the logical structure of QEC codes: single-index transitions parallel the loss of a logical channel at a decoding threshold, while the fully symmetry-broken phase corresponds to the simultaneous collapse of all protected sectors.
More broadly, the breakdown of strong symmetry in SWSSB reflects channel-theoretic irreversibility, suggesting that mixed-state phase transitions can be understood as transitions between distinct channel reversibility classes.

Several interesting directions follow from our work.
First, purification-based cube construction can be extended to larger symmetry groups or higher-form symmetries, potentially uncovering new families of mixed-state topological and symmetry-breaking phases.  
Second, it would be interesting to formalize the connection between purification-based topology and operator-algebraic notions of recoverability in QEC, from which one can investigate the real-time dynamical stability of the phases identified here under generic Lindbladian dynamics.  
Finally, our geometric viewpoint suggests the possibility of experimentally realizing and probing mixed-state anomalies using programmable quantum simulators with built-in measurement and feedback.  
We hope that the framework developed in this work provides a foundation for the systematic understanding of mixed-state phases in higher dimensions.

\begin{acknowledgments}
    This work is supported by the National Natural Science Foundation of China (NSFC) (Grant No. 12475022 and No. 125B2100) and the Quantum Science and Technology - National Science and Technology Major Project (Grant No. 2021ZD0302100).
\end{acknowledgments}

\bibliography{ref}

\appendix
\section{Tensor network representation for a mixed state}\label{Sec: Appendix-A}
In this section, we review the key concept of the LPDO representation for a 1D short-range entangled mixed state~\cite{Verstraete2004, Werner2016, Cheng2021, Guo2024A, Guo2024B}.
We start from a pure state represented by an MPS
\begin{align}
\scalebox{0.8}{
\begin{tikzpicture}[scale=0.75]
\tikzstyle{sergio}=[rectangle,draw=none]
\filldraw[fill=white, draw=black, rounded corners] (-0.75,-0.5)--(0.75,-0.5)--(0.75,0.5)--(-0.75,0.5)--cycle;
\filldraw[fill=white, draw=black, rounded corners] (1.75,-0.5)--(3.25,-0.5)--(3.25,0.5)--(1.75,0.5)--cycle;
\draw[line width=1pt] (0.75,0) -- (1.75,0);
\draw[line width=1pt] (-0.75,0) -- (-1.75,0);
\draw[line width=1pt] (3.25,0) -- (4.25,0);
\draw[line width=2pt, color=red] (0.0168,0.5) -- (0.0168,1);
\draw[line width=2pt, color=red] (2.5168,0.5) -- (2.5168,1);
\path (0,0) node [style=sergio]{\large $\mathsf{A}$};
\path (2.5,0) node [style=sergio]{\large $\mathsf{A}$};
\path (-2.25,0) node [style=sergio]{$\cdots$};
\path (4.75,0) node [style=sergio]{$\cdots$};
\path (-3.25,0) node [style=sergio]{\large $|\psi\rangle=$};
\end{tikzpicture}},
\end{align}
where each local tensor has one physical index with dimension $d_p$ and two virtual indices with dimension $D$.
Such an MPS $\ket{\psi}$ is short-range correlated in terms of the conventional two-point connected correlator
\begin{align}
    \mathcal{C}(O, i, j)\equiv \braket{\psi|O_iO_j|\psi}-\braket{\psi|O_i|\psi}\braket{\psi|O_j|\psi}\sim e^{-|i-j|/\xi}
\label{Equ: C for pure}
\end{align}
iff the linear map formed by the local tensor
\begin{align}
\scalebox{0.8}{
\begin{tikzpicture}[scale=0.75]
\tikzstyle{sergio}=[rectangle,draw=none]
\filldraw[fill=white, draw=black, rounded corners] (0,-0.5)--(1.5,-0.5)--(1.5,0.5)--(0,0.5)--cycle;
\draw[line width=1pt] (1.5,0) -- (2,0);
\draw[line width=1pt] (0,0) -- (-0.5,0);
\draw[line width=2pt, color=red] (0.75,0.5) -- (0.75,1);
\path (0.75,0) node [style=sergio]{\large $\mathsf{A}$};
\path (4.1,0) node [style=sergio]{\large $: \mathbb{C}^D\times \mathbb{C}^D\rightarrow \mathbb{C}^{d_p}$};
\end{tikzpicture}}
\end{align}
is injective~\cite{Perez2007}.

To generalize the MPS concept to the mixed-state regime, we assume that the target mixed state can be locally purified, i.e., each physical index is attached by an ancillary index (with dimension $d_{\kappa}$), as shown in Eq.~\eqref{Equ: Locally purified}.
After tracing out the ancilla, we obtain the LPDO representation for the mixed state in Eq.~\eqref{Equ: LPDO}.
Two injectivity conditions are proposed for an LPDO~\cite{Guo2025B}.
The first is the \emph{weak injectivity} condition that requires the map
\begin{align}
\scalebox{0.8}{
\begin{tikzpicture}[scale=0.75]
\tikzstyle{sergio}=[rectangle,draw=none]
\filldraw[fill=white, draw=black, rounded corners] (0,-0.5)--(1.5,-0.5)--(1.5,0.5)--(0,0.5)--cycle;
\draw[line width=1pt] (1.5,0) -- (2,0);
\draw[line width=1pt] (0,0) -- (-0.5,0);
\draw[line width=2pt, color=red] (0.75,0.5) -- (0.75,1);
\draw[line width=3pt, color=blue] (0.75,-1) -- (0.75,-0.5);
\path (0.75,0) node [style=sergio]{\large $\mathsf{A}$};
\path (4.8,0) node [style=sergio]{\large $: \mathbb{C}^D\times \mathbb{C}^D\rightarrow \mathbb{C}^{d_p}\times \mathbb{C}^{d_{\kappa}}$};
\end{tikzpicture}}
\end{align}
be injective, leading to a short-range correlated mixed state in terms of the linear correlator
\begin{align}
    \mathcal{C}(O, i, j)=\Tr[\rho O_iO_j] - \Tr[\rho O_i]\Tr[\rho O_j]
\end{align}
that naturally extends the definition in Eq.~(\ref{Equ: C for pure}) to a mixed state.
Meanwhile, a more subtle long-range order exists in the mixed state undergoing SWSSB.
To exclude this possibility, one can define the \emph{strong injectivity} condition that requires the injectivity of the following map
\begin{align}
\scalebox{0.8}{
\begin{tikzpicture}[scale=0.75]
\tikzstyle{sergio}=[rectangle,draw=none]
\filldraw[fill=white, draw=black, rounded corners] (-0.25,-1.5)--(1.25,-1.5)--(1.25,-0.5)--(-0.25,-0.5)--cycle;
\filldraw[fill=white, draw=black, rounded corners] (-0.25,-3)--(1.25,-3)--(1.25,-2)--(-0.25,-2)--cycle;
\draw[line width=1pt] (1.25,-1) -- (1.75,-1);
\draw[line width=1pt] (-0.25,-1) -- (-0.75,-1);
\draw[line width=1pt] (1.25,-2.5) -- (1.75,-2.5);
\draw[line width=1pt] (-0.25,-2.5) -- (-0.75,-2.5);
\draw[line width=2pt, color=red] (0.5,-0.5) -- (0.5,0);
\draw[line width=3pt, color=blue] (0.5,-1.5) -- (0.5,-2);
\draw[line width=2pt, color=red] (0.5,-3) -- (0.5,-3.5);
\path (0.5,-1) node [style=sergio]{\large $\mathsf{A}$};
\path (0.5,-2.5) node [style=sergio]{\large $\mathsf{A^*}$};
\path (5.9,-1.75) node [style=sergio]{\large $: \mathbb{C}^D\times \mathbb{C}^D\times  \mathbb{C}^D\times \mathbb{C}^D\rightarrow \mathbb{C}^{d_p}\times \mathbb{C}^{d_{p}}$};
\end{tikzpicture}},
\end{align}
resulting in a short-range correlated mixed state regarding the R\'enyi-$2$ correlator in Eq.~\eqref{Equ: Renyi-2}.

\section{Classification of 1D ASPT phases}\label{Sec: Appendix-B}
Here, we discuss the application of LPDO representation in the classification and construction of mixed-state topological phases.
We first review the classification of 1D SPT phases based on the MPS approach.
Suppose the system symmetry is characterized by the symmetry group $K\times G$, then the local tensor equation of a symmetric MPS reads as~\cite{Chen2011, Schuch2011}
\begin{align}
\scalebox{0.8}{
\begin{tikzpicture}[scale=0.75]
\tikzstyle{sergio}=[rectangle,draw=none]
\filldraw[fill=white, draw=black, rounded corners] (-0.25,-0.5)--(1.25,-0.5)--(1.25,0.5)--(-0.25,0.5)--cycle;
\draw[line width=1pt] (1.25,0) -- (1.75,0);
\draw[line width=1pt] (-0.25,0) -- (-0.75,0);
\draw[line width=2pt, color=red] (0.5,0.5) -- (0.5,2);
\path (0.5,0) node [style=sergio]{\large $\mathsf{A}$};
\path (2.2332,0) node [style=sergio]{$=$};
\filldraw[fill=white, draw=black] (0.5,1.25)circle (10pt);
\path (0.5,1.25) node [style=sergio]{$U_k$};
\filldraw[fill=white, draw=black, rounded corners] (4.25,-0.5)--(5.75,-0.5)--(5.75,0.5)--(4.25,0.5)--cycle;
\draw[line width=1pt] (4.25,0) -- (2.75,0);
\filldraw[fill=white, draw=black] (3.5,0)circle (10pt);
\path (3.5,0) node [style=sergio]{$V^{-1}_k$};
\path (5,0) node [style=sergio]{\large $\mathsf{A}$};
\draw[line width=1pt] (5.75,0) -- (7.25,0);
\filldraw[fill=white, draw=black] (6.5,0)circle (10pt);
\path (6.5,0) node [style=sergio]{$V_k$};
\draw[line width=2pt, color=red] (5,0.5) -- (5,1);
\end{tikzpicture}}
\label{Equ: Symmetry_k}
\end{align}
for $k\in K$, and similar for $g\in G$.
The symmetry transformation on the virtual index forms a projective representation of the total symmetry group $K\times G$ classified by $\mathcal{H}^2[K\times G, U(1)]$, which can be further decomposed as~\cite{Guo2025B}
\begin{align}
    V(k_1)V(k_2)&=\mu_2(k_1, k_2)V(k_1k_2),\label{Equ: mu_2}\\
    V(g_1)V(g_2)&=\nu_2(g_1, g_2)V(g_1g_2),\label{Equ: nu_2}\\
    V(k)V(g) &= n_1(k, g)V(g)V(k).\label{Equ: n_1}
\end{align}
Here, $\mu_2\in \mathcal{H}^2[K, U(1)]$ and $\nu_2\in \mathcal{H}^2[G, U(1)]$ classify the SPT phases solely protected by $K$ and $G$, respectively, while $n_1\in \mathcal{H}^1\left[G,\mathcal{H}^1[K,U(1)]\right]$ describes the $K$-charge carried by $V(g)$, which characterizes the mixed anomaly between $K$ and $G$ on the virtual space.
From the mathematical perspective, it is the MPS formalism for the corollary of Lyndon–Hochschild–Serre spectral sequence~\cite{Brown1982}
\begin{align}
\begin{aligned}
    &\mathcal{H}^2[K\times G, U(1)]\cong \mathcal{H}^2[K, U(1)]\\
    \times \, &\mathcal{H}^1\left[G,\mathcal{H}^1[K, U(1)]\right]\times \mathcal{H}^2[G, U(1)].
\end{aligned}
\end{align}

On the other hand, if we construct a mixed state represented by LPDO that explicitly breaks the symmetry $G$ to a weak one (e.g., by introducing a local charge reservoir as in Eq.~\eqref{Equ: Bell}), the symmetry transformation of the local tensor will change to 
\begin{align}
\scalebox{0.8}{
\begin{tikzpicture}[scale=0.75]
\tikzstyle{sergio}=[rectangle,draw=none]
\filldraw[fill=white, draw=black, rounded corners] (-0.25,-0.5)--(1.25,-0.5)--(1.25,0.5)--(-0.25,0.5)--cycle;
\draw[line width=1pt] (1.25,0) -- (1.75,0);
\draw[line width=1pt] (-0.25,0) -- (-0.75,0);
\draw[line width=2pt, color=red] (0.5,0.5) -- (0.5,2);
\draw[line width=3pt, color=blue] (0.5,-0.5) -- (0.5,-1);
\path (0.5,0) node [style=sergio]{\large $\mathsf{A}$};
\path (2.2332,0) node [style=sergio]{$=$};
\filldraw[fill=white, draw=black] (0.5,1.25)circle (10pt);
\path (0.5,1.25) node [style=sergio]{$U_k$};
\filldraw[fill=white, draw=black, rounded corners] (4.25,-0.5)--(5.75,-0.5)--(5.75,0.5)--(4.25,0.5)--cycle;
\draw[line width=1pt] (4.25,0) -- (2.75,0);
\filldraw[fill=white, draw=black] (3.5,0)circle (10pt);
\path (3.5,0) node [style=sergio]{$V^{-1}_k$};
\path (5,0) node [style=sergio]{\large $\mathsf{A}$};
\draw[line width=1pt] (5.75,0) -- (7.25,0);
\filldraw[fill=white, draw=black] (6.5,0)circle (10pt);
\path (6.5,0) node [style=sergio]{$V_k$};
\draw[line width=2pt, color=red] (5,0.5) -- (5,1);
\draw[line width=3pt, color=blue] (5,-0.5) -- (5,-1);
\end{tikzpicture}}
\end{align}
similar to the pure-state case for $k\in K$, and
\begin{align}
\scalebox{0.8}{
\begin{tikzpicture}[scale=0.75]
\tikzstyle{sergio}=[rectangle,draw=none]
\filldraw[fill=white, draw=black, rounded corners] (-0.25,-0.5)--(1.25,-0.5)--(1.25,0.5)--(-0.25,0.5)--cycle;
\draw[line width=1pt] (1.25,0) -- (1.75,0);
\draw[line width=1pt] (-0.25,0) -- (-0.75,0);
\draw[line width=2pt, color=red] (0.5,0.5) -- (0.5,2);
\draw[line width=3pt, color=blue] (0.5,-0.5) -- (0.5,-1);
\path (0.5,0) node [style=sergio]{\large $\mathsf{A}$};
\filldraw[fill=white, draw=black] (0.5,1.25)circle (10pt);
\path (0.5,1.25) node [style=sergio]{$U_g$};
\path (2.25,0) node [style=sergio]{$=$};
\filldraw[fill=white, draw=black, rounded corners] (4.25,-0.5)--(5.75,-0.5)--(5.75,0.5)--(4.25,0.5)--cycle;
\draw[line width=1pt] (4.25,0) -- (2.75,0);
\filldraw[fill=white, draw=black] (3.5,0)circle (10pt);
\path (3.5,0) node [style=sergio]{$V^{-1}_g$};
\path (5,0) node [style=sergio]{\large $\mathsf{A}$};
\draw[line width=1pt] (5.75,0) -- (7.25,0);
\filldraw[fill=white, draw=black] (6.5,0)circle (10pt);
\path (6.5,0) node [style=sergio]{$V_g$};
\draw[line width=2pt, color=red] (5,0.5) -- (5,1);
\draw[line width=3pt, color=blue] (5,-0.5) -- (5,-2);
\filldraw[fill=white, draw=black] (5,-1.25)circle (10pt);
\path (5,-1.25) node [style=sergio]{$M_g$};
\end{tikzpicture}}
\end{align}
with an additional unitary transformation $M_g$ on the ancilla for $g\in G$.
It can be shown that $\mu_2$ and $n_1$ can be defined in the same way as Eq.~\eqref{Equ: mu_2} and $\eqref{Equ: n_1}$, while any phase structure on the virtual index within the weak symmetry group $G$ will be cancelled out when implementing the symmetry transformation on each side of the density matrix $U_g\rho U_g^{\dagger}$, trivializing the index $\nu_2$ in Eq.~\eqref{Equ: nu_2}.
Therefore, 1D ASPT phases jointly protected by $K(\text{S})\times G(\text{W})$ are classified by $\mu_2\in \mathcal{H}^2[K, U(1)]$ and $n_1\in \mathcal{H}^1\left[G,\mathcal{H}^1[K,U(1)]\right]$.

\section{Tensor-network construction for eight fixed-point purified states}\label{Sec: Appendix-C}
In this section, we explicitly construct the tensor representation for eight fixed-point states belonging to different SPT phases protected by $\Z_2^{\sigma}\times\Z_2^{\tau}\times \Z_2^{\kappa}$, corresponding to No.~1-8 in Table~\ref{Tab: Phases}.
To further obtain the corresponding mixed states after tracing out $\kappa$, one can simply treat the $\kappa$ index as an ancilla, leading to an LPDO representation.

Eight SPT phases are classified by three labels $\mu=(\mu_{\sigma\tau}, \mu_{\tau\kappa}, \mu_{\kappa\sigma})\in \{\pm 1\}^3$, which fall into four categories described by the number of $-1$ in $\mu$.
Different states belonging to the same category can be transformed into each other by simply permuting the physical index.

\subsection{$\mu=(+1, +1, +1)$}
The category with no $-1$ is the trivial product state without any entanglement structure, which can be represented by an MPS with $D=1$
\begin{align}
\scalebox{0.8}{
\begin{tikzpicture}[scale=0.75]
\tikzstyle{sergio}=[rectangle,draw=none]
\filldraw[fill=white, draw=black, rounded corners] (0,-0.5)--(2,-0.5)--(2,0.5)--(0,0.5)--cycle;
\draw[line width=1pt, dotted] (2,0) -- (2.5,0);
\draw[line width=1pt, dotted] (0,0) -- (-0.5,0);
\draw[line width=2pt, color=red] (0.5,0.5) -- (0.5,1);
\draw[line width=2pt, color=red] (1.5,0.5) -- (1.5,1);
\draw[line width=2pt, color=red] (1,0.5) -- (1,1);
\path (1,0) node [style=sergio]{\large $\mathsf{A}$};
\draw[line width=1pt, dotted] (3.5,0) -- (8.25,0);
\draw[line width=2pt, color=red] (4.375,0) -- (4.375,1);
\draw[line width=2pt, color=red] (7.375,0) -- (7.375,1);
\draw[line width=2pt, color=red] (5.875,0) -- (5.875,1);
\path (3,0) node [style=sergio]{$=$};
\filldraw[fill=white, draw=black] (4,-0.375)--(4.75,-0.375)--(4.75,0.375)--(4,0.375)--cycle;
\path (4.375,1.25) node [style=sergio]{$\sigma$};
\path (4.375,0) node [style=sergio]{$\ket{\rightarrow}$};
\filldraw[fill=white, draw=black] (5.5,-0.375)--(6.25,-0.375)--(6.25,0.375)--(5.5,0.375)--cycle;
\path (5.875,1.25) node [style=sergio]{$\tau$};
\path (5.875,0) node [style=sergio]{$\ket{\rightarrow}$};
\filldraw[fill=white, draw=black] (7,-0.375)--(7.75,-0.375)--(7.75,0.375)--(7,0.375)--cycle;
\path (7.375,1.25) node [style=sergio]{$\kappa$};
\path (7.375,0) node [style=sergio]{$\ket{\rightarrow}$};
\end{tikzpicture}}.
\end{align}
Therefore, the symmetry transformation on the virtual index is a one-dimensional representation of the symmetry group, which must be a linear representation and cannot support any nontrivial topological structure.

\subsection{$\mu=(-1, +1, +1)$}
$\ket{\psi^{(-1, +1, +1)}_{\sigma\tau\kappa}}$ is the product state between $\ket{\psi_{\sigma\tau}^{\rm SPT}}$ and $\ket{\psi_{\kappa}^{\rm Trivial}}$, which can be represented by an MPS with $D=2$
\begin{align}
\scalebox{0.8}{
\begin{tikzpicture}[scale=0.75]
\tikzstyle{sergio}=[rectangle,draw=none]
\filldraw[fill=white, draw=black, rounded corners] (0,-0.5)--(2,-0.5)--(2,0.5)--(0,0.5)--cycle;
\draw[line width=1pt] (2,0) -- (2.5,0);
\draw[line width=1pt] (0,0) -- (-0.5,0);
\draw[line width=2pt, color=red] (0.5,0.5) -- (0.5,1);
\draw[line width=2pt, color=red] (1.5,0.5) -- (1.5,1);
\draw[line width=2pt, color=red] (1,0.5) -- (1,1);
\path (1,0) node [style=sergio]{\large $\mathsf{A}$};
\draw[line width=1pt] (3.5,0) -- (8.25,0);
\path (4,1.25) node [style=sergio]{$\sigma$};
\draw[line width=2pt, color=red] (4,0) -- (4,1);
\path (5.75,1.25) node [style=sergio]{$\tau$};
\draw[line width=2pt, color=red] (5.75,0) -- (5.75,1);
\path (7.5,1.25) node [style=sergio]{$\kappa$};
\draw[line width=2pt, color=red] (7.5,-1) -- (7.5,1);
\filldraw[fill=white, draw=black] (4,0)circle (4pt);
\filldraw[fill=white, draw=black] (5.75,0)circle (4pt);
\path (3,0) node [style=sergio]{$=$};
\filldraw[fill=white, draw=black] (4.5,-0.375)--(5.25,-0.375)--(5.25,0.375)--(4.5,0.375)--cycle;
\path (4.875,0) node [style=sergio]{$H$};
\filldraw[fill=white, draw=black] (6.25,-0.375)--(7,-0.375)--(7,0.375)--(6.25,0.375)--cycle;
\path (6.625,0) node [style=sergio]{$H$};
\draw[line width=1pt, dotted] (3.5,-1) -- (8.25,-1);
\filldraw[fill=white, draw=black] (7.125,-1.375)--(7.875,-1.375)--(7.875,-0.625)--(7.125,-0.625)--cycle;
\path (7.5,-1) node [style=sergio]{$\ket{\rightarrow}$};
\end{tikzpicture}},
\end{align}
where
\begin{align}
\scalebox{0.8}{
\begin{tikzpicture}[scale=0.75]
\tikzstyle{sergio}=[rectangle,draw=none]
\draw[line width=2pt, color=red] (7.5,0) -- (7.5,1);
\draw[line width=1pt] (7,0) -- (8,0);
\filldraw[fill=white, draw=black] (7.5,0)circle (4pt);
\path (9,0) node [style=sergio]{\large $=\delta_{i, \alpha, \beta}$};
\path (7.7668,1) node [style=sergio]{$i$};
\path (7.0,0.3) node [style=sergio]{$\alpha$};
\path (8.00,0.3) node [style=sergio]{$\beta$};
\end{tikzpicture}}
\end{align}
is the Kronecker delta function, and
\begin{align}
\scalebox{0.8}{
\begin{tikzpicture}[scale=0.75]
\tikzstyle{sergio}=[rectangle,draw=none]
\draw[line width=1pt] (0,0) -- (1.75,0);
\path (2.25,0) node [style=sergio]{$=$};
\filldraw[fill=white, draw=black] (0.5,-0.375)--(1.25,-0.375)--(1.25,0.375)--(0.5,0.375)--cycle;
\path (0.875,0) node [style=sergio]{$H$};
\path (4,0) node [style=sergio]{$\left[\begin{matrix} 1 & 1 \\ 1 & -1 \end{matrix}\right]/\sqrt{2}$};
\end{tikzpicture}}
\end{align}
is the Hadamard gate.
The symmetry transformations of the local tensor read as
\begin{align}
\scalebox{0.8}{
\begin{tikzpicture}[scale=0.75]
\tikzstyle{sergio}=[rectangle,draw=none]
\filldraw[fill=white, draw=black, rounded corners] (0,-0.75)--(2,-0.75)--(2,0.25)--(0,0.25)--cycle;
\draw[line width=1pt] (2,-0.25) -- (2.5,-0.25);
\draw[line width=1pt] (0,-0.25) -- (-0.5,-0.25);
\draw[line width=2pt, color=red] (0.5,0.25) -- (0.5,1.75);
\draw[line width=2pt, color=red] (1.5,0.25) -- (1.5,1.75);
\draw[line width=2pt, color=red] (1,0.25) -- (1,1.75);
\path (1,-0.25) node [style=sergio]{\large $\mathsf{A}$};
\path (3,-0.25) node [style=sergio]{$=$};
\filldraw[fill=white, draw=black] (0.5,1)circle (10pt);
\path (0.5,1) node [style=sergio]{$\sigma_x$};
\filldraw[fill=white, draw=black, rounded corners] (5,-0.75)--(7,-0.75)--(7,0.25)--(5,0.25)--cycle;
\draw[line width=1pt] (7,-0.25) -- (8.5,-0.25);
\draw[line width=1pt] (5,-0.25) -- (3.5,-0.25);
\filldraw[fill=white, draw=black] (4.25,-0.25)circle (10pt);
\path (4.25,-0.25) node [style=sergio]{$X$};
\filldraw[fill=white, draw=black] (7.75,-0.25)circle (10pt);
\path (7.75,-0.25) node [style=sergio]{$X$};
\draw[line width=2pt, color=red] (5.5,0.25) -- (5.5,0.75);
\draw[line width=2pt, color=red] (6.5,0.25) -- (6.5,0.75);
\draw[line width=2pt, color=red] (6,0.25) -- (6,0.75);
\path (6,-0.25) node [style=sergio]{\large $\mathsf{A}$};
\end{tikzpicture}},
\end{align}
\begin{align}
\scalebox{0.8}{
\begin{tikzpicture}[scale=0.75]
\tikzstyle{sergio}=[rectangle,draw=none]
\filldraw[fill=white, draw=black, rounded corners] (0,-4)--(2,-4)--(2,-3)--(0,-3)--cycle;
\draw[line width=1pt] (2,-3.5) -- (2.5,-3.5);
\draw[line width=1pt] (0,-3.5) -- (-0.5,-3.5);
\draw[line width=2pt, color=red] (0.5,-3) -- (0.5,-1.5);
\draw[line width=2pt, color=red] (1.5,-3) -- (1.5,-1.5);
\draw[line width=2pt, color=red] (1,-3) -- (1,-1.5);
\path (1,-3.5) node [style=sergio]{\large $\mathsf{A}$};
\path (3,-3.5) node [style=sergio]{$=$};
\filldraw[fill=white, draw=black] (1,-2.25)circle (10pt);
\path (1,-2.25) node [style=sergio]{$\tau_x$};
\filldraw[fill=white, draw=black, rounded corners] (5,-4)--(7,-4)--(7,-3)--(5,-3)--cycle;
\draw[line width=1pt] (7,-3.5) -- (8.5,-3.5);
\draw[line width=1pt] (5,-3.5) -- (3.5,-3.5);
\filldraw[fill=white, draw=black] (4.25,-3.5)circle (10pt);
\path (4.25,-3.5) node [style=sergio]{$Z$};
\filldraw[fill=white, draw=black] (7.75,-3.5)circle (10pt);
\path (7.75,-3.5) node [style=sergio]{$Z$};
\draw[line width=2pt, color=red] (5.5,-3) -- (5.5,-2.5);
\draw[line width=2pt, color=red] (6.5,-3) -- (6.5,-2.5);
\draw[line width=2pt, color=red] (6,-3) -- (6,-2.5);
\path (6,-3.5) node [style=sergio]{\large $\mathsf{A}$};
\end{tikzpicture}},
\end{align}
\begin{align}
\scalebox{0.8}{
\begin{tikzpicture}[scale=0.75]
\tikzstyle{sergio}=[rectangle,draw=none]
\path (3,-6.75) node [style=sergio]{$=$};
\filldraw[fill=white, draw=black, rounded corners] (0,-7.25)--(2,-7.25)--(2,-6.25)--(0,-6.25)--cycle;
\draw[line width=1pt] (2,-6.75) -- (2.5,-6.75);
\draw[line width=1pt] (0,-6.75) -- (-0.5,-6.75);
\draw[line width=2pt, color=red] (0.5,-6.25) -- (0.5,-4.75);
\draw[line width=2pt, color=red] (1.5,-6.25) -- (1.5,-4.75);
\draw[line width=2pt, color=red] (1,-6.25) -- (1,-4.75);
\path (1,-6.75) node [style=sergio]{\large $\mathsf{A}$};
\filldraw[fill=white, draw=black] (1.5,-5.5)circle (10pt);
\path (1.5,-5.5) node [style=sergio]{$\kappa_x$};
\filldraw[fill=white, draw=black, rounded corners] (4,-7.25)--(6,-7.25)--(6,-6.25)--(4,-6.25)--cycle;
\draw[line width=1pt] (6,-6.75) -- (6.5,-6.75);
\draw[line width=1pt] (4,-6.75) -- (3.5,-6.75);
\draw[line width=2pt, color=red] (4.5,-6.25) -- (4.5,-5.75);
\draw[line width=2pt, color=red] (5.5,-6.25) -- (5.5,-5.75);
\draw[line width=2pt, color=red] (5,-6.25) -- (5,-5.75);
\path (5,-6.75) node [style=sergio]{\large $\mathsf{A}$};
\end{tikzpicture}}.
\end{align}
Therefore, the corresponding symmetry representations on the virtual index are
\begin{align}
    V_\sigma=X,\quad V_\tau=Z, \quad V_\kappa = I,
\end{align}
characterizing a mixed anomaly between $\sigma$ and $\tau$ spins.

\subsection{$\mu=(+1, -1, +1)$}
The MPS representation of $\ket{\psi_{\sigma\tau\kappa}^{(+1, -1, +1)}}$ can be constructed in a similar way
\begin{align}
\scalebox{0.8}{
\begin{tikzpicture}[scale=0.75]
\tikzstyle{sergio}=[rectangle,draw=none]
\filldraw[fill=white, draw=black, rounded corners] (0,-0.5)--(2,-0.5)--(2,0.5)--(0,0.5)--cycle;
\draw[line width=1pt] (2,0) -- (2.5,0);
\draw[line width=1pt] (0,0) -- (-0.5,0);
\draw[line width=2pt, color=red] (0.5,0.5) -- (0.5,1);
\draw[line width=2pt, color=red] (1.5,0.5) -- (1.5,1);
\draw[line width=2pt, color=red] (1,0.5) -- (1,1);
\path (1,0) node [style=sergio]{\large $\mathsf{A}$};
\draw[line width=1pt] (3.5,0) -- (9.5,0);
\draw[line width=2pt, color=red] (4.25,-1) -- (4.25,1);
\path (4.25,1.25) node [style=sergio]{$\sigma$};
\draw[line width=2pt, color=red] (6,0) -- (6,1);
\path (6,1.25) node [style=sergio]{$\tau$};
\draw[line width=2pt, color=red] (7.75,0) -- (7.75,1);
\path (7.75,1.25) node [style=sergio]{$\kappa$};
\filldraw[fill=white, draw=black] (6,0)circle (4pt);
\filldraw[fill=white, draw=black] (7.75,0)circle (4pt);
\path (3,0) node [style=sergio]{$=$};
\filldraw[fill=white, draw=black] (6.5,-0.375)--(7.25,-0.375)--(7.25,0.375)--(6.5,0.375)--cycle;
\path (6.875,0) node [style=sergio]{$H$};
\filldraw[fill=white, draw=black] (8.25,-0.375)--(9,-0.375)--(9,0.375)--(8.25,0.375)--cycle;
\path (8.625,0) node [style=sergio]{$H$};
\draw[line width=1pt, dotted] (3.5,-1) -- (9.5,-1);
\filldraw[fill=white, draw=black] (3.875,-1.375)--(4.625,-1.375)--(4.625,-0.625)--(3.875,-0.625)--cycle;
\path (4.25,-1) node [style=sergio]{$\ket{\rightarrow}$};
\end{tikzpicture}},
\end{align}
whose symmetry transformations read as
\begin{align}
\scalebox{0.8}{
\begin{tikzpicture}[scale=0.75]
\tikzstyle{sergio}=[rectangle,draw=none]
\path (3,-6.75) node [style=sergio]{$=$};
\filldraw[fill=white, draw=black, rounded corners] (0,-7.25)--(2,-7.25)--(2,-6.25)--(0,-6.25)--cycle;
\draw[line width=1pt] (2,-6.75) -- (2.5,-6.75);
\draw[line width=1pt] (0,-6.75) -- (-0.5,-6.75);
\draw[line width=2pt, color=red] (0.5,-6.25) -- (0.5,-4.75);
\draw[line width=2pt, color=red] (1.5,-6.25) -- (1.5,-4.75);
\draw[line width=2pt, color=red] (1,-6.25) -- (1,-4.75);
\path (1,-6.75) node [style=sergio]{\large $\mathsf{A}$};
\filldraw[fill=white, draw=black] (0.5,-5.5)circle (10pt);
\path (0.5,-5.5) node [style=sergio]{$\sigma_x$};
\filldraw[fill=white, draw=black, rounded corners] (4,-7.25)--(6,-7.25)--(6,-6.25)--(4,-6.25)--cycle;
\draw[line width=1pt] (6,-6.75) -- (6.5,-6.75);
\draw[line width=1pt] (4,-6.75) -- (3.5,-6.75);
\draw[line width=2pt, color=red] (4.5,-6.25) -- (4.5,-5.75);
\draw[line width=2pt, color=red] (5.5,-6.25) -- (5.5,-5.75);
\draw[line width=2pt, color=red] (5,-6.25) -- (5,-5.75);
\path (5,-6.75) node [style=sergio]{\large $\mathsf{A}$};
\end{tikzpicture}},
\end{align}
\begin{align}
\scalebox{0.8}{
\begin{tikzpicture}[scale=0.75]
\tikzstyle{sergio}=[rectangle,draw=none]
\filldraw[fill=white, draw=black, rounded corners] (0,-4)--(2,-4)--(2,-3)--(0,-3)--cycle;
\draw[line width=1pt] (2,-3.5) -- (2.5,-3.5);
\draw[line width=1pt] (0,-3.5) -- (-0.5,-3.5);
\draw[line width=2pt, color=red] (0.5,-3) -- (0.5,-1.5);
\draw[line width=2pt, color=red] (1.5,-3) -- (1.5,-1.5);
\draw[line width=2pt, color=red] (1,-3) -- (1,-1.5);
\path (1,-3.5) node [style=sergio]{\large $\mathsf{A}$};
\path (3,-3.5) node [style=sergio]{$=$};
\filldraw[fill=white, draw=black] (1,-2.25)circle (10pt);
\path (1,-2.25) node [style=sergio]{$\tau_x$};
\filldraw[fill=white, draw=black, rounded corners] (5,-4)--(7,-4)--(7,-3)--(5,-3)--cycle;
\draw[line width=1pt] (7,-3.5) -- (8.5,-3.5);
\draw[line width=1pt] (5,-3.5) -- (3.5,-3.5);
\filldraw[fill=white, draw=black] (4.25,-3.5)circle (10pt);
\path (4.25,-3.5) node [style=sergio]{$X$};
\filldraw[fill=white, draw=black] (7.75,-3.5)circle (10pt);
\path (7.75,-3.5) node [style=sergio]{$X$};
\draw[line width=2pt, color=red] (5.5,-3) -- (5.5,-2.5);
\draw[line width=2pt, color=red] (6.5,-3) -- (6.5,-2.5);
\draw[line width=2pt, color=red] (6,-3) -- (6,-2.5);
\path (6,-3.5) node [style=sergio]{\large $\mathsf{A}$};
\end{tikzpicture}},
\end{align}
\begin{align}
\scalebox{0.8}{
\begin{tikzpicture}[scale=0.75]
\tikzstyle{sergio}=[rectangle,draw=none]
\filldraw[fill=white, draw=black, rounded corners] (0,-0.75)--(2,-0.75)--(2,0.25)--(0,0.25)--cycle;
\draw[line width=1pt] (2,-0.25) -- (2.5,-0.25);
\draw[line width=1pt] (0,-0.25) -- (-0.5,-0.25);
\draw[line width=2pt, color=red] (0.5,0.25) -- (0.5,1.75);
\draw[line width=2pt, color=red] (1.5,0.25) -- (1.5,1.75);
\draw[line width=2pt, color=red] (1,0.25) -- (1,1.75);
\path (1,-0.25) node [style=sergio]{\large $\mathsf{A}$};
\path (3,-0.25) node [style=sergio]{$=$};
\filldraw[fill=white, draw=black] (1.5,1)circle (10pt);
\path (1.5,1) node [style=sergio]{$\kappa_x$};
\filldraw[fill=white, draw=black, rounded corners] (5,-0.75)--(7,-0.75)--(7,0.25)--(5,0.25)--cycle;
\draw[line width=1pt] (7,-0.25) -- (8.5,-0.25);
\draw[line width=1pt] (5,-0.25) -- (3.5,-0.25);
\filldraw[fill=white, draw=black] (4.25,-0.25)circle (10pt);
\path (4.25,-0.25) node [style=sergio]{$Z$};
\filldraw[fill=white, draw=black] (7.75,-0.25)circle (10pt);
\path (7.75,-0.25) node [style=sergio]{$Z$};
\draw[line width=2pt, color=red] (5.5,0.25) -- (5.5,0.75);
\draw[line width=2pt, color=red] (6.5,0.25) -- (6.5,0.75);
\draw[line width=2pt, color=red] (6,0.25) -- (6,0.75);
\path (6,-0.25) node [style=sergio]{\large $\mathsf{A}$};
\end{tikzpicture}}.
\end{align}
The virtual representations are
\begin{align}
    V_{\sigma} = I, \quad V_{\tau} = X, \quad V_{\kappa} = Z.
\end{align}

\subsection{$\mu=(+1, +1, -1)$}
Similarly, the MPS representation in this case reads as
\begin{align}
\scalebox{0.8}{
\begin{tikzpicture}[scale=0.75]
\tikzstyle{sergio}=[rectangle,draw=none]
\filldraw[fill=white, draw=black, rounded corners] (0,-0.5)--(2,-0.5)--(2,0.5)--(0,0.5)--cycle;
\draw[line width=1pt] (2,0) -- (2.5,0);
\draw[line width=1pt] (0,0) -- (-0.5,0);
\draw[line width=2pt, color=red] (0.5,0.5) -- (0.5,1);
\draw[line width=2pt, color=red] (1.5,0.5) -- (1.5,1);
\draw[line width=2pt, color=red] (1,0.5) -- (1,1);
\path (1,0) node [style=sergio]{\large $\mathsf{A}$};
\draw[line width=1pt] (3.5,0) -- (9.25,0);
\path (4,1.25) node [style=sergio]{$\sigma$};
\draw[line width=2pt, color=red] (4,0) -- (4,1);
\path (5.75,1.25) node [style=sergio]{$\tau$};
\draw[line width=2pt, color=red] (5.75,-1) -- (5.75,1);
\path (7.5,1.25) node [style=sergio]{$\kappa$};
\draw[line width=2pt, color=red] (7.5,0) -- (7.5,1);
\filldraw[fill=white, draw=black] (4,0)circle (4pt);
\filldraw[fill=white, draw=black] (7.5,0)circle (4pt);
\path (3,0) node [style=sergio]{$=$};
\filldraw[fill=white, draw=black] (4.5,-0.375)--(5.25,-0.375)--(5.25,0.375)--(4.5,0.375)--cycle;
\path (4.875,0) node [style=sergio]{$H$};
\filldraw[fill=white, draw=black] (8,-0.375)--(8.75,-0.375)--(8.75,0.375)--(8,0.375)--cycle;
\path (8.375,0) node [style=sergio]{$H$};
\draw[line width=1pt, dotted] (3.5,-1) -- (9.25,-1);
\filldraw[fill=white, draw=black] (5.375,-1.375)--(6.125,-1.375)--(6.125,-0.625)--(5.375,-0.625)--cycle;
\path (5.75,-1) node [style=sergio]{$\ket{\rightarrow}$};
\end{tikzpicture}},
\end{align}
with the symmetry transformations
\begin{align}
\scalebox{0.8}{
\begin{tikzpicture}[scale=0.75]
\tikzstyle{sergio}=[rectangle,draw=none]
\filldraw[fill=white, draw=black, rounded corners] (0,-4)--(2,-4)--(2,-3)--(0,-3)--cycle;
\draw[line width=1pt] (2,-3.5) -- (2.5,-3.5);
\draw[line width=1pt] (0,-3.5) -- (-0.5,-3.5);
\draw[line width=2pt, color=red] (0.5,-3) -- (0.5,-1.5);
\draw[line width=2pt, color=red] (1.5,-3) -- (1.5,-1.5);
\draw[line width=2pt, color=red] (1,-3) -- (1,-1.5);
\path (1,-3.5) node [style=sergio]{\large $\mathsf{A}$};
\path (3,-3.5) node [style=sergio]{$=$};
\filldraw[fill=white, draw=black] (0.5,-2.25)circle (10pt);
\path (0.5,-2.25) node [style=sergio]{$\sigma_x$};
\filldraw[fill=white, draw=black, rounded corners] (5,-4)--(7,-4)--(7,-3)--(5,-3)--cycle;
\draw[line width=1pt] (7,-3.5) -- (8.5,-3.5);
\draw[line width=1pt] (5,-3.5) -- (3.5,-3.5);
\filldraw[fill=white, draw=black] (4.25,-3.5)circle (10pt);
\path (4.25,-3.5) node [style=sergio]{$X$};
\filldraw[fill=white, draw=black] (7.75,-3.5)circle (10pt);
\path (7.75,-3.5) node [style=sergio]{$X$};
\draw[line width=2pt, color=red] (5.5,-3) -- (5.5,-2.5);
\draw[line width=2pt, color=red] (6.5,-3) -- (6.5,-2.5);
\draw[line width=2pt, color=red] (6,-3) -- (6,-2.5);
\path (6,-3.5) node [style=sergio]{\large $\mathsf{A}$};
\end{tikzpicture}},
\end{align}
\begin{align}
\scalebox{0.8}{
\begin{tikzpicture}[scale=0.75]
\tikzstyle{sergio}=[rectangle,draw=none]
\path (3,-6.75) node [style=sergio]{$=$};
\filldraw[fill=white, draw=black, rounded corners] (0,-7.25)--(2,-7.25)--(2,-6.25)--(0,-6.25)--cycle;
\draw[line width=1pt] (2,-6.75) -- (2.5,-6.75);
\draw[line width=1pt] (0,-6.75) -- (-0.5,-6.75);
\draw[line width=2pt, color=red] (0.5,-6.25) -- (0.5,-4.75);
\draw[line width=2pt, color=red] (1.5,-6.25) -- (1.5,-4.75);
\draw[line width=2pt, color=red] (1,-6.25) -- (1,-4.75);
\path (1,-6.75) node [style=sergio]{\large $\mathsf{A}$};
\filldraw[fill=white, draw=black] (1,-5.5)circle (10pt);
\path (1,-5.5) node [style=sergio]{$\tau_x$};
\filldraw[fill=white, draw=black, rounded corners] (4,-7.25)--(6,-7.25)--(6,-6.25)--(4,-6.25)--cycle;
\draw[line width=1pt] (6,-6.75) -- (6.5,-6.75);
\draw[line width=1pt] (4,-6.75) -- (3.5,-6.75);
\draw[line width=2pt, color=red] (4.5,-6.25) -- (4.5,-5.75);
\draw[line width=2pt, color=red] (5.5,-6.25) -- (5.5,-5.75);
\draw[line width=2pt, color=red] (5,-6.25) -- (5,-5.75);
\path (5,-6.75) node [style=sergio]{\large $\mathsf{A}$};
\end{tikzpicture}},
\end{align}
\begin{align}
\scalebox{0.8}{
\begin{tikzpicture}[scale=0.75]
\tikzstyle{sergio}=[rectangle,draw=none]
\filldraw[fill=white, draw=black, rounded corners] (0,-0.75)--(2,-0.75)--(2,0.25)--(0,0.25)--cycle;
\draw[line width=1pt] (2,-0.25) -- (2.5,-0.25);
\draw[line width=1pt] (0,-0.25) -- (-0.5,-0.25);
\draw[line width=2pt, color=red] (0.5,0.25) -- (0.5,1.75);
\draw[line width=2pt, color=red] (1.5,0.25) -- (1.5,1.75);
\draw[line width=2pt, color=red] (1,0.25) -- (1,1.75);
\path (1,-0.25) node [style=sergio]{\large $\mathsf{A}$};
\path (3,-0.25) node [style=sergio]{$=$};
\filldraw[fill=white, draw=black] (1.5,1)circle (10pt);
\path (1.5,1) node [style=sergio]{$\kappa_x$};
\filldraw[fill=white, draw=black, rounded corners] (5,-0.75)--(7,-0.75)--(7,0.25)--(5,0.25)--cycle;
\draw[line width=1pt] (7,-0.25) -- (8.5,-0.25);
\draw[line width=1pt] (5,-0.25) -- (3.5,-0.25);
\filldraw[fill=white, draw=black] (4.25,-0.25)circle (10pt);
\path (4.25,-0.25) node [style=sergio]{$Z$};
\filldraw[fill=white, draw=black] (7.75,-0.25)circle (10pt);
\path (7.75,-0.25) node [style=sergio]{$Z$};
\draw[line width=2pt, color=red] (5.5,0.25) -- (5.5,0.75);
\draw[line width=2pt, color=red] (6.5,0.25) -- (6.5,0.75);
\draw[line width=2pt, color=red] (6,0.25) -- (6,0.75);
\path (6,-0.25) node [style=sergio]{\large $\mathsf{A}$};
\end{tikzpicture}},
\end{align}
and the virtual representations
\begin{align}
    V_{\sigma} = X, \quad V_{\tau} = I, \quad V_{\kappa} = Z.
\end{align}

\subsection{$\mu=(-1, -1, +1)$}
Here, we introduce two pairs of mixed anomalies using the combination of delta functions and Hadamard gates, leading to an MPS with $D=4$
\begin{align}
\scalebox{0.8}{
\begin{tikzpicture}[scale=0.75]
\tikzstyle{sergio}=[rectangle,draw=none]
\filldraw[fill=white, draw=black, rounded corners] (0,-1.25)--(2,-1.25)--(2,0.25)--(0,0.25)--cycle;
\draw[line width=1pt] (2,0) -- (2.5,0);
\draw[line width=1pt] (0,0) -- (-0.5,0);
\draw[line width=1pt] (2,-1) -- (2.5,-1);
\draw[line width=1pt] (0,-1) -- (-0.5,-1);
\draw[line width=2pt, color=red] (0.5,0.25) -- (0.5,0.75);
\draw[line width=2pt, color=red] (1.5,0.25) -- (1.5,0.75);
\draw[line width=2pt, color=red] (1,0.25) -- (1,0.75);
\path (1,-0.5) node [style=sergio]{\large $\mathsf{A}$};
\draw[line width=1pt] (3.5,0) -- (9.25,0);
\path (4,1.25) node [style=sergio]{$\sigma$};
\draw[line width=2pt, color=red] (4,0) -- (4,1);
\path (5.75,1.25) node [style=sergio]{$\tau$};
\draw[line width=2pt, color=red] (5.75,-1) -- (5.75,1);
\path (7.5,1.25) node [style=sergio]{$\kappa$};
\draw[line width=2pt, color=red] (7.5,-1) -- (7.5,1);
\filldraw[fill=white, draw=black] (4,0)circle (4pt);
\filldraw[fill=white, draw=black] (5.75,0)circle (4pt);
\path (3,-0.5) node [style=sergio]{$=$};
\filldraw[fill=white, draw=black] (4.5,-0.375)--(5.25,-0.375)--(5.25,0.375)--(4.5,0.375)--cycle;
\path (4.875,0) node [style=sergio]{$H$};
\filldraw[fill=white, draw=black] (6.25,-0.375)--(7,-0.375)--(7,0.375)--(6.25,0.375)--cycle;
\draw[line width=1pt] (3.5,-1) -- (9.25,-1);
\path (6.625,0) node [style=sergio]{$H$};
\filldraw[fill=white, draw=black] (5.75,-1)circle (4pt);
\filldraw[fill=white, draw=black] (7.5,-1)circle (4pt);
\filldraw[fill=white, draw=black] (6.25,-1.375)--(7,-1.375)--(7,-0.625)--(6.25,-0.625)--cycle;
\path (6.625,-1) node [style=sergio]{$H$};
\filldraw[fill=white, draw=black] (8,-1.375)--(8.75,-1.375)--(8.75,-0.625)--(8,-0.625)--cycle;
\path (8.375,-1) node [style=sergio]{$H$};
\end{tikzpicture}},
\end{align}
where
\begin{align}
\scalebox{0.8}{
\begin{tikzpicture}[scale=0.75]
\tikzstyle{sergio}=[rectangle,draw=none]
\draw[line width=2pt, color=red] (7.5,-1) -- (7.5,1);
\draw[line width=1pt] (7,0) -- (8,0);
\filldraw[fill=white, draw=black] (7.5,0)circle (4pt);
\path (9.125,0) node [style=sergio]{\large $=\delta_{i, j, \alpha, \beta}$};
\path (7.75,1) node [style=sergio]{$i$};
\path (7.75,-1) node [style=sergio]{$j$};
\path (7.0,0.3) node [style=sergio]{$\alpha$};
\path (8.00,0.3) node [style=sergio]{$\beta$};
\end{tikzpicture}}
\end{align}
is also the Kronecker delta function.
In this construction, the first virtual bond carries the mixed anomaly between $\Z_2^{\sigma}$ and $\Z_2^{\tau}$, and the second one carries that between $\Z_2^{\tau}$ and $\Z_2^{\kappa}$.
It can be verified that this tensor construction admits the following symmetry transformations
\begin{align}
\scalebox{0.8}{
\begin{tikzpicture}[scale=0.75]
\tikzstyle{sergio}=[rectangle,draw=none]
\filldraw[fill=white, draw=black, rounded corners] (0,-1.25)--(2,-1.25)--(2,0.25)--(0,0.25)--cycle;
\draw[line width=1pt] (2,0) -- (2.5,0);
\draw[line width=1pt] (0,0) -- (-0.5,0);
\draw[line width=1pt] (2,-1) -- (2.5,-1);
\draw[line width=1pt] (0,-1) -- (-0.5,-1);
\draw[line width=2pt, color=red] (0.5,0.25) -- (0.5,1.75);
\draw[line width=2pt, color=red] (1.5,0.25) -- (1.5,1.75);
\draw[line width=2pt, color=red] (1,0.25) -- (1,1.75);
\path (1,-0.5) node [style=sergio]{\large $\mathsf{A}$};
\path (3,-0.5) node [style=sergio]{$=$};
\filldraw[fill=white, draw=black] (0.5,1)circle (10pt);
\path (0.5,1) node [style=sergio]{$\sigma_x$};
\filldraw[fill=white, draw=black, rounded corners] (5,-1.25)--(7,-1.25)--(7,0.25)--(5,0.25)--cycle;
\draw[line width=1pt] (7,0) -- (8.5,0);
\draw[line width=1pt] (5,-1) -- (3.5,-1);
\draw[line width=1pt] (7,0) -- (8.5,0);
\draw[line width=1pt] (7,-1) -- (8.5,-1);
\draw[line width=1pt] (5,0) -- (3.5,0);
\filldraw[fill=white, draw=black] (4.25,0)circle (10pt);
\path (4.25,0) node [style=sergio]{$X$};
\filldraw[fill=white, draw=black] (7.75,0)circle (10pt);
\path (7.75,0) node [style=sergio]{$X$};
\draw[line width=2pt, color=red] (5.5,0.25) -- (5.5,0.75);
\draw[line width=2pt, color=red] (6.5,0.25) -- (6.5,0.75);
\draw[line width=2pt, color=red] (6,0.25) -- (6,0.75);
\path (6,-0.5) node [style=sergio]{\large $\mathsf{A}$};
\end{tikzpicture}},
\end{align}
\begin{align}
\scalebox{0.8}{
\begin{tikzpicture}[scale=0.75]
\tikzstyle{sergio}=[rectangle,draw=none]
\filldraw[fill=white, draw=black, rounded corners] (0,-5)--(2,-5)--(2,-3.5)--(0,-3.5)--cycle;
\draw[line width=1pt] (2,-3.75) -- (2.5,-3.75);
\draw[line width=1pt] (0,-3.75) -- (-0.5,-3.75);
\draw[line width=1pt] (2,-4.75) -- (2.5,-4.75);
\draw[line width=1pt] (0,-4.75) -- (-0.5,-4.75);
\draw[line width=2pt, color=red] (0.5,-3.5) -- (0.5,-2);
\draw[line width=2pt, color=red] (1.5,-3.5) -- (1.5,-2);
\draw[line width=2pt, color=red] (1,-3.5) -- (1,-2);
\path (1,-4.25) node [style=sergio]{\large $\mathsf{A}$};
\path (3,-4.25) node [style=sergio]{$=$};
\filldraw[fill=white, draw=black] (1,-2.75)circle (10pt);
\path (1,-2.75) node [style=sergio]{$\tau_x$};
\filldraw[fill=white, draw=black, rounded corners] (5,-5)--(7,-5)--(7,-3.5)--(5,-3.5)--cycle;
\draw[line width=1pt] (7,-4.75) -- (8.5,-4.75);
\draw[line width=1pt] (5,-4.75) -- (3.5,-4.75);
\filldraw[fill=white, draw=black] (4.25,-4.75)circle (10pt);
\path (4.25,-4.75) node [style=sergio]{$X$};
\filldraw[fill=white, draw=black] (7.75,-4.75)circle (10pt);
\path (7.75,-4.75) node [style=sergio]{$X$};
\draw[line width=1pt] (7,-3.75) -- (8.5,-3.75);
\draw[line width=1pt] (5,-3.75) -- (3.5,-3.75);
\filldraw[fill=white, draw=black] (4.25,-3.75)circle (10pt);
\path (4.25,-3.75) node [style=sergio]{$Z$};
\filldraw[fill=white, draw=black] (7.75,-3.75)circle (10pt);
\path (7.75,-3.75) node [style=sergio]{$Z$};
\draw[line width=2pt, color=red] (5.5,-3.5) -- (5.5,-3);
\draw[line width=2pt, color=red] (6.5,-3.5) -- (6.5,-3);
\draw[line width=2pt, color=red] (6,-3.5) -- (6,-3);
\path (6,-4.25) node [style=sergio]{\large $\mathsf{A}$};
\end{tikzpicture}},
\end{align}
\begin{align}
\scalebox{0.8}{
\begin{tikzpicture}[scale=0.75]
\tikzstyle{sergio}=[rectangle,draw=none]
\filldraw[fill=white, draw=black, rounded corners] (0,-8.75)--(2,-8.75)--(2,-7.25)--(0,-7.25)--cycle;
\draw[line width=1pt] (2,-7.5) -- (2.5,-7.5);
\draw[line width=1pt] (0,-7.5) -- (-0.5,-7.5);
\draw[line width=1pt] (2,-8.5) -- (2.5,-8.5);
\draw[line width=1pt] (0,-8.5) -- (-0.5,-8.5);
\draw[line width=2pt, color=red] (0.5,-7.25) -- (0.5,-5.75);
\draw[line width=2pt, color=red] (1.5,-7.25) -- (1.5,-5.75);
\draw[line width=2pt, color=red] (1,-7.25) -- (1,-5.75);
\path (1,-8) node [style=sergio]{\large $\mathsf{A}$};
\path (3,-8) node [style=sergio]{$=$};
\filldraw[fill=white, draw=black] (1.5,-6.5)circle (10pt);
\path (1.5,-6.5) node [style=sergio]{$\kappa_x$};
\filldraw[fill=white, draw=black, rounded corners] (5,-8.75)--(7,-8.75)--(7,-7.25)--(5,-7.25)--cycle;
\draw[line width=1pt] (7,-7.5) -- (8.5,-7.5);
\draw[line width=1pt] (5,-8.5) -- (3.5,-8.5);
\draw[line width=1pt] (7,-7.5) -- (8.5,-7.5);
\draw[line width=1pt] (7,-8.5) -- (8.5,-8.5);
\draw[line width=1pt] (5,-7.5) -- (3.5,-7.5);
\filldraw[fill=white, draw=black] (4.25,-8.5)circle (10pt);
\path (4.25,-8.5) node [style=sergio]{$Z$};
\filldraw[fill=white, draw=black] (7.75,-8.5)circle (10pt);
\path (7.75,-8.5) node [style=sergio]{$Z$};
\draw[line width=2pt, color=red] (5.5,-7.25) -- (5.5,-6.75);
\draw[line width=2pt, color=red] (6.5,-7.25) -- (6.5,-6.75);
\draw[line width=2pt, color=red] (6,-7.25) -- (6,-6.75);
\path (6,-8) node [style=sergio]{\large $\mathsf{A}$};
\end{tikzpicture}},
\end{align}
with the transformations on the virtual index reading as
\begin{align}
    V_\sigma = X\otimes I, \quad V_\tau = Z\otimes X, \quad V_\kappa = I\otimes Z,
\end{align}
consistent with the topological invariant $\mu=(-1, -1, +1)$.

\subsection{$\mu=(-1, +1, -1)$}
Now we turn to $\ket{\psi_{\sigma\tau\kappa}^{(-1, +1, -1)}}$, whose MPS representation is constructed as
\begin{align}
\scalebox{0.8}{
\begin{tikzpicture}[scale=0.75]
\tikzstyle{sergio}=[rectangle,draw=none]
\filldraw[fill=white, draw=black, rounded corners] (0,-1.25)--(2,-1.25)--(2,0.25)--(0,0.25)--cycle;
\draw[line width=1pt] (2,0) -- (2.5,0);
\draw[line width=1pt] (0,0) -- (-0.5,0);
\draw[line width=1pt] (2,-1) -- (2.5,-1);
\draw[line width=1pt] (0,-1) -- (-0.5,-1);
\draw[line width=2pt, color=red] (0.5,0.25) -- (0.5,0.75);
\draw[line width=2pt, color=red] (1.5,0.25) -- (1.5,0.75);
\draw[line width=2pt, color=red] (1,0.25) -- (1,0.75);
\path (1,-0.5) node [style=sergio]{\large $\mathsf{A}$};
\draw[line width=1pt] (3.5,0) -- (9.25,0);
\path (4,1.25) node [style=sergio]{$\sigma$};
\draw[line width=2pt, color=red] (4,-1) -- (4,1);
\path (5.75,1.25) node [style=sergio]{$\tau$};
\draw[line width=2pt, color=red] (5.75,0) -- (5.75,1);
\path (7.5,1.25) node [style=sergio]{$\kappa$};
\draw[line width=2pt, color=red] (7.5,-1) -- (7.5,1);
\filldraw[fill=white, draw=black] (5.75,0)circle (4pt);
\path (3,-0.5) node [style=sergio]{$=$};
\filldraw[fill=white, draw=black] (4.5,-0.375)--(5.25,-0.375)--(5.25,0.375)--(4.5,0.375)--cycle;
\path (4.875,0) node [style=sergio]{$H$};
\filldraw[fill=white, draw=black] (6.25,-0.375)--(7,-0.375)--(7,0.375)--(6.25,0.375)--cycle;
\draw[line width=1pt] (3.5,-1) -- (9.25,-1);
\path (6.625,0) node [style=sergio]{$H$};
\filldraw[fill=white, draw=black] (4,0)circle (4pt);
\filldraw[fill=white, draw=black] (7.5,-1)circle (4pt);
\filldraw[fill=white, draw=black] (4.5,-1.375)--(5.25,-1.375)--(5.25,-0.625)--(4.5,-0.625)--cycle;
\path (4.875,-1) node [style=sergio]{$H$};
\filldraw[fill=white, draw=black] (8,-1.375)--(8.75,-1.375)--(8.75,-0.625)--(8,-0.625)--cycle;
\path (8.375,-1) node [style=sergio]{$H$};
\filldraw[fill=white, draw=black] (4,-1)circle (4pt);
\end{tikzpicture}},
\end{align}
whose symmetry transformation is expressed as
\begin{align}
\scalebox{0.8}{
\begin{tikzpicture}[scale=0.75]
\tikzstyle{sergio}=[rectangle,draw=none]
\filldraw[fill=white, draw=black, rounded corners] (0,-5)--(2,-5)--(2,-3.5)--(0,-3.5)--cycle;
\draw[line width=1pt] (2,-3.75) -- (2.5,-3.75);
\draw[line width=1pt] (0,-3.75) -- (-0.5,-3.75);
\draw[line width=1pt] (2,-4.75) -- (2.5,-4.75);
\draw[line width=1pt] (0,-4.75) -- (-0.5,-4.75);
\draw[line width=2pt, color=red] (0.5,-3.5) -- (0.5,-2);
\draw[line width=2pt, color=red] (1.5,-3.5) -- (1.5,-2);
\draw[line width=2pt, color=red] (1,-3.5) -- (1,-2);
\path (1,-4.25) node [style=sergio]{\large $\mathsf{A}$};
\path (3,-4.25) node [style=sergio]{$=$};
\filldraw[fill=white, draw=black] (0.5,-2.75)circle (10pt);
\path (0.5,-2.75) node [style=sergio]{$\sigma_x$};
\filldraw[fill=white, draw=black, rounded corners] (5,-5)--(7,-5)--(7,-3.5)--(5,-3.5)--cycle;
\draw[line width=1pt] (7,-4.75) -- (8.5,-4.75);
\draw[line width=1pt] (5,-4.75) -- (3.5,-4.75);
\filldraw[fill=white, draw=black] (4.25,-4.75)circle (10pt);
\path (4.25,-4.75) node [style=sergio]{$X$};
\filldraw[fill=white, draw=black] (7.75,-4.75)circle (10pt);
\path (7.75,-4.75) node [style=sergio]{$X$};
\draw[line width=1pt] (7,-3.75) -- (8.5,-3.75);
\draw[line width=1pt] (5,-3.75) -- (3.5,-3.75);
\filldraw[fill=white, draw=black] (4.25,-3.75)circle (10pt);
\path (4.25,-3.75) node [style=sergio]{$X$};
\filldraw[fill=white, draw=black] (7.75,-3.75)circle (10pt);
\path (7.75,-3.75) node [style=sergio]{$X$};
\draw[line width=2pt, color=red] (5.5,-3.5) -- (5.5,-3);
\draw[line width=2pt, color=red] (6.5,-3.5) -- (6.5,-3);
\draw[line width=2pt, color=red] (6,-3.5) -- (6,-3);
\path (6,-4.25) node [style=sergio]{\large $\mathsf{A}$};
\end{tikzpicture}},
\end{align}
\begin{align}
\scalebox{0.8}{
\begin{tikzpicture}[scale=0.75]
\tikzstyle{sergio}=[rectangle,draw=none]
\filldraw[fill=white, draw=black, rounded corners] (0,-1.25)--(2,-1.25)--(2,0.25)--(0,0.25)--cycle;
\draw[line width=1pt] (2,0) -- (2.5,0);
\draw[line width=1pt] (0,0) -- (-0.5,0);
\draw[line width=1pt] (2,-1) -- (2.5,-1);
\draw[line width=1pt] (0,-1) -- (-0.5,-1);
\draw[line width=2pt, color=red] (0.5,0.25) -- (0.5,1.75);
\draw[line width=2pt, color=red] (1.5,0.25) -- (1.5,1.75);
\draw[line width=2pt, color=red] (1,0.25) -- (1,1.75);
\path (1,-0.5) node [style=sergio]{\large $\mathsf{A}$};
\path (3,-0.5) node [style=sergio]{$=$};
\filldraw[fill=white, draw=black] (1,1)circle (10pt);
\path (1,1) node [style=sergio]{$\tau_x$};
\filldraw[fill=white, draw=black, rounded corners] (5,-1.25)--(7,-1.25)--(7,0.25)--(5,0.25)--cycle;
\draw[line width=1pt] (7,0) -- (8.5,0);
\draw[line width=1pt] (5,-1) -- (3.5,-1);
\draw[line width=1pt] (7,0) -- (8.5,0);
\draw[line width=1pt] (7,-1) -- (8.5,-1);
\draw[line width=1pt] (5,0) -- (3.5,0);
\filldraw[fill=white, draw=black] (4.25,0)circle (10pt);
\path (4.25,0) node [style=sergio]{$Z$};
\filldraw[fill=white, draw=black] (7.75,0)circle (10pt);
\path (7.75,0) node [style=sergio]{$Z$};
\draw[line width=2pt, color=red] (5.5,0.25) -- (5.5,0.75);
\draw[line width=2pt, color=red] (6.5,0.25) -- (6.5,0.75);
\draw[line width=2pt, color=red] (6,0.25) -- (6,0.75);
\path (6,-0.5) node [style=sergio]{\large $\mathsf{A}$};
\end{tikzpicture}},
\end{align}
\begin{align}
\scalebox{0.8}{
\begin{tikzpicture}[scale=0.75]
\tikzstyle{sergio}=[rectangle,draw=none]
\filldraw[fill=white, draw=black, rounded corners] (0,-8.75)--(2,-8.75)--(2,-7.25)--(0,-7.25)--cycle;
\draw[line width=1pt] (2,-7.5) -- (2.5,-7.5);
\draw[line width=1pt] (0,-7.5) -- (-0.5,-7.5);
\draw[line width=1pt] (2,-8.5) -- (2.5,-8.5);
\draw[line width=1pt] (0,-8.5) -- (-0.5,-8.5);
\draw[line width=2pt, color=red] (0.5,-7.25) -- (0.5,-5.75);
\draw[line width=2pt, color=red] (1.5,-7.25) -- (1.5,-5.75);
\draw[line width=2pt, color=red] (1,-7.25) -- (1,-5.75);
\path (1,-8) node [style=sergio]{\large $\mathsf{A}$};
\path (3,-8) node [style=sergio]{$=$};
\filldraw[fill=white, draw=black] (1.5,-6.5)circle (10pt);
\path (1.5,-6.5) node [style=sergio]{$\kappa_x$};
\filldraw[fill=white, draw=black, rounded corners] (5,-8.75)--(7,-8.75)--(7,-7.25)--(5,-7.25)--cycle;
\draw[line width=1pt] (7,-7.5) -- (8.5,-7.5);
\draw[line width=1pt] (5,-8.5) -- (3.5,-8.5);
\draw[line width=1pt] (7,-7.5) -- (8.5,-7.5);
\draw[line width=1pt] (7,-8.5) -- (8.5,-8.5);
\draw[line width=1pt] (5,-7.5) -- (3.5,-7.5);
\filldraw[fill=white, draw=black] (4.25,-8.5)circle (10pt);
\path (4.25,-8.5) node [style=sergio]{$Z$};
\filldraw[fill=white, draw=black] (7.75,-8.5)circle (10pt);
\path (7.75,-8.5) node [style=sergio]{$Z$};
\draw[line width=2pt, color=red] (5.5,-7.25) -- (5.5,-6.75);
\draw[line width=2pt, color=red] (6.5,-7.25) -- (6.5,-6.75);
\draw[line width=2pt, color=red] (6,-7.25) -- (6,-6.75);
\path (6,-8) node [style=sergio]{\large $\mathsf{A}$};
\end{tikzpicture}}.
\end{align}
Consequently, the virtual representation becomes
\begin{align}
    V_{\sigma} = X\otimes X, \quad V_{\tau} = Z\otimes I, \quad V_{\kappa}=I\otimes Z.
\end{align}

\subsection{$\mu=(+1, -1, -1)$}
The MPS representation for this case reads as
\begin{align}
\scalebox{0.8}{
\begin{tikzpicture}[scale=0.75]
\tikzstyle{sergio}=[rectangle,draw=none]
\filldraw[fill=white, draw=black, rounded corners] (0,-1.25)--(2,-1.25)--(2,0.25)--(0,0.25)--cycle;
\draw[line width=1pt] (2,0) -- (2.5,0);
\draw[line width=1pt] (0,0) -- (-0.5,0);
\draw[line width=1pt] (2,-1) -- (2.5,-1);
\draw[line width=1pt] (0,-1) -- (-0.5,-1);
\draw[line width=2pt, color=red] (0.5,0.25) -- (0.5,0.75);
\draw[line width=2pt, color=red] (1.5,0.25) -- (1.5,0.75);
\draw[line width=2pt, color=red] (1,0.25) -- (1,0.75);
\path (1,-0.5) node [style=sergio]{\large $\mathsf{A}$};
\draw[line width=1pt] (3.5,0) -- (9.25,0);
\path (4,1.25) node [style=sergio]{$\sigma$};
\draw[line width=2pt, color=red] (4,-1) -- (4,1);
\path (5.75,1.25) node [style=sergio]{$\tau$};
\draw[line width=2pt, color=red] (5.75,0) -- (5.75,1);
\path (7.5,1.25) node [style=sergio]{$\kappa$};
\draw[line width=2pt, color=red] (7.5,-1) -- (7.5,1);
\filldraw[fill=white, draw=black] (5.75,0)circle (4pt);
\path (3,-0.5) node [style=sergio]{$=$};
\filldraw[fill=white, draw=black] (8,-0.375)--(8.75,-0.375)--(8.75,0.375)--(8,0.375)--cycle;
\path (8.375,0) node [style=sergio]{$H$};
\filldraw[fill=white, draw=black] (6.25,-0.375)--(7,-0.375)--(7,0.375)--(6.25,0.375)--cycle;
\path (6.625,0) node [style=sergio]{$H$};
\draw[line width=1pt] (3.5,-1) -- (9.25,-1);
\filldraw[fill=white, draw=black] (7.5,0)circle (4pt);
\filldraw[fill=white, draw=black] (7.5,-1)circle (4pt);
\filldraw[fill=white, draw=black] (4.5,-1.375)--(5.25,-1.375)--(5.25,-0.625)--(4.5,-0.625)--cycle;
\path (4.875,-1) node [style=sergio]{$H$};
\filldraw[fill=white, draw=black] (8,-1.375)--(8.75,-1.375)--(8.75,-0.625)--(8,-0.625)--cycle;
\path (8.375,-1) node [style=sergio]{$H$};
\filldraw[fill=white, draw=black] (4,-1)circle (4pt);
\end{tikzpicture}},
\end{align}
with the symmetry transformation
\begin{align}
\scalebox{0.8}{
\begin{tikzpicture}[scale=0.75]
\tikzstyle{sergio}=[rectangle,draw=none]
\filldraw[fill=white, draw=black, rounded corners] (0,-8.75)--(2,-8.75)--(2,-7.25)--(0,-7.25)--cycle;
\draw[line width=1pt] (2,-7.5) -- (2.5,-7.5);
\draw[line width=1pt] (0,-7.5) -- (-0.5,-7.5);
\draw[line width=1pt] (2,-8.5) -- (2.5,-8.5);
\draw[line width=1pt] (0,-8.5) -- (-0.5,-8.5);
\draw[line width=2pt, color=red] (0.5,-7.25) -- (0.5,-5.75);
\draw[line width=2pt, color=red] (1.5,-7.25) -- (1.5,-5.75);
\draw[line width=2pt, color=red] (1,-7.25) -- (1,-5.75);
\path (1,-8) node [style=sergio]{\large $\mathsf{A}$};
\path (3,-8) node [style=sergio]{$=$};
\filldraw[fill=white, draw=black] (0.5,-6.5)circle (10pt);
\path (0.5,-6.5) node [style=sergio]{$\sigma_x$};
\filldraw[fill=white, draw=black, rounded corners] (5,-8.75)--(7,-8.75)--(7,-7.25)--(5,-7.25)--cycle;
\draw[line width=1pt] (7,-7.5) -- (8.5,-7.5);
\draw[line width=1pt] (5,-8.5) -- (3.5,-8.5);
\draw[line width=1pt] (7,-7.5) -- (8.5,-7.5);
\draw[line width=1pt] (7,-8.5) -- (8.5,-8.5);
\draw[line width=1pt] (5,-7.5) -- (3.5,-7.5);
\filldraw[fill=white, draw=black] (4.25,-8.5)circle (10pt);
\path (4.25,-8.5) node [style=sergio]{$X$};
\filldraw[fill=white, draw=black] (7.75,-8.5)circle (10pt);
\path (7.75,-8.5) node [style=sergio]{$X$};
\draw[line width=2pt, color=red] (5.5,-7.25) -- (5.5,-6.75);
\draw[line width=2pt, color=red] (6.5,-7.25) -- (6.5,-6.75);
\draw[line width=2pt, color=red] (6,-7.25) -- (6,-6.75);
\path (6,-8) node [style=sergio]{\large $\mathsf{A}$};
\end{tikzpicture}}.
\end{align}
\begin{align}
\scalebox{0.8}{
\begin{tikzpicture}[scale=0.75]
\tikzstyle{sergio}=[rectangle,draw=none]
\filldraw[fill=white, draw=black, rounded corners] (0,-1.25)--(2,-1.25)--(2,0.25)--(0,0.25)--cycle;
\draw[line width=1pt] (2,0) -- (2.5,0);
\draw[line width=1pt] (0,0) -- (-0.5,0);
\draw[line width=1pt] (2,-1) -- (2.5,-1);
\draw[line width=1pt] (0,-1) -- (-0.5,-1);
\draw[line width=2pt, color=red] (0.5,0.25) -- (0.5,1.75);
\draw[line width=2pt, color=red] (1.5,0.25) -- (1.5,1.75);
\draw[line width=2pt, color=red] (1,0.25) -- (1,1.75);
\path (1,-0.5) node [style=sergio]{\large $\mathsf{A}$};
\path (3,-0.5) node [style=sergio]{$=$};
\filldraw[fill=white, draw=black] (1,1)circle (10pt);
\path (1,1) node [style=sergio]{$\tau_x$};
\filldraw[fill=white, draw=black, rounded corners] (5,-1.25)--(7,-1.25)--(7,0.25)--(5,0.25)--cycle;
\draw[line width=1pt] (7,0) -- (8.5,0);
\draw[line width=1pt] (5,-1) -- (3.5,-1);
\draw[line width=1pt] (7,0) -- (8.5,0);
\draw[line width=1pt] (7,-1) -- (8.5,-1);
\draw[line width=1pt] (5,0) -- (3.5,0);
\filldraw[fill=white, draw=black] (4.25,0)circle (10pt);
\path (4.25,0) node [style=sergio]{$X$};
\filldraw[fill=white, draw=black] (7.75,0)circle (10pt);
\path (7.75,0) node [style=sergio]{$X$};
\draw[line width=2pt, color=red] (5.5,0.25) -- (5.5,0.75);
\draw[line width=2pt, color=red] (6.5,0.25) -- (6.5,0.75);
\draw[line width=2pt, color=red] (6,0.25) -- (6,0.75);
\path (6,-0.5) node [style=sergio]{\large $\mathsf{A}$};
\end{tikzpicture}},
\end{align}
\begin{align}
\scalebox{0.8}{
\begin{tikzpicture}[scale=0.75]
\tikzstyle{sergio}=[rectangle,draw=none]
\filldraw[fill=white, draw=black, rounded corners] (0,-5)--(2,-5)--(2,-3.5)--(0,-3.5)--cycle;
\draw[line width=1pt] (2,-3.75) -- (2.5,-3.75);
\draw[line width=1pt] (0,-3.75) -- (-0.5,-3.75);
\draw[line width=1pt] (2,-4.75) -- (2.5,-4.75);
\draw[line width=1pt] (0,-4.75) -- (-0.5,-4.75);
\draw[line width=2pt, color=red] (0.5,-3.5) -- (0.5,-2);
\draw[line width=2pt, color=red] (1.5,-3.5) -- (1.5,-2);
\draw[line width=2pt, color=red] (1,-3.5) -- (1,-2);
\path (1,-4.25) node [style=sergio]{\large $\mathsf{A}$};
\path (3,-4.25) node [style=sergio]{$=$};
\filldraw[fill=white, draw=black] (1.5,-2.75)circle (10pt);
\path (1.5,-2.75) node [style=sergio]{$\kappa_x$};
\filldraw[fill=white, draw=black, rounded corners] (5,-5)--(7,-5)--(7,-3.5)--(5,-3.5)--cycle;
\draw[line width=1pt] (7,-4.75) -- (8.5,-4.75);
\draw[line width=1pt] (5,-4.75) -- (3.5,-4.75);
\filldraw[fill=white, draw=black] (4.25,-4.75)circle (10pt);
\path (4.25,-4.75) node [style=sergio]{$Z$};
\filldraw[fill=white, draw=black] (7.75,-4.75)circle (10pt);
\path (7.75,-4.75) node [style=sergio]{$Z$};
\draw[line width=1pt] (7,-3.75) -- (8.5,-3.75);
\draw[line width=1pt] (5,-3.75) -- (3.5,-3.75);
\filldraw[fill=white, draw=black] (4.25,-3.75)circle (10pt);
\path (4.25,-3.75) node [style=sergio]{$Z$};
\filldraw[fill=white, draw=black] (7.75,-3.75)circle (10pt);
\path (7.75,-3.75) node [style=sergio]{$Z$};
\draw[line width=2pt, color=red] (5.5,-3.5) -- (5.5,-3);
\draw[line width=2pt, color=red] (6.5,-3.5) -- (6.5,-3);
\draw[line width=2pt, color=red] (6,-3.5) -- (6,-3);
\path (6,-4.25) node [style=sergio]{\large $\mathsf{A}$};
\end{tikzpicture}},
\end{align}
and the virtual representations
\begin{align}
    V_{\sigma} = I\otimes X, \quad V_{\tau} = X\otimes I, \quad V_{\kappa} = Z\otimes Z.
\end{align}

\subsection{$\mu=(-1, -1, -1)$}
Finally, we consider the SPT state that exhibits mixed anomaly between each pair of spins.
The local tensor can be constructed as
\begin{align}
\scalebox{0.8}{
\begin{tikzpicture}[scale=0.75]
\tikzstyle{sergio}=[rectangle,draw=none]
\filldraw[fill=white, draw=black, rounded corners] (0,-2.25)--(2,-2.25)--(2,0.25)--(0,0.25)--cycle;
\draw[line width=1pt] (2,0) -- (2.5,0);
\draw[line width=1pt] (0,0) -- (-0.5,0);
\draw[line width=1pt] (2,-2) -- (2.5,-2);
\draw[line width=1pt] (0,-2) -- (-0.5,-2);
\draw[line width=1pt] (2,-1) -- (2.5,-1);
\draw[line width=1pt] (0,-1) -- (-0.5,-1);
\draw[line width=2pt, color=red] (0.5,0.25) -- (0.5,0.75);
\draw[line width=2pt, color=red] (1.5,0.25) -- (1.5,0.75);
\draw[line width=2pt, color=red] (1,0.25) -- (1,0.75);
\path (1,-1) node [style=sergio]{\large $\mathsf{A}$};
\draw[line width=1pt] (3.5,-1) -- (9.25,-1);
\draw[line width=1pt] (3.5,0) -- (9.25,0);
\draw[line width=1pt] (3.5,-2) -- (9.25,-2);
\path (4,1.25) node [style=sergio]{$\sigma$};
\draw[line width=2pt, color=red] (4,-2) -- (4,1);
\path (5.75,1.25) node [style=sergio]{$\tau$};
\draw[line width=2pt, color=red] (5.75,-1) -- (5.75,1);
\path (7.5,1.25) node [style=sergio]{$\kappa$};
\draw[line width=2pt, color=red] (7.5,-2) -- (7.5,1);
\filldraw[fill=white, draw=black] (4,0)circle (4pt);
\filldraw[fill=white, draw=black] (5.75,0)circle (4pt);
\path (3,-0.5) node [style=sergio]{$=$};
\filldraw[fill=white, draw=black] (4.5,-0.375)--(5.25,-0.375)--(5.25,0.375)--(4.5,0.375)--cycle;
\path (4.875,0) node [style=sergio]{$H$};
\filldraw[fill=white, draw=black] (6.25,-0.375)--(7,-0.375)--(7,0.375)--(6.25,0.375)--cycle;
\path (6.625,0) node [style=sergio]{$H$};
\filldraw[fill=white, draw=black] (5.75,-1)circle (4pt);
\filldraw[fill=white, draw=black] (7.5,-1)circle (4pt);
\filldraw[fill=white, draw=black] (4.5,-2.375)--(5.25,-2.375)--(5.25,-1.625)--(4.5,-1.625)--cycle;
\path (4.875,-2) node [style=sergio]{$H$};
\filldraw[fill=white, draw=black] (8,-1.375)--(8.75,-1.375)--(8.75,-0.625)--(8,-0.625)--cycle;
\path (8.375,-1) node [style=sergio]{$H$};
\filldraw[fill=white, draw=black] (7.5,-2)circle (4pt);
\filldraw[fill=white, draw=black] (4,-2)circle (4pt);
\filldraw[fill=white, draw=black] (6.25,-1.375)--(7,-1.375)--(7,-0.625)--(6.25,-0.625)--cycle;
\path (6.625,-1) node [style=sergio]{$H$};
\filldraw[fill=white, draw=black] (8,-2.375)--(8.75,-2.375)--(8.75,-1.625)--(8,-1.625)--cycle;
\path (8.375,-2) node [style=sergio]{$H$};
\end{tikzpicture}},
\end{align}
constituting an MPS with $D=8$.

The corresponding symmetry transformations include
\begin{align}
\scalebox{0.8}{
\begin{tikzpicture}[scale=0.75]
\tikzstyle{sergio}=[rectangle,draw=none]
\filldraw[fill=white, draw=black, rounded corners] (0,-2.25)--(2,-2.25)--(2,0.25)--(0,0.25)--cycle;
\draw[line width=1pt] (2,0) -- (2.5,0);
\draw[line width=1pt] (0,0) -- (-0.5,0);
\draw[line width=1pt] (2,-2) -- (2.5,-2);
\draw[line width=1pt] (0,-2) -- (-0.5,-2);
\draw[line width=1pt] (2,-1) -- (2.5,-1);
\draw[line width=1pt] (0,-1) -- (-0.5,-1);
\draw[line width=2pt, color=red] (0.5,0.25) -- (0.5,1.75);
\draw[line width=2pt, color=red] (1.5,0.25) -- (1.5,1.75);
\draw[line width=2pt, color=red] (1,0.25) -- (1,1.75);
\path (1,-1) node [style=sergio]{\large $\mathsf{A}$};
\path (3,-0.5) node [style=sergio]{$=$};
\filldraw[fill=white, draw=black] (0.5,1)circle (10pt);
\path (0.5,1) node [style=sergio]{$\sigma_x$};
\filldraw[fill=white, draw=black, rounded corners] (5,-2.25)--(7,-2.25)--(7,0.25)--(5,0.25)--cycle;
\draw[line width=1pt] (7,0) -- (8.5,0);
\draw[line width=1pt] (5,-1) -- (3.5,-1);
\draw[line width=1pt] (7,0) -- (8.5,0);
\draw[line width=1pt] (7,-1) -- (8.5,-1);
\draw[line width=1pt] (5,0) -- (3.5,0);
\draw[line width=1pt] (5,-2) -- (3.5,-2);
\draw[line width=1pt] (7,-2) -- (8.5,-2);
\filldraw[fill=white, draw=black] (4.25,0)circle (10pt);
\path (4.25,0) node [style=sergio]{$X$};
\filldraw[fill=white, draw=black] (7.75,0)circle (10pt);
\path (7.75,0) node [style=sergio]{$X$};
\filldraw[fill=white, draw=black] (4.25,-2)circle (10pt);
\path (4.25,-2) node [style=sergio]{$X$};
\filldraw[fill=white, draw=black] (7.75,-2)circle (10pt);
\path (7.75,-2) node [style=sergio]{$X$};
\draw[line width=2pt, color=red] (5.5,0.25) -- (5.5,0.75);
\draw[line width=2pt, color=red] (6.5,0.25) -- (6.5,0.75);
\draw[line width=2pt, color=red] (6,0.25) -- (6,0.75);
\path (6,-1) node [style=sergio]{\large $\mathsf{A}$};
\end{tikzpicture}},
\end{align}
\begin{align}
\scalebox{0.8}{
\begin{tikzpicture}[scale=0.75]
\tikzstyle{sergio}=[rectangle,draw=none]
\filldraw[fill=white, draw=black, rounded corners] (0,-2.25)--(2,-2.25)--(2,0.25)--(0,0.25)--cycle;
\draw[line width=1pt] (2,0) -- (2.5,0);
\draw[line width=1pt] (0,0) -- (-0.5,0);
\draw[line width=1pt] (2,-2) -- (2.5,-2);
\draw[line width=1pt] (0,-2) -- (-0.5,-2);
\draw[line width=1pt] (2,-1) -- (2.5,-1);
\draw[line width=1pt] (0,-1) -- (-0.5,-1);
\draw[line width=2pt, color=red] (0.5,0.25) -- (0.5,1.75);
\draw[line width=2pt, color=red] (1.5,0.25) -- (1.5,1.75);
\draw[line width=2pt, color=red] (1,0.25) -- (1,1.75);
\path (1,-1) node [style=sergio]{\large $\mathsf{A}$};
\path (3,-0.5) node [style=sergio]{$=$};
\filldraw[fill=white, draw=black] (1,1)circle (10pt);
\path (1,1) node [style=sergio]{$\tau_x$};
\filldraw[fill=white, draw=black, rounded corners] (5,-2.25)--(7,-2.25)--(7,0.25)--(5,0.25)--cycle;
\draw[line width=1pt] (7,0) -- (8.5,0);
\draw[line width=1pt] (5,-1) -- (3.5,-1);
\draw[line width=1pt] (7,0) -- (8.5,0);
\draw[line width=1pt] (7,-1) -- (8.5,-1);
\draw[line width=1pt] (5,0) -- (3.5,0);
\draw[line width=1pt] (5,-2) -- (3.5,-2);
\draw[line width=1pt] (7,-2) -- (8.5,-2);
\filldraw[fill=white, draw=black] (4.25,0)circle (10pt);
\path (4.25,0) node [style=sergio]{$Z$};
\filldraw[fill=white, draw=black] (7.75,0)circle (10pt);
\path (7.75,0) node [style=sergio]{$Z$};
\filldraw[fill=white, draw=black] (4.25,-1)circle (10pt);
\path (4.25,-1) node [style=sergio]{$X$};
\filldraw[fill=white, draw=black] (7.75,-1)circle (10pt);
\path (7.75,-1) node [style=sergio]{$X$};
\draw[line width=2pt, color=red] (5.5,0.25) -- (5.5,0.75);
\draw[line width=2pt, color=red] (6.5,0.25) -- (6.5,0.75);
\draw[line width=2pt, color=red] (6,0.25) -- (6,0.75);
\path (6,-1) node [style=sergio]{\large $\mathsf{A}$};
\end{tikzpicture}},
\end{align}
\begin{align}
\scalebox{0.8}{
\begin{tikzpicture}[scale=0.75]
\tikzstyle{sergio}=[rectangle,draw=none]
\filldraw[fill=white, draw=black, rounded corners] (0,-2.25)--(2,-2.25)--(2,0.25)--(0,0.25)--cycle;
\draw[line width=1pt] (2,0) -- (2.5,0);
\draw[line width=1pt] (0,0) -- (-0.5,0);
\draw[line width=1pt] (2,-2) -- (2.5,-2);
\draw[line width=1pt] (0,-2) -- (-0.5,-2);
\draw[line width=1pt] (2,-1) -- (2.5,-1);
\draw[line width=1pt] (0,-1) -- (-0.5,-1);
\draw[line width=2pt, color=red] (0.5,0.25) -- (0.5,1.75);
\draw[line width=2pt, color=red] (1.5,0.25) -- (1.5,1.75);
\draw[line width=2pt, color=red] (1,0.25) -- (1,1.75);
\path (1,-1) node [style=sergio]{\large $\mathsf{A}$};
\path (3,-0.5) node [style=sergio]{$=$};
\filldraw[fill=white, draw=black] (1.5,1)circle (10pt);
\path (1.5,1) node [style=sergio]{$\kappa_x$};
\filldraw[fill=white, draw=black, rounded corners] (5,-2.25)--(7,-2.25)--(7,0.25)--(5,0.25)--cycle;
\draw[line width=1pt] (7,0) -- (8.5,0);
\draw[line width=1pt] (5,-1) -- (3.5,-1);
\draw[line width=1pt] (7,0) -- (8.5,0);
\draw[line width=1pt] (7,-1) -- (8.5,-1);
\draw[line width=1pt] (5,0) -- (3.5,0);
\draw[line width=1pt] (5,-2) -- (3.5,-2);
\draw[line width=1pt] (7,-2) -- (8.5,-2);
\filldraw[fill=white, draw=black] (4.25,-2)circle (10pt);
\path (4.25,-2) node [style=sergio]{$Z$};
\filldraw[fill=white, draw=black] (7.75,-2)circle (10pt);
\path (7.75,-2) node [style=sergio]{$Z$};
\filldraw[fill=white, draw=black] (4.25,-1)circle (10pt);
\path (4.25,-1) node [style=sergio]{$Z$};
\filldraw[fill=white, draw=black] (7.75,-1)circle (10pt);
\path (7.75,-1) node [style=sergio]{$Z$};
\draw[line width=2pt, color=red] (5.5,0.25) -- (5.5,0.75);
\draw[line width=2pt, color=red] (6.5,0.25) -- (6.5,0.75);
\draw[line width=2pt, color=red] (6,0.25) -- (6,0.75);
\path (6,-1) node [style=sergio]{\large $\mathsf{A}$};
\end{tikzpicture}},
\end{align}
or written in the matrix form as
\begin{align}
V_\sigma = X\otimes I\otimes X,\quad V_\tau=Z\otimes X\otimes I,\quad V_\kappa=I\otimes Z\otimes Z,
\end{align}
leading to the topological invariants of $\mu=(-1, -1, -1)$, where three virtual bonds carry the mixed anomaly between $\Z_2^{\sigma}$ and $\Z_2^{\tau}$, that between $\Z_2^{\tau}$ and $\Z_2^{\kappa}$, and that between $\Z_2^{\kappa}$ and $\Z_2^{\sigma}$, respectively.

\section{Analytical solution for Eq.~\eqref{Equ: DQCP} and \eqref{Equ: H_perturb}}\label{Sec: Appendix-D}
\subsection{Kennedy–Tasaki transformation}
The KT transformation maps a $\Z_2^{\sigma}\times \Z_2^{\tau}$ SPT phase to a $\Z_2^{\sigma} \text{ SSB}\times \Z_2^{\tau} \text{ SSB}$ phase.
Specifically, the KT transformation is defined as~\cite{Li2023}
\begin{align}
    \mathcal{N}^{\rm KT}_{\sigma\tau} = \left[\prod_{i=1}^{N-1}\prod_{l=i+1}^{N}(-1)^{\frac{1}{4}(1-\tau_i^x)(1-\sigma_l^x)}\right](-1)^{\frac{1}{4}(1-\prod_i \sigma_i^x)(1-\prod_i \tau_i^x)},
\end{align}
with the following properties~\cite{Li2025}
\begin{align}
    &\mathcal{N}^{\rm KT}_{\sigma\tau}\sigma_i^x = \sigma_i^x\mathcal{N}^{\rm KT}_{\sigma\tau}, \quad \mathcal{N}^{\rm KT}_{\sigma\tau}\tau_i^x = \tau_i^x\mathcal{N}^{\rm KT}_{\sigma\tau},\\
    &\mathcal{N}^{\rm KT}_{\sigma\tau}\sigma_i^z\tau_i^x\sigma_{i+1}^z = \sigma_i^z\sigma_{i+1}^z \mathcal{N}^{\rm KT}_{\sigma\tau}, \\
    &\mathcal{N}^{\rm KT}_{\sigma\tau}\tau_i^z\sigma_{i+1}^x \tau_{i+1}^z = \tau_i^z \tau_{i+1}^z \mathcal{N}^{\rm KT}_{\sigma\tau}.
\end{align}
Consequently, it maps the Hamiltonian in Eq.~\eqref{Equ: SPT} (with $\sigma$ and $\tau$ spins here) to two ferromagnetic Ising chains as
\begin{align}
    \mathcal{N}^{\rm KT}_{\sigma\tau}H_{\sigma\tau}^{\rm SPT} = H_{\sigma\tau}^{\sigma\text{-SSB},\  \tau\text{-SSB}}\mathcal{N}^{\rm KT}_{\sigma\tau},
\end{align}
where
\begin{align}
    H_{\sigma\tau}^{\sigma\text{-SSB},\  \tau\text{-SSB}} = -\sum_i \sigma_i^z\sigma_{i+1}^z - \sum_i \tau_i^z \tau_{i+1}^z.
\end{align}
Under this transformation, the string order parameter in the SPT phase is mapped to the ferromagnetic correlator in the SSB phase as
\begin{align}
    \mathcal{N}^{\rm KT}_{\sigma\tau} \sigma_i^z\left(\prod_{l=i}^{j-1}\tau_{l}^x\right)\sigma_j^z = \sigma_i^z\sigma_j^z \mathcal{N}^{\rm KT}_{\sigma\tau}
\end{align}

\subsection{Solution for Eq.~\eqref{Equ: DQCP}}
Now we are in the position to solve the Hamiltonian in Eq.~\eqref{Equ: DQCP}.
We first apply a DW duality map $U_{\kappa\sigma}^{\rm DW}$ to remove the mixed anomaly between $\Z_2^{\sigma}$ and $\Z_2^{\kappa}$ that are present in both sides of the Hamiltonian, leading to
\begin{align}
    U_{\kappa\sigma}^{\rm DW} H(\lambda_{\tau\kappa}) = H^{\text{map-1}}(\lambda_{\tau\kappa}) U_{\kappa\sigma}^{\rm DW},
\end{align}
where
\begin{align}
    H^{\text{map-1}}(\lambda_{\tau\kappa}) = (1-\lambda_{\tau\kappa})H_{\sigma\tau\kappa}^{(+1, +1, +1)} + \lambda_{\tau\kappa} H_{\sigma\tau\kappa}^{(+1, -1, +1)}.
\end{align}
Next, we perform the KT transformation $\mathcal{N}^{\rm KT}_{\tau\kappa}$ to remove the SPT order of $\Z_2^{\tau}\times \Z_2^{\kappa}$ in $H_{\sigma\tau\kappa}^{(+1, -1, +1)}$, resulting in
\begin{align}
\begin{aligned}
    \mathcal{N}^{\rm KT}_{\tau\kappa} H_{\sigma\tau\kappa}^{(+1, +1, +1)} & = \mathcal{N}^{\rm KT}_{\tau\kappa} \left[-\sum_{i}\left(\sigma_i^x+\tau_i^x+\kappa_i^x\right)\right]\\
    & = \left[-\sum_{i}\left(\sigma_i^x+\tau_i^x+\kappa_i^x\right)\right] \mathcal{N}^{\rm KT}_{\tau\kappa},
\end{aligned}
\end{align}
and
\begin{align}
\begin{aligned}
    \mathcal{N}^{\rm KT}_{\tau\kappa} H_{\sigma\tau\kappa}^{(+1, -1, +1)} & = \mathcal{N}^{\rm KT}_{\tau\kappa} \left[-\sum_{i}\left(\sigma_i^x+\kappa_{i-1}^z\tau_{i}^x\kappa_{i}^z+\tau_i^z\kappa_i^x\tau_{i+1}^z\right)\right]\\
    & = \left[-\sum_{i}\left(\sigma_i^x+\kappa_{i-1}^z\kappa_{i}^z+\tau_i^z\tau_{i+1}^z\right)\right] \mathcal{N}^{\rm KT}_{\tau\kappa}.
\end{aligned}
\end{align}
Finally, we obtain the mapped Hamiltonian as
\begin{align}
    \mathcal{N}^{\rm KT}_{\tau\kappa} H^{\text{map-1}}(\lambda_{\tau\kappa}) = H^{\text{map-2}}(\lambda_{\tau\kappa}) \mathcal{N}^{\rm KT}_{\tau\kappa},
\end{align}
where
\begin{align}
    H^{\text{map-2}}(\lambda_{\tau\kappa}) =& -\sum_i\left[\sigma_i^x + (1-\lambda_{\tau\kappa})(\tau_i^x+\kappa_ i^x) \right. \notag \\
    &+ \left. \lambda_{\tau\kappa}(\tau_i^z\tau_{i+1}^z + \kappa_i^z\kappa_{i+1}^z)\right].
\end{align}
This Hamiltonian has two decoupled transverse-field Ising chains, with a shared critical point at $\lambda_{\tau\kappa} = 0.5$.
Equivalently, the $\Z_2^{\kappa}$ symmetric phase can be viewed as a SSB phase on the dual lattices, and the critical point at $\lambda_{c} = 0.5$ corresponds to the transition between the dual-$\Z_2^{\kappa}$ SSB phase ($\lambda_{\tau\kappa} < 0.5$) and the $\Z_2^{\tau}$ SSB phase ($\lambda_{\tau\kappa} > 0.5$).

Next, we map the order parameters to describe these two SSB phases back to the original model.
The order parameter for the dual-$\Z_2^{\kappa}$ SSB phase is given by $\prod_{l=i}^{j-1}\kappa_l^x$, while the order parameter for the $\Z_2^{\tau}$ SSB phase is given by $\tau_i^z\tau_j^z$.
Under the KT transformation, these two order parameters are mapped to
\begin{align}
    & \prod_{l=i}^{j-1}\kappa_l^x \mathcal{N}^{\rm KT}_{\tau\kappa} = \mathcal{N}^{\rm KT}_{\tau\kappa} \prod_{l=i}^{j-1}\kappa_l^x\\
    & \tau_i^z\tau_j^z \mathcal{N}^{\rm KT}_{\tau\kappa} = \mathcal{N}^{\rm KT}_{\tau\kappa} \tau_i^z\left(\prod_{l=i}^{j-1}\kappa_l^x\right)\tau_j^z.
\end{align}
Subsequently, under the DW duality transformation, these two order parameters are further mapped to
\begin{align}
    & \prod_{l=i}^{j-1}\kappa_l^x U_{\kappa\sigma}^{\rm DW} = U_{\kappa\sigma}^{\rm DW} \sigma_i^z\left(\prod_{l=i}^{j-1}\kappa_l^x\right)\sigma_j^z,\\
    & \tau_i^z\left(\prod_{l=i}^{j-1}\kappa_l^x\right)\tau_j^z U_{\kappa\sigma}^{\rm DW} = U_{\kappa\sigma}^{\rm DW} \sigma_i^z\tau_i^z\left(\prod_{l=i}^{j-1}\kappa_l^z\right)\sigma_j^z\tau_j^z.
\end{align}
These two order parameters correspond to the string order parameters $S(\sigma^z, \kappa^x, i, j)$ and $S(\sigma^z\tau^z, \kappa^x, i, j)$ in the purified model Eq.~\eqref{Equ: DQCP}, respectively.
After tracing out $\kappa$ spins, they further correspond to the order parameters in Eq.~\eqref{Equ: Order Parameter} for the reduced mixed states, characterizing the transitions between two SWSSB phases.
In summary, the order parameters in the original model, the mapped model, and the reduced mixed states are
\begin{align}
    \begin{tabular}{ccc}\hline
        & $\lambda_{\tau\kappa} < 0.5$ & $\lambda_{\tau\kappa} > 0.5$ \\
        \hline
        $H^{\text{map-2}}(\lambda_{\tau\kappa})$ & $\prod_{l=i}^{j-1}\kappa_l^x$ & $\tau_i^z\tau_j^z$ \\
        $H^{\text{map-1}}(\lambda_{\tau\kappa})$ & $\prod_{l=i}^{j-1}\kappa_l^x$ & $\tau_i^z\left(\prod_{l=i}^{j-1}\kappa_l^x\right)\tau_j^z$ \\
        $H(\lambda_{\tau\kappa})$ & \,$\sigma_i^z\left(\prod_{l=i}^{j-1}\kappa_l^x\right)\sigma_j^z$\, & \,$\sigma_i^z\tau_i^z\left(\prod_{l=i}^{j-1}\kappa_l^z\right)\sigma_j^z\tau_j^z$ \\
        Mixed state & \,$\mathcal{C}^{(2)}(\sigma^z, i, j)$\, & \,$\mathcal{C}^{(2)}(\sigma^z\tau^z, i, j)$ \\
        \hline
    \end{tabular},
\end{align}
where the transition between the dual-$\Z_2^{\kappa}$ SSB phase and the $\Z_2^{\tau}$ SSB phase in the mapped model corresponds to the transition between the $\sigma$-SWSSB phase and the $\sigma\tau$-preserved SWSSB phase in the reduced mixed state of the original model in Eq.~\eqref{Equ: DQCP}.

\subsection{Solution for Eq.~\eqref{Equ: H_perturb}}
The Hamiltonian in Eq.~\eqref{Equ: H_perturb} can be solved similarly.
The sequential application of $U_{\kappa\sigma}^{\rm DW}$ and $\mathcal{N}^{\rm KT}_{\tau\kappa}$ leads to
\begin{align}
    \mathcal{N}^{\rm KT}_{\tau\kappa} U_{\kappa\sigma}^{\rm DW} H_{\rm Perturb}(\lambda_{\tau\kappa}) = H_{\rm Perturb}^{\text{map-2}}(\lambda_{\tau\kappa}) \mathcal{N}^{\rm KT}_{\tau\kappa} U_{\kappa\sigma}^{\rm DW},
\end{align}
where only the transverse field of $\tau$ spins is amplified, i.e.,
\begin{align}
    H_{\rm Perturb}^{\text{map-2}}(\lambda_{\tau\kappa})
    =& -\sum_i\left[\sigma_i^x + (1-\lambda_{\tau\kappa})(2\tau_i^x+\kappa_i^x)\right. \notag \\
    &+\left. \lambda_{\tau\kappa}(\tau_i^z\tau_{i+1}^z + \kappa_i^z\kappa_{i+1}^z)\right].
\end{align}
The critical point for $\Z_2^{\tau}$ SSB transition is thus shifted to $\lambda_c = 2/3$.
Consequently, the critical point between the trivial and $\sigma\tau$-preserved SWSSB phases is also shifted to $\lambda_c = 2/3$ in the reduced mixed state of the perturbed model in Eq.~\eqref{Equ: H_perturb}, which is consistent with the numerical results shown in Fig.~\ref{Fig: Perturb}.
On the other hand, the critical point for $\sigma$-SWSSB to trivial transition remains unchanged at $\lambda_c = 0.5$, with the corresponding order parameter $M_{\sigma}^{(2)}$ remaining unchanged.

\section{Numerical calculation of topological invariants $\mu$}\label{Sec: Appendix-E}
In this section, we discuss how to numerically calculate the topological invariants $\mu$ for a given MPS representation of an SPT state.
Given an MPS representation of an SPT state, the central task is to calculate the virtual representations $V_k$ of the symmetry group $K$.
To achieve this, we can follow the procedure described in Ref.~\cite{Pollmann2012}.
First, we construct the transfer matrix $\mathbb{E}$ of the MPS, which is defined as
\begin{align}
\scalebox{0.8}{
\begin{tikzpicture}[scale=0.75]
\tikzstyle{sergio}=[rectangle,draw=none]
\filldraw[fill=white, draw=black, rounded corners] (-0.25,-0.5)--(1.25,-0.5)--(1.25,0.5)--(-0.25,0.5)--cycle;
\filldraw[fill=white, draw=black, rounded corners] (-0.25,1.5)--(1.25,1.5)--(1.25,2.5)--(-0.25,2.5)--cycle;
\draw[line width=1pt] (-0.25,0) -- (-0.75,0);
\draw[line width=1pt] (1.25,2) -- (1.75,2);
\draw[line width=1pt] (1.25, 0) -- (1.75,0);
\draw[line width=1pt] (-0.25,2) -- (-0.75,2);
\draw[line width=2pt, color=red] (0.5,0.5) -- (0.5,1.5);
\path (0.5,0) node [style=sergio]{\large $\mathsf{A}^{*}$};
\path (0.5,2) node [style=sergio]{\large $\mathsf{A}$};
\path (-1.5,1) node [style=sergio]{\large $\mathbb{E}\,=$};
\end{tikzpicture}.}
\end{align}
The dominant eigenvector of the transfer matrix $\mathbb{E}$ determines the fixed point of the MPS, where the normalization condition requires the dominant eigenvalue to be equal to 1, i.e., $\mathbb{E}r = r$.
The corresponding tensor equation is graphically represented as
\begin{align}
\scalebox{0.8}{
\begin{tikzpicture}[scale=0.75]
\tikzstyle{sergio}=[rectangle,draw=none]
\filldraw[fill=white, draw=black, rounded corners] (-0.25,-0.5)--(1.25,-0.5)--(1.25,0.5)--(-0.25,0.5)--cycle;
\filldraw[fill=white, draw=black, rounded corners] (-0.25,1.5)--(1.25,1.5)--(1.25,2.5)--(-0.25,2.5)--cycle;
\draw[line width=1pt] (-0.25,0) -- (-0.75,0);
\draw[line width=1pt] (1.25,2) -- (1.75,2) -- (1.75,0) -- (1.25, 0);
\filldraw[fill=white, draw=black] (1.75,1)circle (10pt);
\path (1.75,1) node [style=sergio]{$r$};
\draw[line width=1pt] (-0.25,2) -- (-0.75,2);
\draw[line width=2pt, color=red] (0.5,0.5) -- (0.5,1.5);
\path (0.5,0) node [style=sergio]{\large $\mathsf{A}^{*}$};
\path (0.5,2) node [style=sergio]{\large $\mathsf{A}$};
\path (2.5,1) node [style=sergio]{\large $=$};
\draw[line width=1pt] (3,2) -- (3.5,2) -- (3.5,0) -- (3,0);
\filldraw[fill=white, draw=black] (3.5,1)circle (10pt);
\path (3.5,1) node [style=sergio]{$r$};
\end{tikzpicture}.}
\end{align}
Next, we insert the symmetry transformation $U_k$ on the (internal) physical index of the transfer matrix, leading to a new transfer matrix $\mathbb{E}_k$ as
\begin{align}
\scalebox{0.8}{
\begin{tikzpicture}[scale=0.75]
\tikzstyle{sergio}=[rectangle,draw=none]
\filldraw[fill=white, draw=black, rounded corners] (-0.25,-0.5)--(1.25,-0.5)--(1.25,0.5)--(-0.25,0.5)--cycle;
\filldraw[fill=white, draw=black, rounded corners] (-0.25,2)--(1.25,2)--(1.25,3)--(-0.25,3)--cycle;
\draw[line width=1pt] (-0.25,0) -- (-0.75,0);
\draw[line width=1pt] (1.25,2.5) -- (1.75,2.5);
\draw[line width=1pt] (1.25, 0) -- (1.75,0);
\draw[line width=1pt] (-0.25,2.5) -- (-0.75,2.5);
\draw[line width=2pt, color=red] (0.5,0.5) -- (0.5,2);
\path (0.5,0) node [style=sergio]{\large $\mathsf{A}^{*}$};
\path (0.5,2.5) node [style=sergio]{\large $\mathsf{A}$};
\path (-1.75,1.25) node [style=sergio]{\large $\mathbb{E}_k\,=$};
\filldraw[fill=white, draw=black] (0.5,1.25)circle (10pt);
\path (0.5,1.25) node [style=sergio]{$U_k$};
\end{tikzpicture}.}
\end{align}
By instituting the symmetry transformation in Eq.~\eqref{Equ: Symmetry_k}, we can derive the following equation
\begin{align}
\scalebox{0.8}{
\begin{tikzpicture}[scale=0.75]
\tikzstyle{sergio}=[rectangle,draw=none]
\filldraw[fill=white, draw=black, rounded corners] (-0.25,-0.5)--(1.25,-0.5)--(1.25,0.5)--(-0.25,0.5)--cycle;
\filldraw[fill=white, draw=black, rounded corners] (-0.25,2)--(1.25,2)--(1.25,3)--(-0.25,3)--cycle;
\draw[line width=1pt] (-0.25,0) -- (-0.75,0);
\draw[line width=1pt] (1.25,2.5) -- (1.75,2.5);
\draw[line width=1pt] (1.25, 0) -- (1.75,0);
\draw[line width=1pt] (-0.25,2.5) -- (-0.75,2.5);
\draw[line width=2pt, color=red] (0.5,0.5) -- (0.5,2);
\path (0.5,0) node [style=sergio]{\large $\mathsf{A}^{*}$};
\path (0.5,2.5) node [style=sergio]{\large $\mathsf{A}$};
\filldraw[fill=white, draw=black] (0.5,1.25)circle (10pt);
\path (0.5,1.25) node [style=sergio]{$U_k$};
\filldraw[fill=white, draw=black, rounded corners] (4.25,-0.5)--(5.75,-0.5)--(5.75,0.5)--(4.25,0.5)--cycle;
\filldraw[fill=white, draw=black, rounded corners] (4.25,2)--(5.75,2)--(5.75,3)--(4.25,3)--cycle;
\draw[line width=1pt] (4.25,0) -- (2.75,0);
\draw[line width=1pt] (5.75,2.5) -- (7.25,2.5);
\draw[line width=1pt] (5.75,0) -- (7.25,0);
\draw[line width=1pt] (4.25,2.5) -- (2.75,2.5);
\draw[line width=2pt, color=red] (5,0.5) -- (5,2);
\path (5,0) node [style=sergio]{\large $\mathsf{A}^{*}$};
\path (5,2.5) node [style=sergio]{\large $\mathsf{A}$};
\path (2.25,1.25) node [style=sergio]{\large $=$};
\filldraw[fill=white, draw=black] (6.5,2.5)circle (10pt);
\path (6.5,2.5) node [style=sergio]{$V_k$};
\filldraw[fill=white, draw=black] (3.5,2.5)circle (10pt);
\path (3.5,2.5) node [style=sergio]{$V_k^{-1}$};
\end{tikzpicture}.}
\end{align}
It means that these two transfer matrices $\mathbb{E}_k$ and $\mathbb{E}$ are related by a similarity transformation
\begin{align}
    \mathbb{E}_k = (V_k^{-1}\otimes I) \mathbb{E} (V_k \otimes I).
\end{align}
Therefore, the dominant eigenvector of $\mathbb{E}_k$ is given by $r_k = (V_k^{-1}\otimes I) r$, which satisfies $\mathbb{E}_k r_k = r_k$.
This relation can be graphically verified as
\begin{align}
\scalebox{0.8}{
\begin{tikzpicture}[scale=0.75]
\tikzstyle{sergio}=[rectangle,draw=none]
\filldraw[fill=white, draw=black, rounded corners] (-0.25,-0.5)--(1.25,-0.5)--(1.25,0.5)--(-0.25,0.5)--cycle;
\filldraw[fill=white, draw=black, rounded corners] (-0.25,2)--(1.25,2)--(1.25,3)--(-0.25,3)--cycle;
\draw[line width=1pt] (-0.25,0) -- (-0.75,0);
\draw[line width=1pt] (1.25,2.5) -- (2.75,2.5) -- (2.75,0) -- (1.25,0);
\draw[line width=1pt] (-0.25,2.5) -- (-0.75,2.5);
\draw[line width=2pt, color=red] (0.5,0.5) -- (0.5,2);
\path (0.5,0) node [style=sergio]{\large $\mathsf{A}^{*}$};
\path (0.5,2.5) node [style=sergio]{\large $\mathsf{A}$};
\filldraw[fill=white, draw=black] (0.5,1.25)circle (10pt);
\path (0.5,1.25) node [style=sergio]{$U_k$};
\filldraw[fill=white, draw=black] (2.75,1.25)circle (10pt);
\path (2.75,1.25) node [style=sergio]{$r$};
\filldraw[fill=white, draw=black, rounded corners] (5.5,-0.5)--(7,-0.5)--(7,0.5)--(5.5,0.5)--cycle;
\filldraw[fill=white, draw=black, rounded corners] (5.5,2)--(7,2)--(7,3)--(5.5,3)--cycle;
\draw[line width=1pt] (5.5,0) -- (4,0);
\draw[line width=1pt] (7,2.5) -- (7.5,2.5) -- (7.5,0) -- (7,0);
\draw[line width=1pt] (5.5,2.5) -- (4,2.5);
\draw[line width=2pt, color=red] (6.25,0.5) -- (6.25,2);
\path (6.25,0) node [style=sergio]{\large $\mathsf{A}^{*}$};
\path (6.25,2.5) node [style=sergio]{\large $\mathsf{A}$};
\filldraw[fill=white, draw=black] (2,2.5)circle (10pt);
\path (2,2.5) node [style=sergio]{$V_k^{-1}$};
\filldraw[fill=white, draw=black] (4.75,2.5)circle (10pt);
\path (4.75,2.5) node [style=sergio]{$V_k^{-1}$};
\filldraw[fill=white, draw=black] (7.5,1.25)circle (10pt);
\path (7.5,1.25) node [style=sergio]{$r$};
\draw[line width=1pt] (8.75,2.5) -- (10.25,2.5) -- (10.25,0) -- (8.75,0);
\filldraw[fill=white, draw=black] (10.25,1.25)circle (10pt);
\path (10.25,1.25) node [style=sergio]{$r$};
\filldraw[fill=white, draw=black] (9.5,2.5)circle (10pt);
\path (9.5,2.5) node [style=sergio]{$V_k^{-1}$};
\path (8.25,1.3125) node [style=sergio]{\large $=$};
\path (3.5,1.25) node [style=sergio]{\large $=$};
\end{tikzpicture},}
\end{align}
i.e., $(V_k^{-1}\otimes I) r$ is indeed the eigenvector of $\mathbb{E}_k$ with unit eigenvalue.
Therefore, the virtual representation $V_k$ satisfies the following relation
\begin{align}
\scalebox{0.8}{
\begin{tikzpicture}[scale=0.75]
\tikzstyle{sergio}=[rectangle,draw=none]
\draw[line width=1pt] (1,2.5) -- (3.625,2.5);
\filldraw[fill=white, draw=black] (2.875,2.5)circle (10pt);
\path (2.875,2.5) node [style=sergio]{$r$};
\filldraw[fill=white, draw=black] (1.75,2.5)circle (10pt);
\path (1.75,2.5) node [style=sergio]{$V_k^{-1}$};
\path (4,2.5) node [style=sergio]{\large $=$};
\draw[line width=1pt] (4.375,2.5) -- (5.875,2.5);
\filldraw[fill=white, draw=black] (5.125,2.5)circle (10pt);
\path (5.125,2.5) node [style=sergio]{$r_k$};
\draw[line width=1pt] (7,2.5) -- (8.5,2.5);
\filldraw[fill=white, draw=black] (7.75,2.5)circle (10pt);
\path (7.75,2.5) node [style=sergio]{$V_k$};
\path (8.875,2.5) node [style=sergio]{\large $=$};
\draw[line width=1pt] (9.25,2.5) -- (11.875,2.5);
\filldraw[fill=white, draw=black] (10,2.5)circle (10pt);
\path (10,2.5) node [style=sergio]{$r$};
\filldraw[fill=white, draw=black] (11.125,2.5)circle (10pt);
\path (11.125,2.5) node [style=sergio]{$r_k^{-1}$};
\path (6.5,2.5) node [style=sergio]{\large $\Rightarrow$};
\end{tikzpicture}.}
\end{align}
In this way, we can numerically obtain the virtual representation $V_k$ for each symmetry transformation $U_k$ by solving the eigenvalue problems of both $\mathbb{E}$ and $\mathbb{E}_k$.
Finally, we can calculate the topological invariants $\mu$ by evaluating the commutation relations between these virtual representations $V_k$ as shown in Eqs.~\eqref{Equ: TI-1}-\eqref{Equ: TI-3}.
\end{document}